\documentclass[12pt]{article}
\usepackage{graphicx}
\usepackage{amsmath}
\usepackage{amsfonts}
\usepackage{amsthm}
\usepackage{enumerate,setspace}
\usepackage{xr}
\usepackage{bbm,comment}
\usepackage{authblk}
\usepackage{dcolumn,booktabs}
\usepackage{siunitx}
\newcommand\mc[1]{\multicolumn{1}{c}{#1}}
\usepackage[natbibapa]{apacite}
\usepackage{listings}
\DeclareMathOperator{\diag}{diag}
\newtheorem{result}{Result}
\usepackage{xcolor}
\usepackage{hyperref}
\usepackage{enumitem}
\usepackage{lstbayes}

\DeclareMathOperator{\tr}{tr}

\newcommand{\beginsupplement}{%
        \setcounter{section}{0}
        \renewcommand{\thesection}{S\arabic{section}}%
        \setcounter{table}{0}
        \renewcommand{\thetable}{S\arabic{table}}%
        \setcounter{figure}{0}
        \renewcommand{\thefigure}{S\arabic{figure}}%
     }

\lstdefinestyle{mystyle}{
    backgroundcolor=\color{backcolour},   
    commentstyle=\color{codegreen},
    keywordstyle=\color{funcolor},
    numberstyle=\tiny\color{codegray},
    stringstyle=\color{codepurple},
    basicstyle=\ttfamily\footnotesize,
    breakatwhitespace=false,         
    breaklines=false,                 
    captionpos=b,                    
    keepspaces=true,                    
    numbersep=5pt,                  
    showspaces=false,                
    showstringspaces=false,
    showtabs=false,                  
    tabsize=2
}

\newcommand{\blind}{0}

\definecolor{comcolor}{rgb}{0,0.5,0}
\definecolor{funcolor}{rgb}{0,0.4,1}
\definecolor{concolor}{rgb}{0,0,1}

\definecolor{codegreen}{rgb}{0,0.6,0}
\definecolor{codegray}{rgb}{0.5,0.5,0.5}
\definecolor{codepurple}{rgb}{0.58,0,0.82}
\definecolor{backcolour}{rgb}{0.93,0.93,0.93}

\begin{document}

\def\spacingset#1{\renewcommand{\baselinestretch}%
{#1}\small\normalsize} \spacingset{1}

\if0\blind
{
  \title{\bf Fast Bayesian Functional Principal Components Analysis}
  \author[1]{Joseph Sartini \thanks{The DASH4D Trial is funded by a grant from the Sheikh Khalifa Stroke Institute at Johns Hopkins University School of Medicine. The DASH4D-CGM study is funded by NIH/NIDDK grant R01 DK128900. Dr.~Selvin was also supported by NIH/NHLBI grant K24 HL152440. Mr.~Sartini was supported by Grant Number T32 HL007024 from the NIH/NHLBI. The content is solely the responsibility of the authors and does not necessarily represent the official views of the National Institutes of Health. Abbott Diabetes Care provided continuous glucose monitoring systems for this investigator-initiated research.}}
  \author[1]{Xinkai Zhou}
  \author[2]{Liz Selvin}
  \author[1]{Scott Zeger}
  \author[1]{Ciprian M. Crainiceanu}
  \affil[1]{Department of Biostatistics, Johns Hopkins University}
  \affil[2]{Department of Epidemiology, Johns Hopkins University}
  \maketitle
} \fi

\if1\blind
{
  \bigskip
  \bigskip
  \bigskip
  \begin{center}
    {\LARGE\bf Fast Bayesian Functional Principal Components Analysis}
\end{center}
  \medskip
} \fi

\bigskip

\begin{abstract}
Functional Principal Components Analysis (FPCA) is a widely used analytic tool for dimension reduction of functional data. Traditional implementations of FPCA estimate the principal components from the data, then treat these estimates as fixed in subsequent analyses. To account for the uncertainty of PC estimates, we propose FAST,  a fully-Bayesian FPCA with three core components: (1) projection of eigenfunctions onto an orthonormal spline basis; (2) efficient sampling of the orthonormal spline coefficient matrix using a parameter expansion scheme based on polar decomposition; and (3) ordering eigenvalues during sampling. Extensive simulation studies show that FAST is very stable and performs better compared to existing methods. FAST is motivated by and applied to a study of the variability in mealtime glucose from the Dietary Approaches to Stop Hypertension for Diabetes Continuous Glucose Monitoring (DASH4D CGM) study. All relevant STAN code and simulation routines are available as supplementary material.
\end{abstract}

\noindent%
\vfill

\newpage
\spacingset{1.2} 

\section{Introduction}\label{subsec:innovation}

Functional principal component analysis (FPCA) is a common analytic approach for high-dimensional applications. FPCA approximates the covariance between all possible pairs of observations by identifying a small set of orthonormal principal components (FPCs). Using FPCs to provide a reduced-rank representation of functional data, rather than fixed (spline) bases \citep{shi_analysis_1996, rice_nonparametric_2001}, provides a data-driven representation based on the leading modes of variability \citep{james_principal_2000}. While the means of estimating these FPCs can vary, e.g. polynomial smoothing \citep{yao_functional_2005}, kernel smoothing \citep{staniswalis_nonparametric_1998}, or sandwich covariance smoothing \citep{xiao_fast_2016}, the key idea remains consistent. Conditioning on the FPCs, functional  models become  mixed effects models with a small number of random effects and inference can be conducted using existing software \citep{reiss_functional_2007, liu_functional_2012,di_multilevel_2009, zhou_reduced_2010, shou_structured_2015}. Frequentist implementations of FPCA follow a two-step approach: (1) diagonalizing a smooth estimator of the observed functional covariance; and then (2) conducting  inference conditional on the FPC estimates \citep{xiao_fast_2016, crainiceanu_bayesian_2010, peng_geometric_2009}. This conditional approach may lead to invalid inference in smaller samples and/or when eigenfunctions explain smaller variance \citep{goldsmith_PCA_2013}. FPC forms may be falsely identified, or FPC scores with substantial uncertainty may be used in downstream regression.

There are several Bayesian FPCA implementations that model jointly  the FPCs and their scores. An early fully-Bayesian FPCA was developed by \cite{goldsmith_generalized_2015}, who modeled both the fixed and random effects functions using unconstrained splines and applied post-processing to obtain orthonormal FPC samples. \cite{jauch_monte_2021} directly sampled the FPC matrix using a parameter expansion scheme based on polar decomposition. \cite{suarez_bayesian_2017} proposed an empirical Bayes approach based on spline expansion which models the coefficient covariance structure. A recent approach used variational inference based on message passing \citep{nolan_bayesian_2023}. \cite{boland_central_2023} implemented Bayesian FPCA using latent factor methods, also providing useful visualization tools for posterior FPC uncertainty. In the context of sparse functional data, \cite{ye_functional_2024} projected the FPCs onto an orthonormal spline basis and used Gibbs sampling approach of \cite{hoff_simulation_2009} to sample the spline weights such that FPC orthonormality was preserved. Recently \cite{zhang_robust_2025}  proposed a robust implementation based on sequential Monte Carlo designed to handle outliers and sparse observations. 

In this paper, we introduce Fast Bayesian FPCA, or FAST for short, which fuses ideas introduced \cite{ye_functional_2024} and \cite{jauch_monte_2021} and expands them to create a self-contained methodology that is computationally stable, scalable, and extensible to a variety of other data structures and  models. FAST leverages orthogonal spline expansion to reduce the dimension of sampled FPC parameters to the corresponding weights, which are then efficiently sampled using a parameter expansion scheme based on polar decomposition. The dimensionality reduction of the posterior space by spline expansion combined with stabilizing order constraints of FPCA ensures that the approach has excellent convergence and mixing properties while being computationally efficient. All core components of FAST are easy to implement in  Bayesian software such as {\ttfamily STAN} \citep{carpenter_stan_2017} and can be extended to include covariates, multilevel, sparse, and multivariate data.

This research is motivated by the Dietary Approaches to Stop Hypertension for Diabetes Continuous Glucose Monitoring (DASH4D CGM) clinical trial. DASH4D CGM was a crossover feeding trial which included continuous glucose monitoring (CGM) during meals that were controlled and provided by the study. Participants visited the study site three times per week to provide blood samples and receive their assigned meals, all while wearing a CGM device. When participants consumed a meal at the study center, study staff recorded the start times. Here we focus on CGM data collected after the observed meals, with dysregulation of the corresponding glycemic response being known to be associated with adverse cardiovascular outcomes \citep{hershon_importance_2019, garber_postprandial_2012}. 

Analyses of postprandial CGM glucose with verified meal times and composition are rare, and previous studies have not considered the continuous functional measurements obtained from CGM \citep{barua_discordance_2022,rohling_determination_2019}. This paper introduces novel statistical methods designed to quantify the five-hour CGM curves from one hour before to four hours after each meal. The main results of the trial are reported elsewhere \citep{fang_dash4d_2025, pilla_dietary_2025}, so this paper focuses on the variability of the CGM trajectories by diet and not on treatment effects. Specifically, it applies FPCA to the CGM curves aggregated by participant and Multilevel FPCA (MFPCA) to the individual meal response curves. The number of participants in this study is relatively small because of participant burden and data collection cost, which may result in substantial FPCA uncertainty.

The rest of the paper is organized as follows. Section~\ref{sec:methods} presents the modeling approach for FAST. Section~\ref{section:STAN} provides the {\ttfamily STAN} code. Section~\ref{sec:simulation} compares FAST with existing implementations using simulation. Section~\ref{sec:data_analysis} demonstrates the utility of FAST in estimating FPCA for the motivating data from the DASH4D clinical trial, with appropriate uncertainty quantification. Section~\ref{sec:discussion} provides a short discussion.

\section{Methods}\label{sec:methods}

\subsection{FPCA model}\label{subsec:model}
The functional data structure has the form $Y_i(t), i=1,\ldots, N$ for $t \in \{t_1, \ldots, t_M\}\in [0,1]$, which can be thought of as functions observed on a finite grid. Assuming that $Y_i(t)$ are generated by the same Gaussian Process and truncating the decomposition  \citep{karhunen_uber_1947,kosambi_statistics_1943,  loeve_probability_1978} produces the standard FPCA model: 
\begin{equation}
Y_{i}(t) = \mu(t) + \sum_{k = 1}^\infty \xi_{ik}\phi_k(t) \approx \mu(t) + \sum_{k = 1}^K \xi_{ik}\phi_k(t) + \epsilon_i(t)\;. \label{EQ:Defn}
\end{equation}
Here $\mu(t)$ is the population mean, $\phi_k(t) \in L^2([0,1])$ for $k=1,\ldots,K$ are orthonormal eigenfunctions of the covariance operator of $Y_i(t)$, $\xi_{ik}\sim N(0, \lambda_k)$, $\lambda_k$ are the eigenvalues corresponding to $\phi_k(t)$, $\epsilon_i(t) \sim N(0, \sigma^2)$ is the error process, and $\epsilon_i(t)$ and $\xi_{ik}$ are mutually independent over $i$ and $k$.

For this paper, we fix the number of FPCs, $K$, which controls the rank of the smooth covariance operator. Choosing $K$ is a difficult problem discussed extensively in the literature \citep{crainiceanu2024book, li_selecting_2013, yao_functional_2005}. For the purpose of this paper,  $K$ is chosen based on the proportion of total variability explained. This can either be estimated using frequentist FPCA or our FAST procedure using a sequence of $K$s.

Model~\eqref{EQ:Defn} is a first line approach in the analysis of high dimensional data \citep{crainiceanu2024book} because in many applications $K\ll M$ and model~\eqref{EQ:Defn} provides a practical, low-dimensional approximation to the original high-dimensional data. Moreover, conditional on the FPCs, $\phi_k(\cdot)$, the model becomes a mixed effects model with uncorrelated random effects. In this paper we treat the eigenfunctions $\phi_k(t)$ as parameters, estimate them from the data, and account for their uncertainty. FAST provides a stable approach to simulating the posterior distribution of all parameters, including the eigenfunctions.

\subsection{Orthonormal spline expansion}\label{subsec:osplines}

All functional components are expanded using a $Q-$dimensional orthonormal spline basis, such that $Q \geq K$. Orthonormality is defined with respect to the scalar product on $L^2([0,1])$, i.e. $\langle f, g \rangle = \int_0^1 f(x)g(x)dx$. Denote by $\mathbf{B}(t) = \{B_1(t),\ldots, B_Q(t)\}$ the spline basis and by $\mathbf{B} \in \mathbb{R}^{M \times Q}$ the matrix representation of the spline evaluated over an $M$-dimensional grid. Each column of $\mathbf{B}$ corresponds to a spline basis and each row corresponds to a sampling point. We model $\mu(t) = \mathbf{B}(t)w_\mu$ and $\phi_k(t) = \mathbf{B}(t)\psi_k$, where $w_{\mu}$ and $\psi_k$ are $Q$-dimensional vectors of spline coefficients. In the data space, these become $\{\mu(t_1), \ldots, \mu(t_M)\}^t = \mathbf{B}w_\mu$ and $\{\phi_k(t_1), \ldots, \phi_k(t_M)\}^t = \mathbf{B}\psi_k$. 

We will show that the $\phi_k(t)$ functions are orthonormal if and only if $\boldsymbol{\Psi}^t \boldsymbol{\Psi} = I_{K\times K}$, where $\mathbf{\Psi} = [\psi_1| \ldots, |\psi_K]$ is the $Q \times K$ dimensional matrix obtained by column binding the $\psi_k$. Indeed, as $\mathbf{B}(t)$ is an orthonormal spline basis, $\langle B_i(t), B_j(t)\rangle = 1$ if $i = j$ and $0$ otherwise. Therefore, $\langle \phi_k(t), \phi_{k'}(t) \rangle$ is equal to
\begin{equation*}
       \int_0^1 [\mathbf{B}(t)\psi_k][\mathbf{B}(t)\psi_{k'}] dt
     = \sum_{i = 1}^Q \sum_{j = 1}^Q \psi_{k,i} \psi_{k',j} \bigl[\int_0^1 B_i(t)B_j(t) dt \bigr] = \psi_{k}^t \psi_{k'}\;.
\end{equation*}
The quantity $\psi_{k}^t \psi_{k'} = 1$ when $k = k'$ and $0$ otherwise if and only if $\boldsymbol{\Psi}$ is orthonormal. Therefore, the eigenfunctions $\phi_k(t)$ are orthonormal in $L^2([0,1])$ if and only if the low-dimensional spline coefficients are orthonormal in $\mathbb{R}^Q$. This is a crucial point of our methodology, as the orthonormality of the eigenfunctions (in infinite or high dimensional spaces) is controlled by  orthonormality  of vectors in a small dimensional space. Because it is much easier to sample from a small dimensional space, this projection is one of the main reasons the FAST method is stable and computationally efficient.

In functional data analysis, orthonormality in the theoretical $L^2([0,1])$ functional space is often confounded with orthonormality in the  $\mathbb{R}^M$ observed vector space. For a fixed $M$ and equally-spaced observations, the inner product on $\mathbb{R}^M$ is the Riemann sum approximation (up to the scaling factor $1/M$) of the functional inner product. However, this approximation depends on the density of the observation in the functional domain  and needs to be re-adjusted when the number of sampling points is changed. FAST can be used in the vector space as well by choosing the  matrix $\mathbf{B}$ such that $\mathbf{B}^t \mathbf{B} = \mathbf{I}_Q$. The resulting FPC matrix $\mathbf{B}\boldsymbol{\Psi}$ is then a product of orthonormal matrices. To maintain robustness to different $M$ and sampling regimes, we focus here on functional orthonormality. However, our software implementation can use vector orthormality, as well.

The basis dimension $Q$ must be sufficiently large to capture the maximum complexity of the subject-specific functions. We follow here the standard recommendations for penalized splines and use $Q \in [20, 40]$ \citep{ruppert_selecting_2002}, which performed  well in our tests (see Supplement Sections~\ref{supp:Q_Choice}). The choice of the basis matrix $\mathbf{B}$ can substantially impact the computational stability and efficiency of FAST. While, in theory, any spline matrix $\mathbf{B}$ which is  orthonormal in $L^2[0,1]$ could be used, in practice some approaches have better computational properties and interpretability. We choose as our basis `Splinets', an orthonormalized B-spline basis introduced by \cite{liu_splinets_2020} to maintain temporal localization. We augment their basis with a slope and intercept, appropriately orthonormalized. 

\subsection{Priors}\label{subsec:priors}

We add smoothing penalties on $\mu(t)$ and the FPCs $\phi_k(t)$ to the target posterior, which have form $\alpha \int f^2(t) dt + (1-\alpha) \int \{f''(t)\}^2 dt$ for generic function $f(\cdot)$. This penalty, used by \cite{goldsmith_generalized_2015}, contains the fixed weighting parameter $\alpha$, where $\alpha=0$ corresponds to the famous penalized spline prior \citep{cravenwahba1979,osullivan1986}. If $f(t) = \mathbf{B}(t)\theta$, then the penalty is a quadratic form of $\theta$ with associated matrix $\mathbf{P}_\alpha$. Supplement Section~\ref{supp:Penalty} covers the derivation of $\mathbf{P}_{\alpha}$. We use $\alpha = 0.1$ for consistency with existing implementations, but FAST can handle any $\alpha \in [0,1]$. Supplement Section~\ref{supp:Alpha_Choice} indicates robustness to reasonable choice of $\alpha$. The quadratic penalties are equivalent to Normal priors when $\mathbf{P}_\alpha$ is non-degenerate \citep{brumback1999,ruppert2003semiparametric,wood2006}. The penalties $g_\mu(w_\mu), g_{\phi_k}(\psi_k)$ have the following forms:

\begin{center}
$ g_\mu(w_{\mu}) = h_\mu^{\text{R}(\mathbf{P}_\alpha)/2}\exp\left\{-\frac{h_{\mu}}{2} w^t_{\mu}\mathbf{P}_{\alpha} w_{\mu}\right\}\;,\;\;
    g_{\phi_k}(\psi_k) = h_k^{\text{R}(\mathbf{P}_\alpha)/2}\exp\left\{-\frac{h_k}{2} \psi^t_k\mathbf{P}_{\alpha}\psi_k\right\}\;, 
$
\end{center}

\noindent where $\text{R}(\mathbf{P}_\alpha)$ is the rank of the penalty $\mathbf{P}_\alpha$ and $h_\mu,h_k>0$ are the smoothing parameters for $\mu(\cdot)$ and $\phi_(\cdot)$, respectively. FAST uses separate parameters for each functional component, in-line with \cite{ye_functional_2024} and \cite{goldsmith_generalized_2015}.

We assume that the inverse variance components, including the mean smoothing parameter, have Gamma priors with the shape and rate parameters equal to $0.001$. For the FPC smoothing parameters, we choose shape parameters of $0.01$ and rate parameters of equal to the trace of $\mathbf{P}_\alpha/2$ plus $0.01$ to ensure a proper prior (see Result~\ref{Thm:Prior}). For discussion on the choice of Gamma priors for inverse variance components, see \cite{crainiceanuwinbugs,crainiceanu_bayesian_2010}.

Consider the joint prior $g(\boldsymbol{\Psi}, \mathbf{H}) = g_\psi(\boldsymbol{\Psi}|\mathbf{H}) g_h(\mathbf{H})$ on the FPCs ($\Psi$) and the smoothing parameters $\mathbf{H} = \diag(h_1, \ldots, h_K)$. Here $g_\psi(\cdot)$ is the conditional prior of $\boldsymbol{\Psi}$ given the smoothing parameter matrix $\mathbf{H}$ and $g_h(\cdot)$ is the product of independent Gamma priors on $h_k$. We define the conditional prior $g_\psi(\boldsymbol{\Psi}|\mathbf{H})$ as follows:
\begin{equation*}
g_\psi(\boldsymbol{\Psi}|\mathbf{H}) \propto \frac{\prod_{i = 1}^K h_i^{{\rm R}(\mathbf{P}_\alpha)/2}}{{\rm Vol}(\mathcal{V}_{K,Q})} \times \text{\rm exp}\{-{\rm trace}(\mathbf{H \boldsymbol{\Psi}^t \mathbf{P}_\alpha}\boldsymbol{\Psi}/2)\}\;,
\end{equation*}

\noindent where ${\rm Vol}(\mathcal{V}_{K, Q})$ is the volume of the Stiefel manifold $\mathcal{V}_{K, Q}$ and ``trace($\boldsymbol{A}$)" denotes the trace of the matrix $\boldsymbol{A}$. The conditional density $g_\psi(\boldsymbol{\Psi}|\mathbf{H})$ is a combination of a uniform and smoothing prior on individual PCs. Implementing this prior in Bayesian software is straightforward. 

\begin{result} \label{Thm:Prior}
The prior distribution $g(\boldsymbol{\Psi}, \mathbf{H}) = g_\psi(\boldsymbol{\Psi}|\mathbf{H}) g_h(\mathbf{H})$ is proper if $\beta_\psi > {\rm trace}(\mathbf{P}_\alpha)/2$, where $\beta_\psi$ is the rate parameter for the Gamma priors on the smoothing parameters $h_k$.
\end{result}

Result~\ref{Thm:Prior} provides sufficient conditions under which the joint prior $g(\boldsymbol{\Psi}, \mathbf{H})$ is proper; proof can be found in Supplement Section~\ref{supp:Prior_Proper}. 

\subsection{Polar decomposition}\label{subsec:polar_decomp}

The conditional posterior distributions for all parameters except $\boldsymbol{\Psi}$ have analytical forms; see Supplement Section~\ref{supp:CondPost}. The $Q \times K$ dimensional matrix parameter $\boldsymbol{\Psi}$ is orthonormal and, hence, belongs to the Stiefel manifold \citep{james1976}, which has a finite volume and is denoted as $\mathcal{V}_{K,Q}$ \citep{chikuse_statistics_2003, MARDIA1977468}. Our prior on $\mathbf{\Psi}$, conditional on the smoothing parameters $\{h_1,\ldots, h_K\}$, is a prior on $\mathcal{V}_{K,Q}$.

\begin{result} \label{Thm:EF}
The conditional posterior distribution of the eigenfunction matrix $\mathbf{\Psi}$ is:
$$ f(\mathbf{\Psi} |{\rm others}) \propto \text{\rm etr}\left[\frac{\mathbf{\Xi}(\mathbf{Y} - 1_N^t \otimes \mathbf{B}w_{\mu})^t\mathbf{B}\mathbf{\Psi}}{\sigma^2} - \frac{\Xi \Xi^t \boldsymbol{\Psi}^t \mathbf{B}^t \mathbf{B} \boldsymbol{\Psi}}{2\sigma^2} -\frac{\mathbf{H}\mathbf{\Psi}^t\mathbf{P}\mathbf{\Psi}}{2} \right] \mathbbm{1}(\mathbf{\Psi} \in \mathcal{V}_{K,Q})\;,$$
    where $\mathbf{H} = \diag(h_1, \ldots, h_K)$, $\mathbf{\Xi}$ is a $N\times K$ dimensional matrix with the $(i,k)$ entry equal to $\xi_{i,k}$, and the matrix $\mathbf{Y}$ is the $N\times M$ dimensional matrix with the row $i$ equal to $Y_i^t$. The quantity $1_N$ is the N-dimensional column vector of ones, the symbol $\otimes$ denotes the Kronecker product of matrices, and ``etr" denotes the exponential of the matrix trace.
\end{result}

The distribution in Result~\ref{Thm:EF} does not follow a known form, though it is close to the generalized Langevin-Bingham family \citep{hoff_simulation_2009}. We sample this parameter indirectly using a parameter expansion based on the polar decomposition. The polar decomposition of the $Q\times K$ dimensional matrix  $\mathbf{X}$ is $\mathbf{X} = \mathbf{U P}$, where $\mathbf{U} \in \mathbb{R}^{Q \times K}$ is orthonormal (on the Stiefel manifold $\mathcal{V}_{K,Q}$) and $\mathbf{P} \in \mathbb{R}^{K \times K}$ is positive semi-definite. The polar decomposition is unique when $\mathbf{X}$ is full rank $K$. Crucially, if $\mathbf{X}$ has independent $N(0,1)$ entries, then the matrix $\mathbf{U}$ is uniformly distributed over the Stiefel manifold $\mathcal{V}_{K,Q}$ \citep{chikuse_statistics_2003}. Therefore, the uniform component of the prior on $\boldsymbol{\Psi}$ can be incorporated by imposing independent, entry-wise $N(0,1)$ priors on  $\mathbf{X}$. This is a particular case of the parameter expansion strategy used by \cite{jauch_monte_2021,park_bayesian_2025,meng_bayesian_2024}.

FAST performs polar decomposition using the eigendecomposition $\mathbf{X}^t\mathbf{X}=\mathbf{ZDZ}^t$, defining $\mathbf{\Psi}=\mathbf{XZD}^{-1/2}\mathbf{Z}^t$ and $\mathbf{P}=\mathbf{\Psi}^t\mathbf{X}$. It can be easily verified that $\mathbf{\Psi}^t\mathbf{\Psi}=I_Q$. Using this decomposition to sample $\mathbf{\Psi}$ works well because: (1) the element-wise independent $N(0,1)$ priors on $\mathbf{X}$ are computationally simple; (2) the matrix $\mathbf{X} \in \mathbb{R}^{Q \times K}$ is much smaller dimensional than matrix of eigenfunctions $\mathbf{\Phi} \in \mathbb{R}^{M \times K}$ when $Q<<M$; and (3) deterministically calculating $\mathbf{\Phi}$ from $\mathbf{X}$ can be accomplished efficiently using standard linear algebra routines.

\subsection{Alignment, Estimation, and Convergence}\label{subsec:practical}

There are two main sources of non-identifiability for the eigenfunctions and scores in FPCA: sign flipping and ordering. This problem is exacerbated by the complex geometry of the Stiefel manifold. To address these issues, FAST incorporates both constrained MCMC sampling and post-processing steps. During sampling, strict ordering of the $\lambda_k$ is imposed. In post-processing, we first re-order FPCs/scores according to the score sample variances, then align the signs to maximize vector correlation between each $\phi_k(\cdot)$ sample and the corresponding first FPC sample after burn-in. This approach is different from existing Bayesian implementations \citep{goldsmith_generalized_2015, jauch_monte_2021}, but it is computationally simple and performed well in our testing.

An additional problem is that the point-wise means of the FPC samples are not guaranteed to be orthonormal. To address that, we take the element-wise mean of the spline coefficient $\boldsymbol{\Psi}$ matrix and orthonormalize the results. The resulting $\widehat{\boldsymbol{\Psi}}$ will be orthonormal, and thus produce orthonormal FPC estimate $\widehat{\Phi} = \mathbf{B}\widehat{\boldsymbol{\Psi}}$ as described in Section~\ref{subsec:osplines}.

To monitor convergence of FAST we use the Gelman-Rubin RHat statistic for all model parameters \citep{gelman_inference_1992}. For functional components, we have RHat values at all observation points. This is done after post-processing the PCs and scores.

\subsection{Extensions}\label{section:extension}
FAST can be extended to multilevel, structural, and longitudinal functional data \citep{di_multilevel_2009,greven_longitudinal_2010,shou_structured_2015}, which are becoming increasingly common in applications. We demonstrate how to extend FAST to fit a multilevel model. The data structure is of the form $Y_{ij}(t), i =1,\ldots, N$, $j=1,\ldots,J_i$, $t \in T = \{t_1, \ldots, t_M\}$, where there are $J_i$ observations for each group $i$. A model for such data is MFPCA \citep{di_multilevel_2009,Zipunnikov_HDMFPCA}

\begin{equation}
    Y_{ij}(t) = \mu(t) + \eta_j(t) + \sum_{k = 1}^{K_1} \xi_{ik}\phi_k^{(1)}(t) + \sum_{l = 1}^{K_2}\zeta_{ijl}\phi_{l}^{(2)}(t) + \epsilon_{ij}(t)\;, \label{EQ:Multilevel}
\end{equation}
where $\mu(t)$ is the population mean and $\eta_j(t)$ are visit-specific deviations from the population mean. The subject-level (level 1) deviations from the visit-specific mean are modeled by $U_i(t)=\sum_{k = 1}^{K_1} \xi_{ik}\phi_k^{(1)}(t)$, where $\phi_k^{(1)}(t)$, $k = 1,\ldots, K_1$ are the first level eigenfunctions, $\xi_{ik} \sim N(0, \lambda^{(1)}_k)$, and $\lambda_k^{(1)}$ are the eigenvalues corresponding to $\phi_k^{(1)}(t)$. The subject/visit-specific (level 2) deviations from the subject-specific mean are modeled by $W_{ij}(t)=\sum_{l = 1}^{K_2}\zeta_{ijl}\phi_{l}^{(2)}(t)$, where $\phi_l^{(2)}(t)$, $l = 1,\ldots, K_2$ are the second level eigenfunctions, $\zeta_{ijl} \sim N(0, \lambda^{(2)}_l)$, and $\lambda^{(2)}_l$ are the eigenvalues corresponding to $\phi_l^{(2)}(t)$. Finally, $\epsilon_{ij}(t) \sim N(0, \sigma^2)$ is the error process. All functional components, $\epsilon_{ij}(t), U_i(t)$ and $W_{ij}(t)$, are assumed to be mutually independent.

The fixed effects, population mean $\mu(t)$ and the visit-specific deviations $\eta_j(t)$, are modeled using any penalized splines. The subject-specific $U_i(t)=\sum_{k = 1}^{K_1} \xi_{ik}\phi_k^{(1)}(t)$ and the visit-specific $W_{ij}(t)=\sum_{l = 1}^{K_2}\zeta_{ijl}\phi_{l}^{(2)}(t)$ can be modeled separately using the same techniques described in Sections~\ref{subsec:osplines}, ~\ref{subsec:polar_decomp}, and ~\ref{subsec:practical}. This is the first Bayesian MFPCA implementation directly modeling the FPCs.

\section{Bayesian implementation in STAN}\label{section:STAN}

In this section, we provide the {\ttfamily STAN} implementation of FAST. {\ttfamily STAN} is a flexible probabilistic programming language which facilitates sampling from user-specified Bayesian models using Hamiltonian Monte Carlo \citep{betancourt_conceptual_2018, carpenter_stan_2017}. We begin with the {\ttfamily data} section, which declares constants and imports the data, spline basis, and penalty matrix. We also calculate the trace of $\mathbf{P}_\alpha$ in the {\ttfamily transformed data} section for the FPC smoothing parameter priors. All notation is as used in Section~\ref{sec:methods}.

\singlespacing
\begin{lstlisting}[language=Stan]
data {
  int N;   // Number of time series
  int M;   // Number of observations
  int Q;   // Number of spline bases
  int K;   // Number of Eigenfunctions
  
  matrix[N, M] Y;  // Functional data
  matrix[M, Q] B;  // Orthogonal spline basis
  matrix[Q, Q] P;  // Penalty matrix for splines
}
transformed data {
  real tr_P = trace(P_alpha);  // Trace of penalty
}
\end{lstlisting}
\doublespacing

In the {\ttfamily parameters} block, we model the matrix $\mathbf{X}$ ({\ttfamily X}). Through the parameter expansion described in Section~\ref{subsec:polar_decomp}, we indirectly model $\mathbf{\Psi}$ using the polar decomposition of $\mathbf{X}$. This block also contains the eigenvalues $\lambda_k$ ({\ttfamily lambda}) stored using a {\ttfamily positive\_ordered} type, ordering the eigenvalues and corresponding eigenfunctions. The matrix {\ttfamily Scores} is the $N \times K$ score matrix $\mathbf{\Xi}$ introduced in Result~\ref{Thm:EF}.

\singlespacing
\begin{lstlisting}[language=Stan]
parameters{
  real<lower=0> sigma2;    // Error in observation
  
  // Mean structure
  vector[Q] w_mu;        // Population mean parameters
  real<lower=0> h_mu;    // Population mean smoothing parameter

  // Covariance structure
  positive_ordered[K] lambda;  // Eigenvalues
  vector<lower=0>[K] H;        // EF smoothing parameters
  matrix[Q, K] X;              // Unconstrained EF weights (X)
  matrix[N, K] Scores;         // EF scores (xi)
}
\end{lstlisting}
\doublespacing

The {\ttfamily transformed parameters} section contains deterministic functions of the data and model parameters. The key component of this block is the calculation of $\mathbf{\Psi}$ ({\ttfamily Psi}) as the orthogonal component of the polar decomposition $\mathbf{X}=\mathbf{\Psi}\mathbf{P}$. More precisely, $\mathbf{\Psi}=\mathbf{XZD}^{-1/2}\mathbf{Z}^t$, where $\mathbf{Z}$ ({\ttfamily evec\_XtX}) is the $K\times K$ dimensional matrix of eigenvectors of $\mathbf{X}^t\mathbf{X}$ and $\mathbf{D}$ ({\ttfamily diag(eval\_XtX)}) is a diagonal matrix containing the corresponding $K$ eigenvalues. Because $\mathbf{X} \in \mathbb{R}^{Q \times K}$ is low dimensional, all these operations are efficient.

\singlespacing
\begin{lstlisting}[language=Stan]
// Calculate observed parameters from latent
transformed parameters{
  // Population mean
  vector[M] mu = (B * w_mu)';

  // Polar decomposition
  matrix[Q,K] Psi;
  {
    matrix[K,K] evec_XtX = eigenvectors_sym(X'*X);
    vector[K] eval_XtX = eigenvalues_sym(X'*X);
    Psi = X*evec_XtX*diag_matrix(1/sqrt(eval_XtX))*evec_XtX';
  }
}
\end{lstlisting}
\doublespacing

The model proceeds with the specification of priors within the {\ttfamily model} section. The key components are the priors on $\mathbf{\Psi}$ ({\ttfamily Psi}). We first include the smoothing component directly through the penalties described in Section~\ref{subsec:priors}. Then, we incorporate the uniform component by specifying independent $N(0,1)$ priors for each entry of the latent matrix $\mathbf{X}$ ({\ttfamily X}). Combined with the polar decomposition in the {\ttfamily transformed parameters} block, this induces the uniform component of the prior on $\mathbf{\Psi}$. FAST thus models the FPCs through the small-dimensional $\mathbf{X}, \boldsymbol{\Psi} \in \mathbb{R}^{Q \times K}$, both constrained by the size of the spline basis and number of FPCs. This dimensionality reduction likely contributes to the much improved computational efficiency. All other priors are as described in Section~\ref{subsec:priors}.

\singlespacing
\begin{lstlisting}[language=Stan]
model{
  // (Inverse) variance priors
  H (*$\sim$*) gamma(0.01, tr_P/2 + 0.01);
  h_mu (*$\sim$*) gamma(0.001, 0.001);
  lambda (*$\sim$*) inv_gamma(0.001, 0.001); 
  sigma2 (*$\sim$*) inv_gamma(0.001, 0.001);
  
  // Mean function smoothing prior
  target += Q/2.0 * log(h_mu) - h_mu / 2.0 * w_mu' * P * w_mu;

  // FPC smoothing prior component
  for(i in 1:K){
    target += Q/2.0 * log(H[i]) - H[i] / 2.0 * Psi[,i]' * P * Psi[,i]; 
  }
  
  // FPC uniform prior component
  to_vector(X) (*$\sim$*) normal(0,1);

  // FPCA-based Score priors
  for(i in 1:K){
    to_vector(Scores[,k]) (*$\sim$*) normal(0, sqrt(lambda[k])); 
  } (*$\cdots$*)
}
\end{lstlisting}
\doublespacing

We next present the likelihood contribution. This  portion of the {\ttfamily model} block  computes the random deviations $\theta_i(t)=\sum_{k=1}^K\xi_{ik}\phi_k(t) = \sum_{k=1}^K\xi_{ik}\{\mathbf{B}(t)\psi_{k}\}$. If $\mathbf{\Theta}$ ({\ttfamily Theta}) is the $N\times M$ dimensional matrix with $(i,m)$ entry equal to $\theta_i(t_m)$, it can be shown that $\mathbf{\Theta}=\mathbf{\Xi}(\mathbf{B}\mathbf{\Psi})^t$. This value is computed, used, and discarded to conserve memory. The model finally incorporates the observed data contribution, following the FPCA data likelihood in Section~\ref{subsec:model}, using a vectorized distributional statement for computational efficiency.

\singlespacing
\begin{lstlisting}[language=Stan]
model{ (*$\cdots$*)
  { // Data Likelihood
    matrix[N,M] Theta = Scores * (B * Psi)'; // FPC linear combination
    to_vector(Y) (*$\sim$*) normal(to_vector(rep_matrix(mu, N) + Theta),
                          sqrt(sigma2));
  }
}
\end{lstlisting}
\doublespacing

This implementation suggests how to extend FAST  to more complex structures. For MFPCA, we sample two  spline coefficient matrices $\mathbf{\Psi}_1$ and $\mathbf{\Psi}_2$. For each matrix, we independently model a latent matrix: $\mathbf{X}_1$ for $\mathbf{\Psi}_1$ and $\mathbf{X}_2$ for $\mathbf{\Psi}_2$. We assign independent $N(0,1)$ priors to the entries of both $\mathbf{X}_1$ and $\mathbf{X}_2$. The $\mathbf{\Psi}_1$ and $\mathbf{\Psi}_2$ matrices are obtained via separate polar decompositions in {\ttfamily transformed parameters}. All other implementation components are straightforward, just requiring careful accounting of indices. We present both the implementation and simulation evaluation in Supplement Section~\ref{supp:simMulti}. 

\section{Simulations}\label{sec:simulation}

We consider the following two simulation scenarios for model~\eqref{EQ:Defn}, where $f_k(t)$ are taken to be the Legendre polynomials on $[0,1]$:

\noindent\textit{(S1)} Emulation of CGM data from Section~\ref{sec:data_analysis}
\begin{itemize}[label={},noitemsep,topsep=0pt]
    \item $\mu(t) = 140 - 20 \times f_2(t)$,  $\sigma^2 = 4$, $\lambda_k = \{2250, 450, 150\}$
    \item $\phi_k(t)=\{f_0(t), \sqrt{\frac{84}{31}}[f_1(t) - 0.5 f_3(t)], -\sqrt{5}f_2(t)\}$
\end{itemize}
\noindent\textit{(S2)} Canonical FPCA example from \cite{xiao_fast_2016}
\begin{itemize}[label={}, noitemsep,topsep=0pt]
    \item $\mu(t) = 0$, $\sigma^2 = 0.35$, $\lambda_k=\{1, 0.5, 0.25\}$
    \item $\phi_k(t)=\{\sqrt{2}\sin(2\pi t), \sqrt{2}\cos(4 \pi t), \sqrt{2} \sin(4 \pi t)\}$
\end{itemize}

For each scenario we generate $200$ datasets with $N = 50$ functions observed at $M = 50$ points distributed along $[0,1]$ according to quadrature. FAST is compared to the Generalized Function-on-Scalar Regression of  \cite{goldsmith_generalized_2015} (labeled ``GFSR"), modeling FPCs in the observed data space using parameter expansion based upon polar decomposition from \cite{jauch_monte_2021} (labeled ``POLAR"), and Variational Message Passing from \cite{nolan_bayesian_2023} (labeled ``VMP"). While all comparators fit a Bayesian FPCA, they do not all address uncertainty quantification for the same set of parameters. Neither POLAR nor VMP address construction of credible intervals for the functional components. Further, POLAR does not model the mean, instead subtracting point-wise means in pre-processing. 

For the MCMC-based approaches, we performed 1500 sampling iterations and discarded the first 1000 as burn-in. This was chosen based upon the convergence properties of FAST and kept uniform between methods. 

To obtain the orthonormal FPCs from the unconstrained samples of GFSR, we leverage the post-processing rotation suggested by \cite {goldsmith_generalized_2015}. This involves calculating the SVD of the unconstrained FPC matrix sample and taking the right singular vectors. The scores are correspondingly transformed by the left singular vectors and diagonal matrix of singular values. This selection of a common rotation addresses the permutation- and rotation-based non-identifiability of the FPCs and scores. With a suitable alignment of sign, such as described in Section~\ref{subsec:practical}, the rotated posterior FPC/score samples can be used to form corresponding credible intervals.

Similar to FAST, both POLAR and GFSR require procedures to produce orthonormal (in the vector sense) FPC estimates. For POLAR, \cite{jauch_monte_2021} recommends taking the right singular vectors of the mean smoothed data. As for GFSR, we take the point-wise mean of the non-orthogonal latent principal component matrix, then apply the SVD-based rotation \cite{goldsmith_generalized_2015} describes. VMP provides only FPC estimates. 

POLAR, GFSR, and VMP all produce FPC estimates/samples which are orthonormal in the vector sense. So, we rescale the FPC estimates for these methods to have norm 1 in $L^2([0,1])$. As all methods are subject to the identifiability issues described in Section~\ref{subsec:practical}, we applied the post-processing steps described in that section to align posterior samples and estimates. FPC/score signs were then aligned with those of the true underlying FPCs for the purposes of evaluation. 

\begin{figure}[!ht]
\centering
\includegraphics[width=12cm]{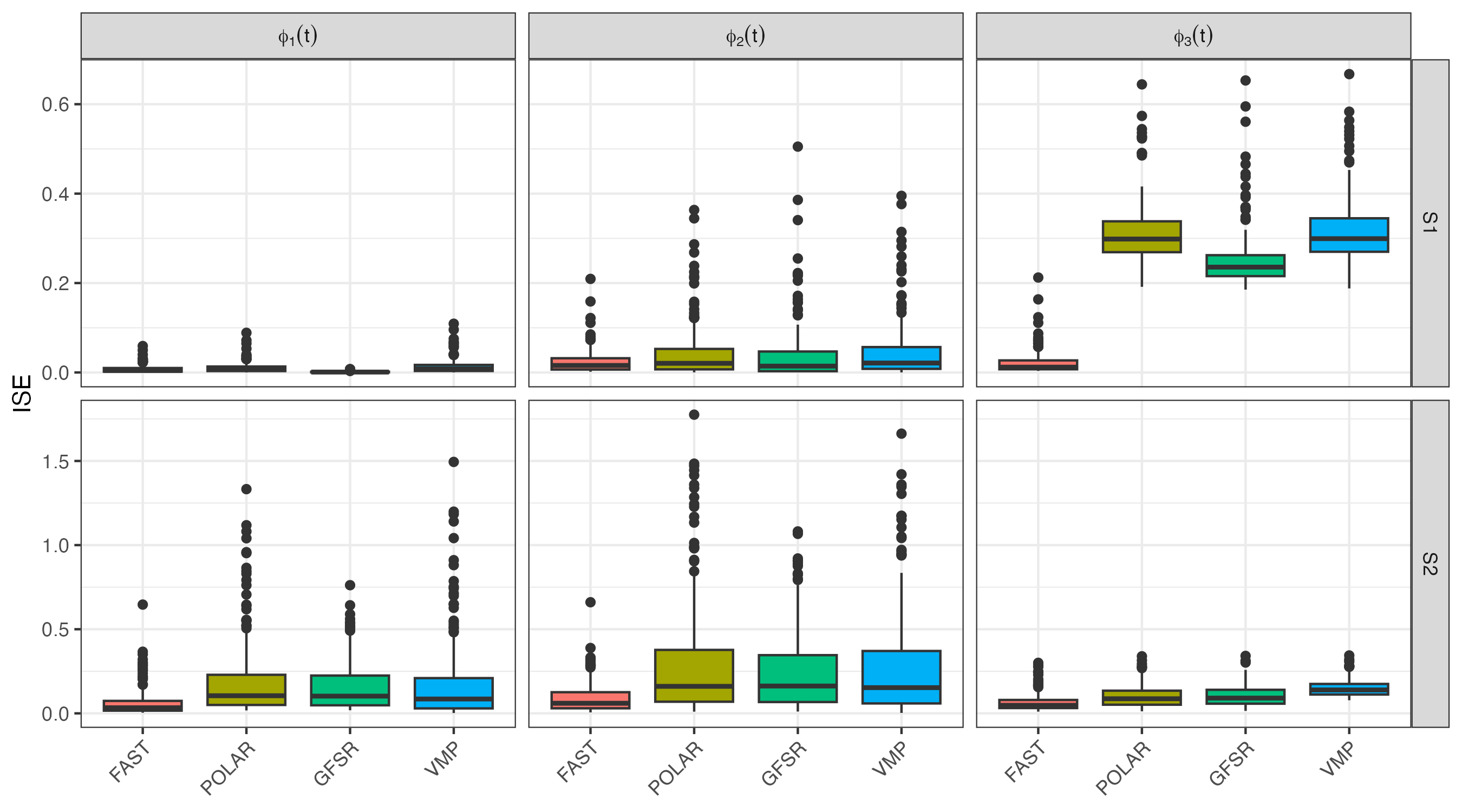}
\caption{ISE Boxplots for FAST, POLAR, GFSR, and VMP. Row indicates simulation scenario, S1 then S2. Each column corresponds to a particular eigenfunction.}
\label{fig:FPCA_FPC_ISE}
\end{figure}

We first focus on the Integrated Square Error (ISE) for FPCs $\phi_k(t)$. If $\widehat{\phi}_k^b(t)$ is the posterior estimate of $\phi_k(t)$ for simulated data set $b\in \{1,\ldots,B\}$,  ISE is defined as: 

\begin{center}
${\rm ISE}^b\{\phi_k(\cdot)\}=\int_0^1\{\widehat{\phi}_k^b(t)-\phi_k(t)\}^2dt \approx \sum_{m = 1}^M w(t_m)\{\widehat{\phi}_k^b(t_m) - \phi_k(t_m)\}^2\;,$
\end{center}

\noindent where $w(\cdot)$ are the corresponding Gaussian quadrature weights at $t_m$, $m=1,\ldots,M$. Combining the ${\rm ISE}^b$ values over simulations provides a $B$-dimensional vectors for each method and each principal component. Figure~\ref{fig:FPCA_FPC_ISE} displays the boxplots of these ISE by principal component (column) for scenario S1 (first row) and scenario S2 (second row) and method. Results indicate that FAST consistently outperforms existing methods.

We also estimate pointwise posterior coverage probabilities of the true FPCs $\phi_k(t)$, $k=1,\ldots,K$. For each simulated data set $b\in\{1,\ldots,B\}$, we obtain the equal-tail $95$\% credible intervals at the observed time points $t_m$, $m=1,\ldots,M$. The proportion of times these credible intervals cover the true function is calculated over the $t_m$ for each combination of $b$ and $k$. These proportions are then collected in a vector of length $B=200$ for each function $\phi_k(t)$. VMP and POLAR are excluded from this comparison, as they do not address uncertainty quantification for the FPCs.

\begin{figure}[!ht]
\centering
\includegraphics[width=12cm]{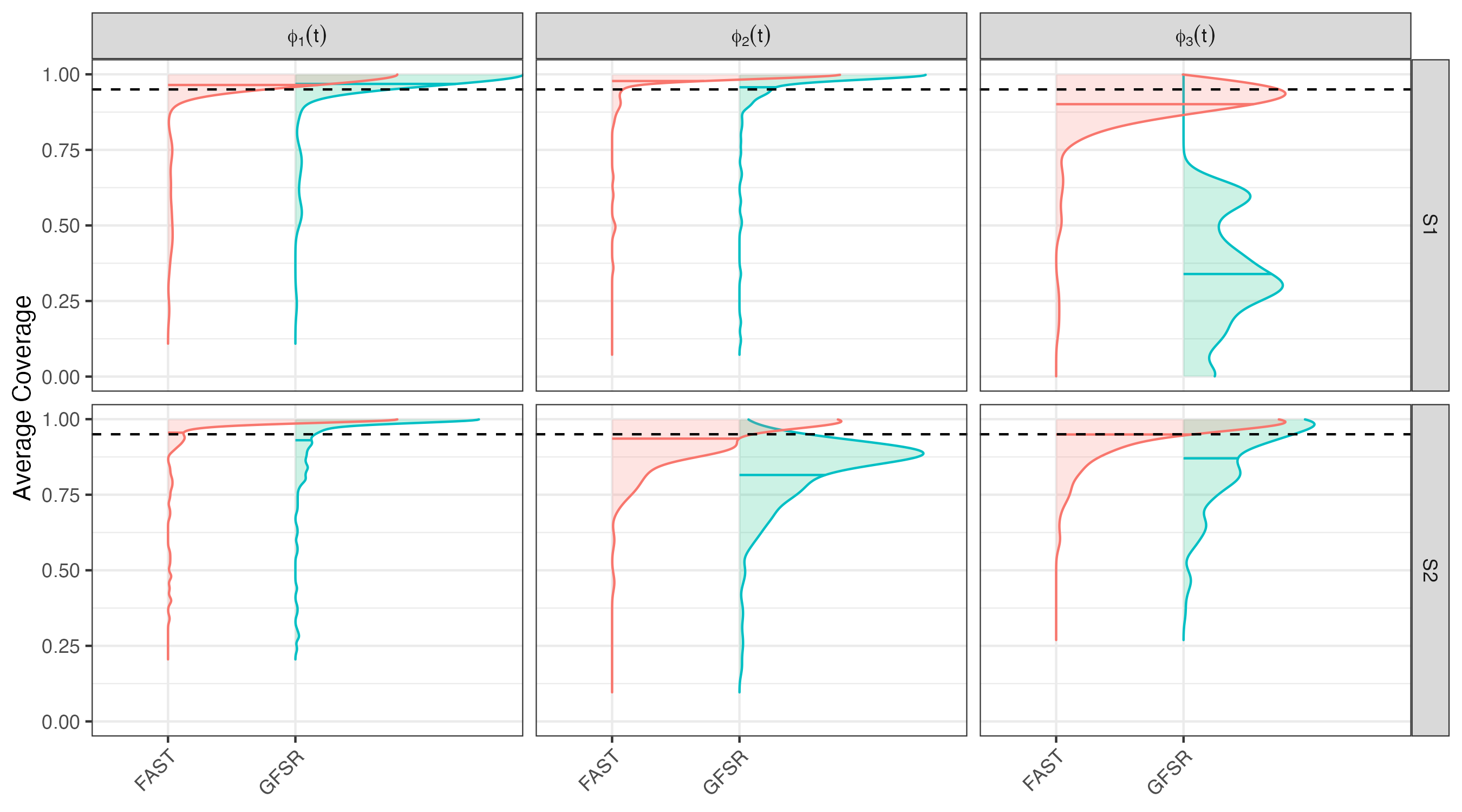}
\caption{Kernel smoother of $95$\% credible interval coverage probabilities of the true FPCs for FAST,  POLAR, and GFSR. Row indicates simulation scenario, S1 then S2. Each column corresponds to a particular eigenfunction. Distribution means: horizontal solid lines; nominal $95$\% level: horizontal dotted lines.}
\label{fig:FPCA_FPC_COV}
\end{figure}

Figure~\ref{fig:FPCA_FPC_COV} displays the kernel smooth of the estimated point-wise coverage proportions for FAST and GFSR using the same structure as Figure~\ref{fig:FPCA_FPC_ISE}. FAST performs well in all scenarios, consistently out-performing GFSR. 

Figures~\ref{fig:FPCA_FPC_ISE} and ~\ref{fig:FPCA_FPC_COV} indicate that the advantage of FAST is particularly striking for $\phi_3(t)$ in S1. In this case, the mean can be written as a linear combination of $\phi_1(t)$ and $\phi_3(t)$: $\mu(t) = 140\phi_1(t) + 4\sqrt{5}\phi_3(t)$. This choice makes the problem more challenging in finite samples, when information may leak between the mean and covariance; see, for example, \cite{zhou_prediction_2025} for a more detailed discussion. 

For inference on the mean, $\mu(t)$, we only compared to GFSR, as neither the VMP nor POLAR perform inference on this component. Results indicate that FAST produces smaller ISE and better coverage probability; see Supplement Section~\ref{supp:Sim_FE} for more details.

We also compared estimation accuracy and posterior coverage of credible intervals for the scores $\xi_k$. Accuracy is calculated as mean square error aggregated by FPC, $k$, within each simulation, $b$. Coverage was  calculated as the proportion of times the equal-tail $95$\% credible intervals cover the corresponding true scores; see results in Supplement Section~\ref{supp:Sim_Scores}. FAST outperforms the other methods and has near nominal coverage. 

Table~\ref{tab:CompTime_Scale} provides computation times in minutes on a personal laptop (MacBook Pro, 3.49 GHz and 32GB RAM) as both $N$ (number of subjects) and $M$ (number of observation points) scale up for scenario S1. The left column provides times for $N = 50$ and $M \in \{50, 100, 250, 500\}$, while the right provides times for $M = 50$ and $N\in \{50, 100, 250, 500\}$. 

\begin{table}[!ht]
\centering
\begin{tabular}{cSSSS|cSSSS}
 \hline
 \multicolumn{5}{c|}{Fixed $N=50$} & \multicolumn{5}{c}{Fixed $M = 50$} \\
 \hline
 $M$ & \mc{FAST} & \mc{GFSR} & \mc{POLAR} & \multicolumn{1}{c|}{VMP} & $N$ & \mc{FAST} & \mc{GFSR} & \mc{POLAR} & \mc{VMP} \\
 \hline
 50 & 3.5 & 24.3 & 9.2 & 0.2 & 50 & 3.5 & 24.3 & 9.2 & 0.2 \\ 
 100 & 6.8 & 39.5 & 28.5 & 1.0 & 100 & 4.3 & 43.2 & 15.4 & 0.6 \\ 
 250 & 18.3 & 82.8 & \textemdash & 1.2 & 250 & 5.3 & 47.2 & 29.0 & 0.6 \\
 500 & 37.6 & 151.0 & \textemdash & 5.2 & 500 & 14.7 & 27.8 & 52.1 & 0.7 \\
 \hline
\end{tabular}
\caption{Computation times (in minutes) for each combination of scale and method (simulation S1). All methods evaluated on the same simulated dataset for each $N,M$ combination.}
\label{tab:CompTime_Scale}
\end{table}

FAST and GFSR scale approximately linearly with $M$, though GFSR has a substantially higher slope. FAST and POLAR scale linearly in $N$, while GFSR seems to be more performant with higher $N$, possibly due to the increased information better specifying the posterior geometry. We could not fit POLAR for $M \geq 250$ because it uses too much memory. VMP is the fastest, which is not unexpected given the approximation inherent to the technique. All MCMC-based methods were run for 1500 iterations with the first 1000 discarded as burn-in; however, the convergence characteristics were not uniform. FAST and POLAR exhibit lower RHats ($\leq 1.05$) compared to GFSR ($\leq 1.25$), which may require a longer warm-up. The computation time of FAST is relatively robust to different $Q$ and $K$ values (see Supplement Section~\ref{supp:timing_sensitivity}).

Simulations for the multilevel scenario provide  similar results; see Supplement Section~\ref{supp:simMulti}. POLAR was not included because it does not have a multilevel implementation. 

\section{Data analysis: The DASH4D CGM Study}\label{sec:data_analysis}

Dietary Approaches to Stop Hypertension for Diabetes (DASH4D, NCT04286555) is a nutritional trial designed to assess how blood pressure and glucose respond to combinations of DASH-style and low-sodium diets in persons with type 2 diabetes (T2D). The study team prepared four diets: DASH4D diet with lower sodium, DASH4D diet with higher sodium, Comparison diet with lower sodium, and Comparison diet with higher sodium. Weight was held constant by adjusting calorie level. Diet effects were observed through a single-site, 4-period crossover design \citep{pilla_dietary_2025}. Each period consisted of 5 weeks of feeding a diet followed by a $\geq1$-week break (median 2 weeks). Approval for DASH4D was granted by the Johns Hopkins University School of Medicine institutional review board, and all study participants provided informed written consent. There were $N = 105$ randomized T2D participants recruited from the Baltimore area, of which $N = 65$ had meal timing data. Participants with meal timing data had a median age of $68$ years, were $66$\% female, and had the following race distribution: $6.2$\% Asian, $87.7$\% Black, and $6.2$\% White. 

The DASH4D CGM sub-study was conducted to evaluate the impact of the dietary intervention on glucose assessed by CGM. Participants wore the Abbott Freestyle Libre Pro (Abbott Diabetes Care), placed near the middle of the third week and worn into the fifth week for each feeding period. The CGM devices were placed on the back of the upper arm (approved location) by trained technicians. These devices record interstitial glucose every $15$ minutes. The Libre Pro is a masked CGM system, so participants were not aware of the glucose measurements \citep{wang_design_2025}. 

\subsection{Mealtime CGM glucose}\label{subsec:CGM_glucose}

During feeding periods, participants ate meals at the study site 3 days each week. Clinic staff observed consenting participants to document meal timing. We used these timestamps to extract CGM data starting $1$ hour before and ending $4$ hours after each meal began. There were $\leq 20$ such $5-$hour periods per participant. The final dataset was comprised of 768 meals over 65 individuals. This data is not yet available as trial results are forthcoming.

\begin{figure}[!ht]
\centering
\includegraphics[width=14cm,height=9cm]{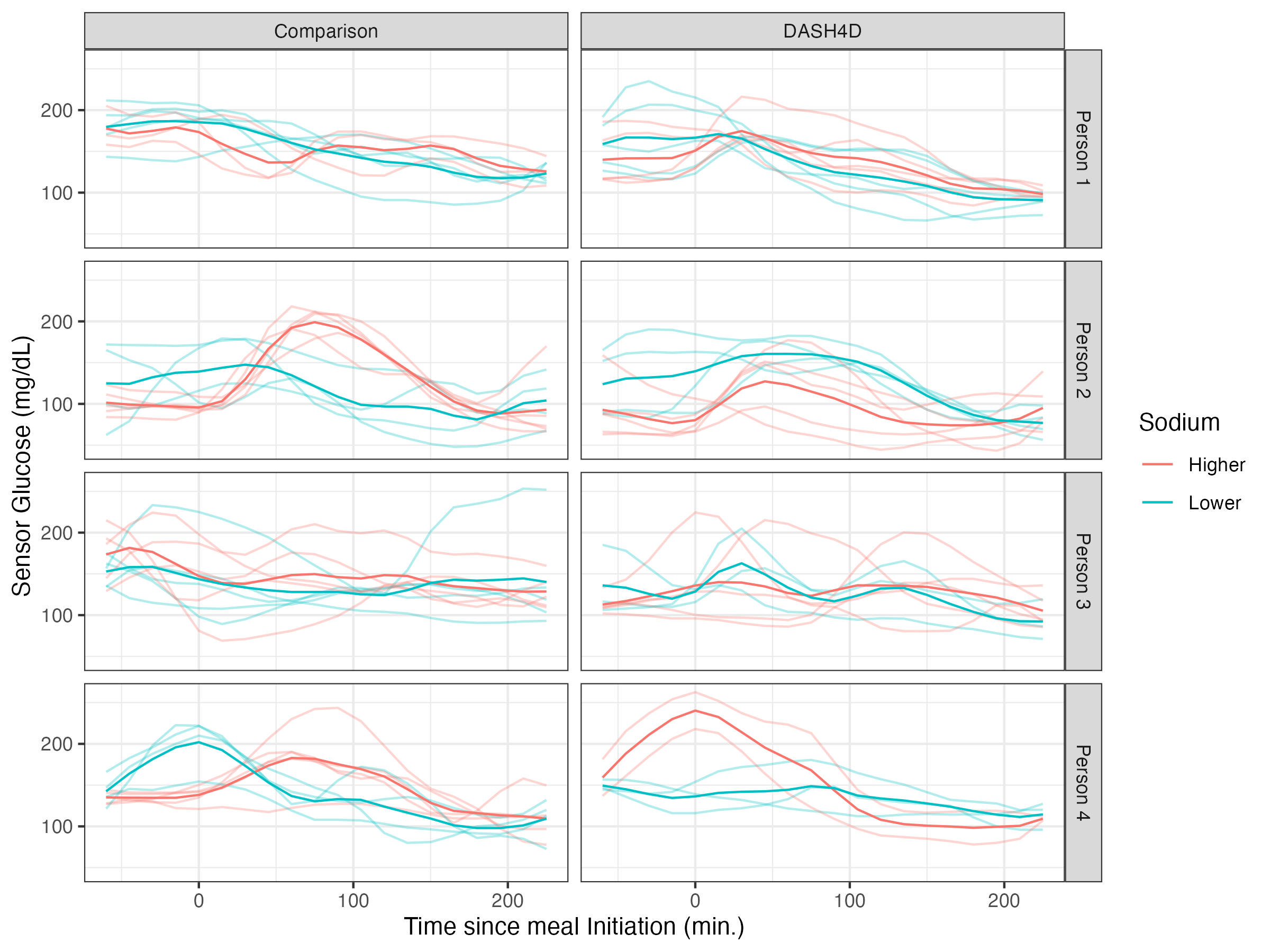}
\caption{Mealtime CGM for $4$ participants (one per row), for the Comparison (left column) and DASH4D (right column) diets. Sodium level within diet type is indicated by color. X-axis: time from meal initiation, y-axis: CGM in mg/dL. Person- and diet-specific average curves are darker, while meal-specific curves are lighter.}
\label{fig:Motivating_ML}
\end{figure}

Figure~\ref{fig:Motivating_ML} provides a visualization of mealtime CGM for four randomly-sampled study participants. The x-axis in each panel corresponds to time in minutes from meal initiation; the y-axis is the recorded glucose. Each row corresponds to a participant. The first column is for Comparison meals, and the second is for DASH4D meals. Curve color corresponds to sodium level, and darker curves are within-person averages. Given this data structure, one could model the average average curves within person using a single-level FPCA or the meal-specific trajectories using a two-level FPCA to capture both the participant-specific averages (level 1) and the meal-specific deviations (level 2) within each diet. We undertake these analyses in Sections~\ref{subsec:Bayes_FPCA_data} and ~\ref{subsec:Bayes_MFPCA_data}, respectively.

\subsection{Bayesian FPCA of average CGM within diets}\label{subsec:Bayes_FPCA_data}
We fit FAST to the participant average CGM curves for each diet separately (darker curves in Figure~\ref{fig:Motivating_ML}) using a spline basis of dimension $Q = 20$ and $K = 3$ FPCs. This number of FPCs explained $\geq 95\%$ of the variability across diets (see Supplemental Figure~\ref{supp:K_PVE}) The number of curves per-diet ranges between $36$ and $49$, so accounting for variability in FPC estimates is of particular importance \citep{goldsmith_PCA_2013}. 

\begin{figure}[!ht]
\centering
\includegraphics[width=16cm,height=8cm]{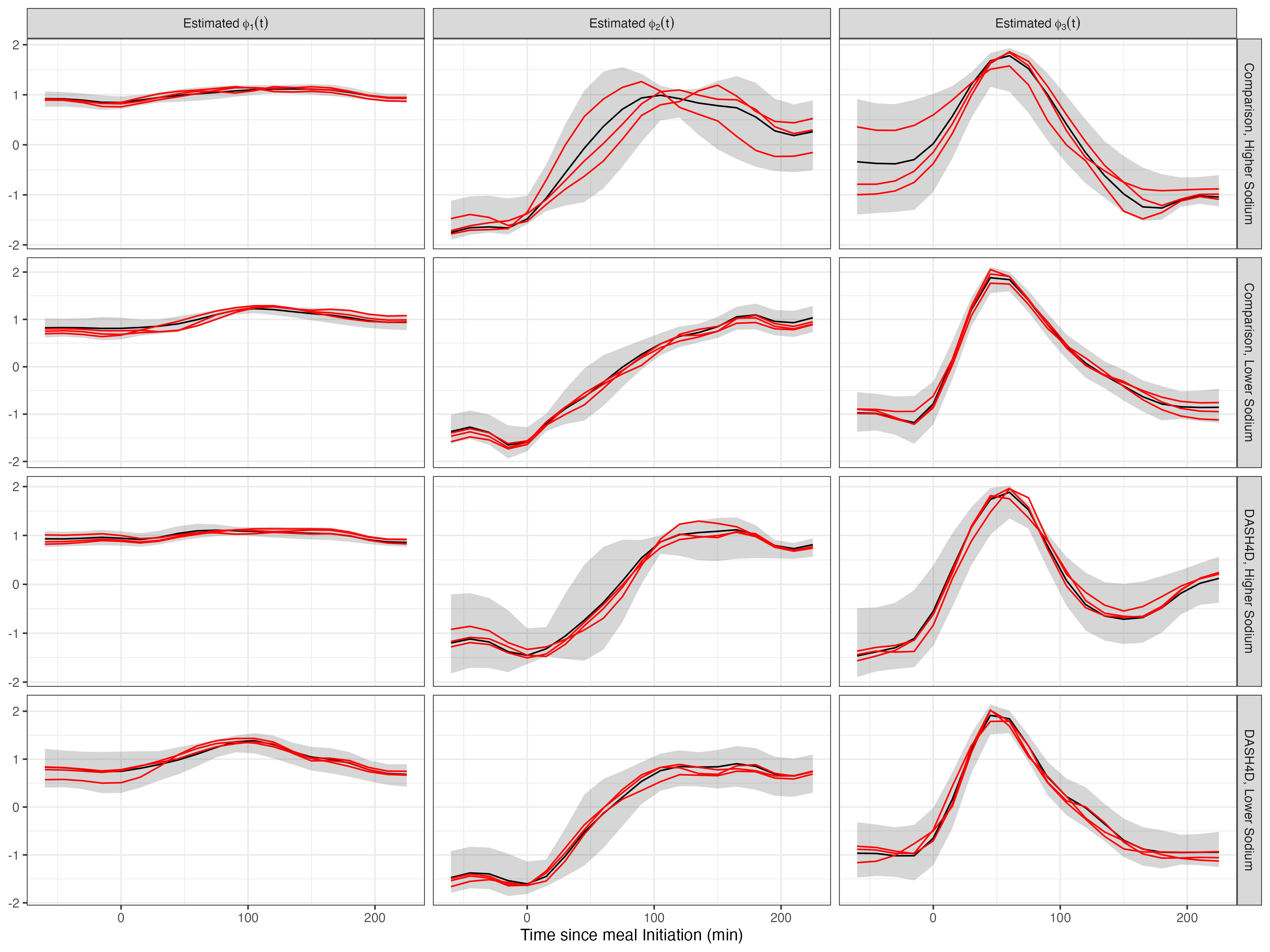}
\caption{Bayesian FPCA results for the first three FPCs (columns) for each of the four diets (rows). X-axis: time relative to meal initiation. Black curves: posterior mean; red curves: three FPC posterior samples; shaded areas: point-wise $95$\% credible intervals.}
\label{fig:EF_SL}
\end{figure}

Figure~\ref{fig:EF_SL} displays results for Bayesian FPCA inference for the first three PCs (columns) for the four diets (rows). For all panels the x-axis is the time from the start of the meal, the black curves are the posterior means, the red curves are three samples from the posterior of the PCs, and the shaded areas are point-wise, equal-tail $95$\% credible intervals. The red curves correspond to the same sample iterations across PCs.

\begin{figure}[!ht]
\centering
\includegraphics[width=14cm,height=5cm]{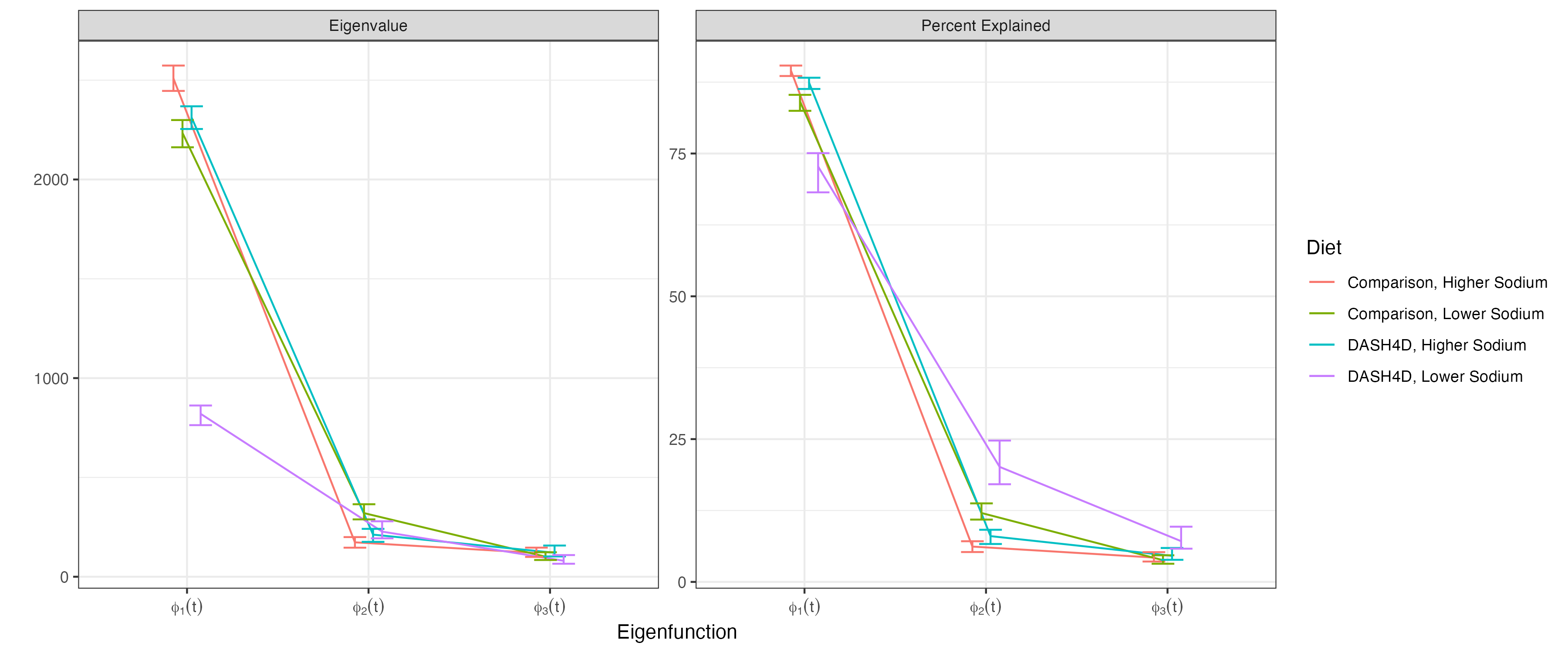}
\caption{Eigenvalues (left panel) and percent-variability (right panel) estimates and $95$\% credible intervals for each of the four diets (one color corresponds to one diet).}
\label{fig:EV_SL}
\end{figure}

Figure~\ref{fig:EV_SL} provides the eigenvalues (left panel) and percent variance explained (right panel) for the eigenfunctions displayed in Figure~\ref{fig:EF_SL}. Diet is indicated by color.

Posterior eigenfunction estimates $\widehat{\phi}_k(t)$ are qualitatively similar across diets in Figure~\ref{fig:EF_SL}. The primary modes of variability $\widehat{\phi}_1(t)$ are nearly constant with a small curvature. Higher scores on these components correspond to higher CGM responses over the entire mealtime period. The sample-to-sample variability in the shape of $\phi_1(t)$ is low. Moreover, $\phi_1(t)$ explains $70$ to $85$\% of the observed variability. This quantifies the importance  of the mean postprandial glucose level \citep{garber_postprandial_2012}. The second FPCs, $\widehat{\phi}_2(t)$, generally exhibit an $S$ pattern. Therefore, individuals with a higher positive score on  $\widehat{\phi}_2(t)$ have lower glucose in the pre-prandial and immediate postprandial period with a larger response one to four hours after eating. The third FPCs, $\widehat{\phi}_3(t)$, are negative during the hour before the meal, increase to a peak positive value about $50$ to $70$ minutes after food intake, and decrease back to negative values about $150-200$ minutes after the meal. Therefore, individuals with higher scores on $\widehat{\phi}_3(t)$ tend to have a to lower glucose before the meal and a rapid glucose increase after the meal. Sample-to-sample variability of both the $\phi_2(t)$ and $\phi_3(t)$ is higher than that of $\phi_1(t)$, likely due to the smaller signal.

The variability of glucose response in the DASH4D/lower sodium diet is much lower than for the other three diets. Figure~\ref{fig:EV_SL} indicates that $\widehat{\lambda}_1(t)$ is $\approx3$ times smaller for this diet compared to the others. This diet, a primary interest of the DASH4D study, appears to be associated with a more stable glycemic response between participants. 

FAST produces reasonable results, qualitatively consistent with standard FPCA, across all four diets. Each model takes $10-40$ seconds (depending on the diet) to fit on a personal laptop using $2000$ iterations with the first $1000$ discarded as burn-in. We employ the routine from Section~\ref{subsec:practical} to assess convergence of the scores, $\xi_{ik}$, and sampled points along the domain for each FPC, $\phi_k(t_m)$. Final Gelman-Rubin statistics were $< 1.05$. Despite differences in number of meals per participant, results were similar when fitting FPCA to randomly sampled meals; see Supplement Section~\ref{supp:Random_FPCA_Data}. 

\subsection{Bayesian MFPCA of individual CGM curves}\label{subsec:Bayes_MFPCA_data}
So far, we have ignored the variability of meal-specific CGM curves around the subject-specific mean. This could raise questions about accounting for all observed variability, different number of meals per person, and representativeness of the average curves. To address these problems, 
we fit the multilevel extension of FAST described in Section~\ref{section:extension} for each diet separately, with meals nested within participants. We use a rich spline basis of dimension $Q = 20$, $K_1 = 2$ FPCs at the subject level, and $K_2 = 3$ FPCs at the meal level. These choices explain $\geq 80\%$ variability at both levels. The number of observed meals ranges from $147$ to $219$ for each diet, with at most five meals per participant. There is no reason to assume that meal/visit order has effect, so we set $\eta_j(t) = 0$ for all $j$.

\begin{figure}[!ht]
\centering
\includegraphics[width=16cm]{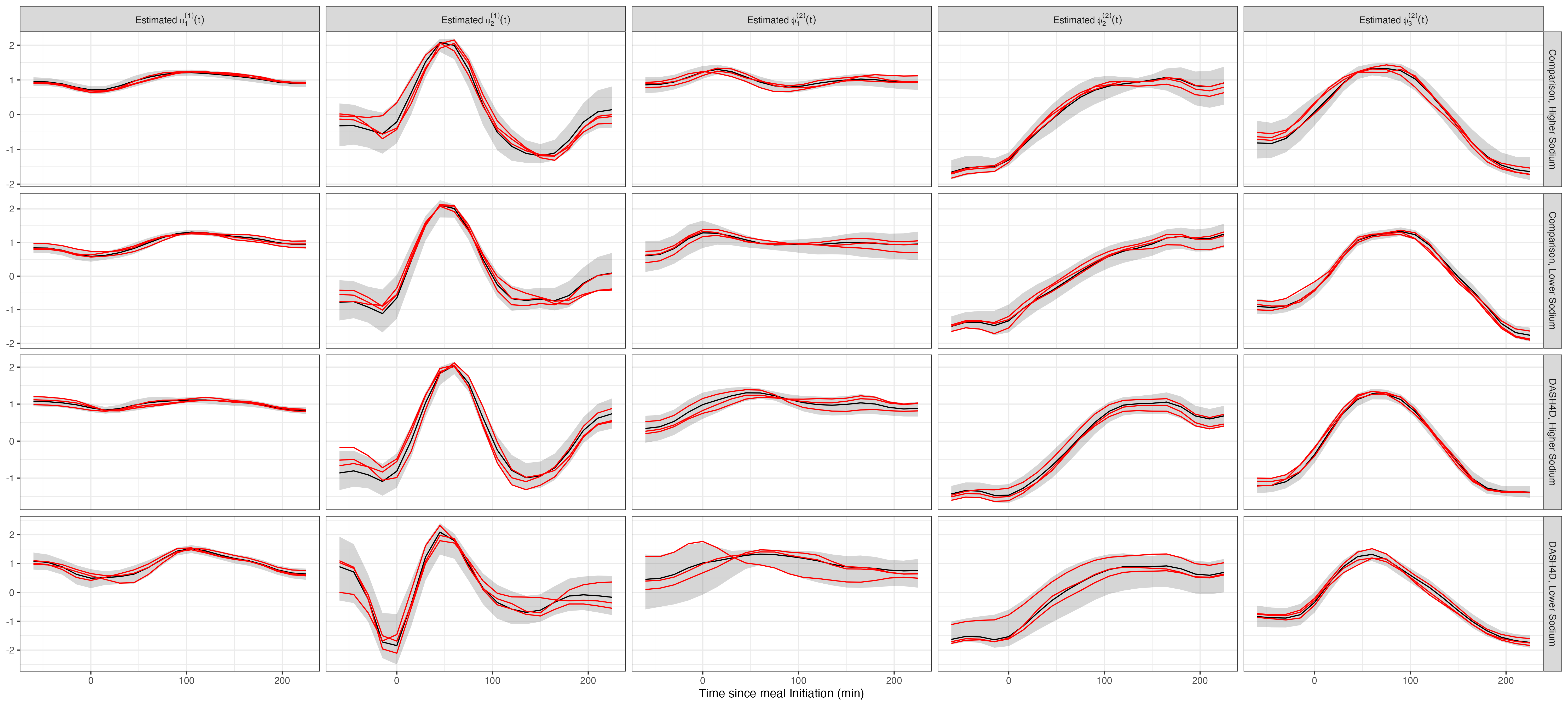}
\caption{Bayesian MFPCA results for the first two PCs at the subject level and first three PCs at the meal level. Each column corresponds to an FPC and each row corresponds to a diet. X-axis: time from the meal start. Plotting conventions are consistent with Figure~\ref{fig:EF_SL}.}
\label{fig:EF_ML}
\end{figure}

Figure~\ref{fig:EF_ML} displays the Bayesian MFPCA results, with row indicating diet and column indicating FPC. The first two columns correspond to the subject level, while the last three correspond to the meal level. Plotting conventions are consistent with those in Figure~\ref{fig:EF_SL}.

\begin{figure}[!ht]
\centering
\includegraphics[width=16cm,height=6cm]{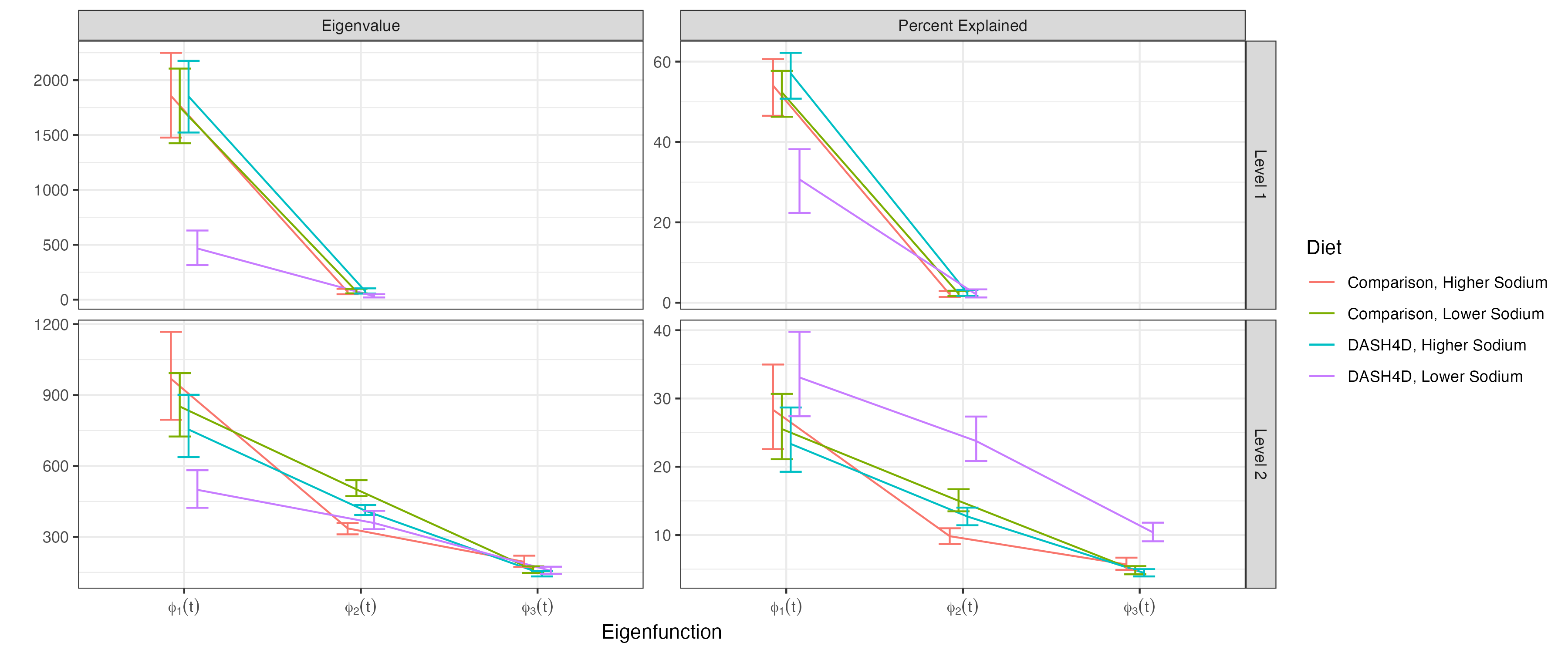}
\caption{Eigenvalues (left panels), percent-variability explained (right panels) within level (first row:  level 1; second row: level 2) and diet (each diet shown in one color). All plotting conventions are consistent with Figure~\ref{fig:EV_SL}.}
\label{fig:EV_ML}
\end{figure}

Figure~\ref{fig:EV_ML} provides the estimated variability (left panels) and percent variability explained (right panels) within level and diet. The first row of panels corresponds to the subject (level 1) while the second row of panels corresponds to the meal (level 2). Each color corresponds to a diet. As in the single level analyses, Figure~\ref{fig:EF_ML} indicates qualitative similarity of estimates $\widehat{\phi}_k^{(1)}(t), \widehat{\phi}_l^{(2)}(t)$ across diets. The primary mode of variability within each diet is $\widehat{\phi}^{(1)}_1(t)$, a nearly constant function. Higher subject scores on this component correspond to higher subject-specific glucose over the entire observation window. The sample-to-sample variability in the shape of $\phi^{(1)}_1(t)$ is consistently low, indicating confidence in this functional form. There is some heterogeneity in the percent variability explained by this component, but it is $\geq 30$\% for every diet. The first eigenfunction at the second level, $\widehat{\phi}^{(2)}_1(t)$, is also nearly constant with a small curvature and low sampling variability. This FPC explains between $25-35$\% of the total variability according to Figure~\ref{fig:EV_ML}, indicating that a constant shift is the main source of variation in glucose responses both among and within participants. The second eigenfunction at the meal level, $\widehat{\phi}^{(2)}_2(t)$, explains $10-25$\% of the total variability and has an $S$ pattern. As in the single-level case, larger scores on $\widehat{\phi}^{(2)}_2(t)$ correspond to lower pre- and immediately post-prandial glucose, then increasingly larger response one to four hours after the meal. The eigenfunctions $\widehat{\phi}^{(1)}_2(t), \widehat{\phi}^{(2)}_3(t)$ capture higher-order glycemic reaction patterns and explain much smaller proportions of variability ($\leq10$\%).

As in the single level case, the DASH4D/lower sodium diet exhibits less variability among subject-specific glucose response (compare first row, first column in Figure~\ref{fig:EV_ML} to the first panel in Figure~\ref{fig:EV_SL}). This diet also has lower meal-to-meal variability, evident in the second row, first column of Figure~\ref{fig:EV_ML}. Across all diets, the second eigenvalues at the subject level are very small, particularly compared to the first two eigenvalues at the meal level (compare the first row, first column and second row, first column panels in Figure~\ref{fig:EV_ML}). This suggests that most variability can be partitioned in a constant vertical shift at the subject level and more complex meal-to-meal variation.

Each FAST MFPCA model takes $1-3$ minutes to run $2000$ iterations on a standard laptop, with the first $1000$ discarded as burn-in. We again utilize the routine described in Section~\ref{subsec:practical} to assess convergence of the scores and FPCs at both levels. This requires aligning each level separately. Final Gelman-Rubin statistics were $< 1.05$. 

\section{Discussion}\label{sec:discussion}

Standard implementations of FPCA use a two-step procedure: (1) diagonalize  a smooth estimate of the observed data covariance; and (2) conduct inference conditional on the FPC estimates from the first step. This does not account for sampling variability in the FPCs, which could be quite large for low to moderate  signal-to-noise. Our approach is a single-step joint Bayesian model that treats PCs as parameters in the model, which combines the orthogonal spline expansion from \cite{ye_functional_2024}, the parameter expansion  based on polar decomposition from \cite{jauch_monte_2021}, and stabilizing constraints to efficiently jointly model all FPCA model components. Combining the right ingredients led to FAST, a stable, replicable, and easy to implement methodology. The central element of FAST is expressing the FPCs in terms of an orthonormal spline basis. This is different from the approach in \cite{ye_functional_2024}, as this spline basis: (1) is orthonormal in $L^2([0,1])$, rather than in a vector sense; and (2) retains the computational efficiency of B-splines \citep{liu_splinets_2020}. This spline expansion simplifies the problem of sampling the high dimensional FPCs to sampling a low dimensional spline coefficients.
The FPC spline coefficient matrix must be orthonormal to produce the corresponding FPC samples. This is accomplished using the polar decomposition parameter expansion of \cite{jauch_monte_2021}, applying the technique to the coefficient matrix rather than the data level FPCs.

FAST performs as well or better than all existing methods, while being highly computational efficient and having excellent mixing properties. Moreover, the implementation of FAST within {\ttfamily STAN}, or equivalent software, allows for a widespread use of methods and requires only small adjustments for when tailoring the model for other applications. 

\clearpage

\begin{singlespace}

\bibliographystyle{apalike}

\bibliography{BFPCA}
\end{singlespace}

\pagebreak

\section*{SUPPLEMENTAL MATERIAL}
\beginsupplement

\section{Penalty matrix $\mathbf{P}_{\alpha}$}\label{supp:Penalty}

Recall we chose a posterior addition which mixes second- and zero-order penalties of the form $\alpha \int f^2(t) dt + (1-\alpha) \int \{f''(t)\}^2 dt$. For spline parameters $\theta$ such that $f(t) \approx \mathbf{B}(t)\theta$, there exist unique penalty matrices $\mathbf{P}_0, \mathbf{P}_2$ such that $\int f^2(t) dt \approx \theta^t \mathbf{P}_0\theta$ and $\int \{f''(t)\}^2 dt \approx \theta^t\mathbf{P}_2\theta$ \citep{cravenwahba1979,kimeldorfwahba1970,osullivan1986,Wahba1983}. 

We proceed by defining both of the penalty components $\mathbf{P}_0, \mathbf{P}_2$ separately. Let $\mathbf{B}(t) = [b_1(t)|\ldots,|b_Q(t)]$ represent the basis functions, chosen to be orthonormal in $L^2([0,1])$. First, we define the zero-order penalty $\mathbf{P}_0$ element-wise. This derivation leverages the orthonormal definition of the $b_i(t)$.
\begin{align*}
    (\mathbf{P}_0)_{ij} & = \int_0^1 \int_0^1 b_i(t) b_j(t) dt\\
    & = \begin{cases} 1 \text{ when } i = j\\
    0 \text{ otherwise} \end{cases}
\end{align*}
The resulting matrix $\mathbf{P}_0 = \mathbf{I}_Q$ by the definition of the basis $\mathbf{B}(t)$. 

Next, we define the more central ``wiggliness" penalty $\mathbf{P}_2$. For this penalty, based on the squared second derivative, we introduce the second derivatives of the basis $\mathbf{B}(T)$: $\mathbf{B}''(t) = [b_1''(t)|\ldots,|b_Q''(t)]$. We are able to quickly retrieve these derivatives using the properties of B-splines, from which the default Splinet basis is constructed \citep{liu_splinets_2020}. The elements of $\mathbf{P}_2$ are as follows.
\begin{align*}
    (\mathbf{P}_2)_{ij} & = \int_0^1 \int_0^1 b''_i(t) b_j''(t) dt
\end{align*}
We approximate these quantities using numerical integration.

To ensure uniformity of penalization scale between $\mathbf{P}_0$ and $\mathbf{P}_2$, we scale both raw penalties such that their leading eigenvalues are 1. We then follow \cite{goldsmith_generalized_2015} and define $\mathbf{P}_{\alpha} = \alpha \mathbf{P}_0 + (1-\alpha) \mathbf{P}_2$. This allows us to write the final penalty using a single quadratic form: 
\begin{align*}
    \alpha \theta^t \mathbf{P}_0 \theta + (1- \alpha) \theta^t \mathbf{P}_2 \theta &  = \theta^t(\alpha \mathbf{P}_0 + (1-\alpha) \mathbf{P}_2) \theta\\
    & = \theta^t\mathbf{P}_{\alpha}\theta
\end{align*}

Considering the form of $\mathbf{P}_\alpha$, it becomes clear that $\mathbf{P}_0$ acts to ensure the non-singularity of the final penalty $\mathbf{P}_{\alpha}$ when $\alpha > 0$, similar to adding a ridge penalty in the context of regression.

\section{Joint Prior on $h_k$ and $\boldsymbol{\Psi}$}\label{supp:Prior_Proper}

We derive sufficient conditions under which the joint prior distribution on the smoothing parameters $h_k$ and the FPC spline coefficient matrix $\boldsymbol{\Psi}$ is proper. This is equivalent to showing under what assumptions it can be integrated to a constant, finite value. We assume here that ${\rm R}(\mathbf{P}_\alpha) = Q$, as this is true for the tested implementation. Letting $K$ refer to the dimension of the FPC basis, $\Gamma(x| a, b)$ refer to a Gamma distribution with expectation $a/b$ evaluated at $x$, $\mathbf{H} = {\rm diag}(h_1, \ldots, h_K)$, and ${\rm etr}(\cdot)$ refer to the exponential of the matrix trace, we proceed:
\begin{align*}
    g(\boldsymbol{\Psi}, \mathbf{H}) & = g_\psi(\boldsymbol{\Psi}|\mathbf{H})g_h(\mathbf{H})\\
    & \propto \frac{{\rm etr}(-\mathbf{H}\boldsymbol{\Psi}^T \mathbf{P}_\alpha \boldsymbol{\Psi}/2)}{{\rm Vol}(\mathcal{V}_{K,Q})} \times \prod_{i = 1}^K h_i^{Q/2} \Gamma(h_i|\alpha_\psi, \beta_\psi)
\end{align*}
With the form of the density defined, we now integrate over the Stiefel manifold $\boldsymbol{\Psi} \in \mathcal{V}_{K, Q}$ and the Gamma-distributed smoothing parameters $h_k \geq 0$. Here, we use that $g_\psi(\boldsymbol{\Psi}|\mathbf{H})$ is proportional to a Matrix Bingham distribution with arguments $\mathbf{H}$ and $-\mathbf{P}_\alpha/2$ (using commutativity of scalar and matrix multiplication). This distribution has normalizing constant $\Phi_{K, Q}(\mathbf{H}, -\mathbf{P}_\alpha/2)^{-1}$, where $\Phi$ is the hypergeometric function of two matrix arguments. This function has the following series definition:
$$\Phi_{K, Q}(\mathbf{H}, -\mathbf{P}_\alpha/2) = \sum_{k = 0}^\infty \frac{1}{k!} \sum_{|\kappa| = k} \frac{C_\kappa(\mathbf{H})C_\kappa(-\mathbf{P}_\alpha/2)}{C(I_K)}$$
\noindent where $C_\kappa()$ denotes the zonal polynomial related to partition $\kappa$ and the inner sum is over all partitions $\kappa$ of weight $k$ and length $\leq K$. Using this normalizing constant, we can integrate the joint prior density of the smoothing parameters and FPC spline coefficients as follows, letting $V = {\rm Vol}(\mathcal{V}_{K,Q})$. Differentials are omitted for notational brevity.
\begin{align*}
     \int_0^{\infty} \cdots \int_0^\infty \int_{\boldsymbol{\Psi} \in \mathcal{V}_{K,Q}} g(\boldsymbol{\Psi}, \mathbf{H}) & = \int_0^\infty \cdots \int_0^\infty \int_{\boldsymbol{\Psi} \in \mathcal{V}_K,Q} \frac{{\rm etr}(-\mathbf{H}\boldsymbol{\Psi}^T \mathbf{P}_\alpha \boldsymbol{\Psi}/2)}{V}  \times \prod_{i = 1}^K h_i^{Q/2} \Gamma(h_i| \alpha_\psi, \beta_\psi)\\
     & = \frac{1}{V}\int_0^\infty \cdots \int_0^\infty \Phi_{K,Q}(\mathbf{H}, -\mathbf{P}_\alpha/2) \times \prod_{i = 1}^K h_i^{Q/2} \Gamma(h_i| \alpha_\psi, \beta_\psi)\\
     & = \frac{1}{V}\int_0^\infty \cdots \int_0^\infty \sum_{k = 0}^\infty \frac{1}{k!} \sum_{|\kappa| = k} \frac{C_\kappa(\mathbf{H})C_\kappa(-\mathbf{P}_\alpha/2)}{C_\kappa(I_K)} \times \prod_{i = 1}^K h_i^{Q/2} \Gamma(h_i| \alpha_\psi, \beta_\psi)
\end{align*}

At this point, we recall that the zonal polynomials are homogeneous (order equal to the weight $|\kappa|$), symmetric polynomials in the eigenvalues of the argument matrix. For positive definite matrix $\mathbf{P}_\alpha$, the eigenvalues of $-\mathbf{P}_\alpha/2$ will be negative. The negative sign contribution from each such eigenvalue will be be present in every factor within each term of the polynomial, with cumulative power corresponding to the uniform order $|\kappa|$. This implies that $C_\kappa(-\mathbf{P}_\alpha/2) = (-1)^{|\kappa|}C_\kappa(\mathbf{P}_\alpha/2)$. Using this fact, the integral of interest becomes the following.
\begin{align*}
    \int_0^{\infty} \cdots \int_0^\infty \int_{\mathcal{V}_{K,Q}} g(\boldsymbol{\Psi}, \mathbf{H}) & = \frac{1}{V}\int_0^\infty \cdots \int_0^\infty \sum_{k = 0}^\infty \frac{(-1)^k}{k!} \sum_{|\kappa| = k} \frac{C_\kappa(\mathbf{H})C_\kappa(\mathbf{P}_\alpha/2)}{C_\kappa(I_K)} \times \prod_{i = 1}^K h_i^{Q/2} \Gamma(h_i| \alpha_\psi, \beta_\psi)
\end{align*}

We now recall the combinatorial definition of zonal polynomials. In particular, each zonal polynomial $C_\kappa(A)$ for partition $\kappa$ and matrix $A \in m\times m$ can be represented as a linear combination of the monomial symmetric functions $M_\lambda(\cdot)$ in the eigenvalues of $A$ $(\sigma_1 \ldots, \sigma_m)$:  
$$C_\kappa(A) = \sum_{\lambda \leq \kappa} c_{ \kappa, \lambda} M_\lambda(\sigma_1,\ldots, \sigma_m)$$
\noindent These monomial symmetric functions $M_\lambda(\cdot)$ correspond to partitions $\lambda$ which have equal weight to $(|\lambda| = |\kappa|)$ and are before $\kappa$ according to lexicographic ordering. All coefficients $c_{\kappa, \lambda}$ are non-negative. The monomial symmetric functions are defined as follows for partition $\lambda$ of length $\mathcal{L}$.
$$M_\lambda(\sigma_1, \ldots, \sigma_m) = \sum_{p \in P}\sigma_{p_1}^{\lambda_1}\cdots \sigma_{p_\mathcal{L}}^{\lambda_{\mathcal{L}}}$$
\noindent where the summation is taken over all distinct permutations $p \in P$ of $\mathcal{L}$ elements drawn from $\{1,\ldots, m\}$ \citep{bagyan_complete_2024}. By this definition, it follows that the zonal polynomial of a positive semi-definite matrix will be positive when the length of the partition is at most the rank of the matrix, taking zero-value when the length is greater. Note that the summation over partitions $\kappa$ above requires that all have length $\leq K < Q$.

As $\mathbf{P}_\alpha/2$ and $I_K$ are positive definite, it follows that $C_\kappa(\mathbf{P}_\alpha/2) > 0$ and $C_\kappa(I_K) > 0$. Similarly, $\mathbf{H}$ being positive semi-definite implies that $C_\kappa(\mathbf{H}) \geq 0$. This, combined with the non-negativity of each factor $\frac{h_i^{Q/2}}{k!} \Gamma(h_i|\alpha_\psi, \beta_\psi)$, implies that the sign of the series indexed by $k$ is entirely decided by the $(-1)^k$ term. Using this fact, we can break the summation into positive and negative portions.
\begin{align*}
    \int_0^{\infty} \cdots \int_0^\infty \int_{\mathcal{V}_{K,Q}} g(\boldsymbol{\Psi}, \mathbf{H}) & = \frac{1}{V}\int_0^\infty \cdots \int_0^\infty \sum_{k = 0}^\infty \frac{1}{(2k)!} \sum_{|\kappa| = 2k} \frac{C_\kappa(\mathbf{H})C_\kappa(\mathbf{P}_\alpha/2)}{C_\kappa(I_K)} \times \prod_{i = 1}^K h_i^{Q/2} \Gamma(h_i| \alpha_\psi, \beta_\psi)\\
    & - \frac{1}{V}\int_0^\infty \cdots \int_0^\infty \sum_{k = 0}^\infty \frac{1}{(2k+1)!} \sum_{|\kappa| = 2k+1} \frac{C_\kappa(\mathbf{H})C_\kappa(\mathbf{P}_\alpha/2)}{C_\kappa(I_K)} \times \prod_{i = 1}^K h_i^{Q/2} \Gamma(h_i| \alpha_\psi, \beta_\psi)
\end{align*}

We will now apply Tonelli's theorem to each part, once for each Gamma-distributed smoothing parameter. To verify that Tonelli's theorem applies, we consider the necessary assumptions: (1) the measure spaces integrated over must be $\sigma-$finite, (2) the integrand must be non-negative and, (3) the integrand must be measurable. Each application of Tonelli's theorem here applies to a function $f(h_j, k)$ as defined for the first and second terms below. We use here that zonal polynomials are polynomial functions of the eigenvalues, which are just the diagonal elements for a diagonal matrix such as $\mathbf{H}$.
\begin{align*}
    \text{Term 1: } f(h_j, k) & = \frac{1}{(2k)!} \sum_{|\kappa| = 2k} \frac{C_\kappa(h_1, \ldots, h_j, \ldots, h_K)C_\kappa(\mathbf{P}_\alpha/2)}{C_\kappa(I_K)} \times \prod_{i = 1}^K h_i^{Q/2} \Gamma(h_i| \alpha_\psi, \beta_\psi)\\
    \text{Term 2: } f(h_j, k) & =\frac{1}{(2k+1)!} \sum_{|\kappa| = 2k+1} \frac{C_\kappa(h_1, \ldots, h_j, \ldots, h_K)C_\kappa(\mathbf{P}_\alpha/2)}{C_\kappa(I_K)} \times \prod_{i = 1}^K h_i^{Q/2} \Gamma(h_i| \alpha_\psi, \beta_\psi)
\end{align*}

We consider the $f(h_j, k)$ for both terms concurrently. In these functions, $h_j \in (\mathbb{R}^+, \mathbf{B}(\mathbb{R}^+), \gamma)$ for probability measure $\gamma$ corresponding to the Gamma distribution and $k \in (\mathbb{N}, 2^{\mathbb{N}}, \mu)$ for counting measure $\mu$. To assumption (1), probability measures and the counting measure on $\mathbb{N}$ are both known to be $\sigma-$finite by definition. To assumption (2), we have already demonstrated that $f(h_j, k)$ in both cases is non-negative using the properties of zonal polynomials for positive semi-definite matrices. For the final assumption (3), note that it suffices to show that the $f(h_j, k)$ are Borel-measurable for fixed $k$ (see \cite{bartle_product_1995} chapter 10). This follows directly from the fact that each term will reduce to an exponential polynomial in $h_j$, which are continuous and thus measurable. With this in mind, we proceed:
\begin{align*}
     \int_0^{\infty} \cdots \int_0^\infty \int_{\mathcal{V}_{K,Q}} g(\boldsymbol{\Psi}, \mathbf{H}) & = \frac{1}{V} \sum_{k = 0}^\infty \frac{1}{(2k)!} \sum_{|\kappa| = 2k} \frac{C_\kappa(\mathbf{P}_\alpha/2)}{C_\kappa(I_K)} \times \int_0^\infty \cdots \int_0^\infty C_\kappa(\mathbf{H})\prod_{i = 1}^K h_i^{Q/2} \Gamma(h_i| \alpha_\psi, \beta_\psi)\\
    & - \frac{1}{V} \sum_{k = 0}^\infty \frac{1}{(2k+1)!} \sum_{|\kappa| = 2k+1} \frac{C_\kappa(\mathbf{P}_\alpha/2)}{C_\kappa(I_K)} \times \int_0^\infty \cdots \int_0^\infty C_\kappa(\mathbf{H})\prod_{i = 1}^K h_i^{Q/2} \Gamma(h_i| \alpha_\psi, \beta_\psi)
\end{align*}

At this point, we are able to combine the difference back into a single sum, resulting in the following:
\begin{align*}
    \int_0^{\infty} \cdots \int_0^\infty \int_{\mathcal{V}_{K,Q}} g(\boldsymbol{\Psi}, \mathbf{H}) & = \frac{1}{V} \sum_{k = 0}^\infty \frac{(-1)^k}{k!} \sum_{|\kappa| = k} \frac{C_\kappa(\mathbf{P}_\alpha/2)}{C_\kappa(I_K)} \times \int_0^\infty \cdots \int_0^\infty C_\kappa(\mathbf{H})\prod_{i = 1}^K h_i^{Q/2} \Gamma(h_i| \alpha_\psi, \beta_\psi)
\end{align*}

To prove that this alternating series converges, it is sufficient to prove absolute convergence. We define the series of the absolute values below using $S_A(\alpha_\psi, \beta_\psi)$ -- a function of the smoothing prior parameters $\alpha_\psi, \beta_\psi$. If this series converges, then the integral $\int_0^{\infty} \cdots \int_0^\infty \int_{\mathcal{V}_{K,Q}} g(\boldsymbol{\Psi}, \mathbf{H})$ will be finite as desired.
\begin{align*}
    S_A(\alpha_\psi, \beta_\psi) & = \frac{1}{V} \sum_{k = 0}^\infty \frac{1}{k!} \sum_{|\kappa| = k} \frac{C_\kappa(\mathbf{P}_\alpha/2)}{C_\kappa(I_K)} \times \int_0^\infty \cdots \int_0^\infty C_\kappa(\mathbf{H})\prod_{i = 1}^K h_i^{Q/2} \Gamma(h_i| \alpha_\psi, \beta_\psi)
\end{align*}

We can now apply inequality $5.26$ from \cite{bagyan_complete_2024}, derived from \cite{faraut_analysis_1994}. This step will provide an upper bounding series for the non-negative $S_A(\alpha_\psi, \beta_\psi)$ which is easier to work with. This inequality states that, for $\Sigma$ being a $d \times d$ positive definite matrix with ordered eigenvalues $\sigma_1 \geq \ldots \geq \sigma_d$, it follows that 
$$C_\kappa(\Sigma) \leq C_\kappa (I_d) \prod_{j = 1 }^d \sigma_j^{\kappa_j}$$
\noindent where the partition $\kappa$ is permitted to contain zero entries. The diagonal matrix $\mathbf{H}$ will have eigenvalues equal to its diagonal elements $h_k$, which will be non-zero with probability 1, meaning $\mathbf{H}$ will be positive definite with probability 1. This almost-sure inequality of the integrands will produce the proper inequality when taking the requisite expectation/integral.
\begin{align*}
    S_A(\alpha_\psi, \beta_\psi) & \leq  \frac{1}{V} \sum_{k = 0}^\infty \frac{1}{k!} \sum_{|\kappa| = k} \frac{C_\kappa(I_K)C_\kappa(\mathbf{P}_\alpha/2)}{C_\kappa(I_K)} \times \int_0^\infty \cdots \int_0^\infty \prod_{i = 1}^K h_i^{\kappa_i + Q/2} \Gamma(h_i| \alpha_\psi, \beta_\psi)\\
    & = \frac{1}{V} \sum_{k = 0}^\infty \frac{1}{k!} \sum_{|\kappa| = k} C_\kappa(\mathbf{P}_\alpha/2) \prod_{i = 1}^K \mathbb{E}[h_i^{\kappa_i + Q/2}]
\end{align*}

We have reduced the integral of interest to the independent product of moments of each smoothing coefficient. Applying the known form of these moments in terms of the shape $\alpha_\psi$ and rate $\beta_\psi$ results in the following:
\begin{align*}
    S_A(\alpha_\psi, \beta_\psi) & \leq \frac{1}{V} \sum_{k = 0}^\infty \frac{1}{k!} \sum_{|\kappa| = k} C_\kappa(\mathbf{P}_\alpha/2) \prod_{i = 1}^K \frac{\Gamma(\alpha_\psi + \kappa_i + Q/2)}{\beta_\psi^{\kappa_i + Q/2}\Gamma(\alpha_\psi)}\\
    & = \frac{1}{V}\sum_{k = 0}^\infty \frac{1}{k! \times \beta_\psi^{k + KQ/2}} \sum_{|\kappa| = k} C_\kappa(\mathbf{P}_\alpha/2) \prod_{i = 1}^K \prod_{j = 1}^{\kappa_i + Q/2}(\alpha_\psi + j - 1)
\end{align*}

We will now consider the last factor in the above series, for which we provide the shorthand $Pr(\kappa) = \prod_{i = 1}^K \prod_{j = 1}^{\kappa_i + Q/2}(\alpha_\psi + j - 1)$. We endeavor to demonstrate that $Pr(\kappa)$ is maximized by the partition which places all of the weight on a single entry, denoted $\kappa = (k)$ in literature. Towards this end, we will demonstrate that $Pr(\kappa)$ for $\kappa \neq (k)$ is always strictly increased by moving weight to the first entry from the last non-zero entry. Before proceeding, it is important to recall that each $\kappa$ is defined such that the entries are non-decreasing: $\kappa_1 \geq \kappa_2 \geq \ldots \geq \kappa_K \geq 0$.

Consider an arbitrary integer partition $\kappa \neq (k)$ with length (number of non-zero components) $\mathcal{L}(\kappa)$. We compare $Pr(\kappa)$ with $Pr(\kappa')$ for $\kappa'_1 = \kappa_1 + 1, \kappa'_i = \kappa_i$ when $1 < i < \mathcal{L}(\kappa)$, and $\kappa'_{\mathcal{L}(\kappa)} = \kappa_{\mathcal{L}(\mathcal{\kappa})} - 1$. This $\kappa'$ still has the same weight, and should satisfy the ordering requirement, so it is a well-defined partition. We can now calculate $\frac{Pr(\kappa')}{Pr(\kappa)}$ as follows. This ratio will always be well-defined, as $Pr(\cdot)$ will always be a product of positive factors.
\begin{align*}
    \frac{Pr(\kappa')}{Pr(\kappa)} & = \frac{\prod_{i = 1}^K\prod_{j = 1}^{\kappa'_i + Q/2}(\alpha_\psi + j - 1)}{\prod_{i = 1}^K\prod_{j = 1}^{\kappa_i + Q/2}(\alpha_\psi + j - 1)}\\
    & = \frac{\prod_{j = 1}^{(\kappa_1 + 1) +Q/2}(\alpha_\psi + j - 1)}{\prod_{j =1}^{\kappa_1 + Q/2}(\alpha_\psi + j - 1)} \times \frac{\prod_{j = 1}^{(\kappa_{\mathcal{L}(\kappa)} - 1) +Q/2}(\alpha_\psi + j - 1)}{\prod_{j =1}^{\kappa_{\mathcal{L}(\kappa)} + Q/2}(\alpha_\psi + j - 1)}\\
    & = \frac{\alpha_\psi + \kappa_1 + Q/2}{\alpha_\psi + \kappa_{\mathcal{L}(\kappa)} + Q/2 - 1}
\end{align*}

As $\kappa_1 \geq \kappa_{\mathcal{L(\kappa)}}$ inherently, it follows that the above ratio will be $> 1$, so $Pr(\cdot)$ is strictly increased by moving weight to the first entry. Then, beginning at any arbitrary partition, transitioning weight from the last non-zero entry in the partition to the first will always increase $Pr(\cdot)$. As any partition for can be related to $(k)$ through a series of such "transitions", it follows from the transitive property that $Pr[(k)] > Pr(\kappa)$ for arbitrary partition $\kappa$.

Using the above result, we establish another bounding Series on $S_A(\alpha_\psi, \beta_\psi)$. Through this step, we are able to move the $Pr(\cdot)$ term out of the summation over partitions of weight $k$, allowing use of the identity $\sum_{|\kappa| = k} C_\kappa(A) = {\rm tr}(A)^k$ \citep{bagyan_complete_2024}.
\begin{align*}
    S_A(\alpha_\psi, \beta_\psi) & \leq \frac{1}{V}\sum_{k = 0}^\infty \frac{1}{k! \times \beta_\psi^{k + KQ/2}} \sum_{|\kappa| = k} C_\kappa(\mathbf{P}_\alpha/2) Pr(\kappa)\\
    & < \frac{1}{V}\sum_{k = 0}^\infty \frac{1}{k! \times \beta_\psi^{k + KQ/2}} \sum_{|\kappa| = k} C_\kappa(\mathbf{P}_\alpha/2) Pr[(k)]\\
    & = \frac{1}{V}\sum_{k = 0}^\infty \frac{1}{k! \times \beta_\psi^{k + KQ/2}} \left[\sum_{|\kappa| = k} C_\kappa(\mathbf{P}_\alpha/2) \right] Pr[(k)]\\
    & = \frac{1}{V}\sum_{k = 0}^\infty \frac{1}{k! \times \beta_\psi^{k + KQ/2}} {\rm tr}(\mathbf{P}_\alpha/2)^k Pr[(k)]\\
    & = \frac{1}{V}\sum_{k = 0}^\infty \frac{1}{k! \times \beta_\psi^{k + KQ/2} \times 2^k} {\rm tr}(\mathbf{P}_\alpha)^k Pr[(k)]
\end{align*}

We can now discern under what conditions this bounding series, and thus the original series by comparison test, converge. This is accomplished below using the ratio test.
\begin{align*}
    a_k & =  \frac{1}{k! \times \beta_\psi^{k + KQ/2} \times 2^k} {\rm tr}(\mathbf{P}_\alpha)^k Pr[(k)]\\
    \implies \frac{a_{k+1}}{a_k} & = \frac{\frac{1}{(k+1)! \times \beta_\psi^{(k+1) + KQ/2} \times 2^{k+1}} {\rm tr}(\mathbf{P}_\alpha)^{(k+1)} Pr[(k+1)]}{\frac{1}{k! \times \beta_\psi^{k + KQ/2} \times 2^k} {\rm tr}(\mathbf{P}_\alpha)^k Pr[(k)]}\\
    & = \frac{{\rm tr}(\mathbf{P}_\alpha)}{2(k+1)\beta_\psi} \times \frac{\prod_{j = 1}^{(k+1) + Q/2}(\alpha_\psi + j - 1) \times \prod_{i = 2}^{K}\prod_{p = 1}^{Q/2}(\alpha_{\psi} + p - 1)}{\prod_{j' = 1}^{k + Q/2}(\alpha_\psi + j' - 1) \times \prod_{i' = 2}^{K}\prod_{p' = 1}^{Q/2}(\alpha_{\psi} + p' - 1)}\\
    & = \frac{{\rm tr}(\mathbf{P}_\alpha)(\alpha_\psi + k + Q/2)}{2(k+1)\beta_\psi}
\end{align*}

Taking the limit as $k \rightarrow \infty$ provides the following.
\begin{align*}
    \lim_{k \rightarrow \infty} \frac{a_{k+1}}{a_k} & = \lim_{k \rightarrow \infty} \frac{{\rm tr}(\mathbf{P}_\alpha)(\alpha_\psi + k + Q/2)}{2(k+1)\beta_\psi}\\
    & = \frac{{\rm tr}(\mathbf{P}_\alpha)}{2\beta_\psi} \lim_{k \rightarrow \infty} \frac{\alpha_\psi + k + Q/2}{k + 1}\\
    & = \frac{{\rm tr}(\mathbf{P}_\alpha)}{2\beta_\psi}
\end{align*}

By the ratio test, the above series $a_k$ converges if and only if $\lim_{k \rightarrow \infty} \frac{a_{k + 1}}{a_k} < 1$. This occurs precisely when $ \frac{{\rm tr}(\mathbf{P}_\alpha)}{2\beta_\psi} < 1 \implies {\rm tr}(\mathbf{P}_\alpha)/2 < \beta_\psi$. This informs our choice of the prior parameter $\beta_\psi$, as it must be greater than the half of the trace (or half of the sum of the eigenvalues) for the penalty matrix. This makes some intuitive sense, as the smoothing parameters having overly heavy upper tails, combined with there being non-trivial regions of the manifold $\mathcal{V}_{K,Q}$ over which penalization is quite extreme (quantified through the eigenvalues themselves), could result in an unbounded integrated penalty. 

\section{Joint and conditional posteriors}\label{supp:CondPost}

We begin by enumerating the model likelihood and parameter priors explicitly. We use $\Gamma(a,b)$ to denote the gamma distribution with shape $a$ and rate $b$, $\Gamma^{-1}(a,b)$ to denote the inverse gamma distribution with shape $a$ and scale $b$, $N(\mu, \sigma^2)$ to indicate the normal distribution with mean $\mu$ and variance $\sigma^2$, and $MVN(\boldsymbol{\mu}, \boldsymbol{\Sigma})$ to denote the multivariate normal distribution with mean $\boldsymbol{\mu}$ and variance-covariance $\boldsymbol{\Sigma}$. Throughout, $I_A$ denotes the identity matrix of dimension $A$.

\vspace{0.5em}
\underline{Variance Component Priors:}

Prior distributions of the variance and smoothing parameter components are as follows.
\begin{align*}
    \sigma^2 & \sim \Gamma^{-1}(\alpha_\sigma,\beta_\sigma)\\
    \lambda_k & \sim \Gamma^{-1}(\alpha_\lambda,\beta_\lambda) \quad \forall k = 1,\ldots,K\\
    h_\mu & \sim \Gamma(\alpha_\mu,\beta_\mu) \\
     h_k & \sim \Gamma(\alpha_\psi,\beta_\psi) \quad \forall k = 1,\ldots,K
\end{align*}
We set nearly uninformative priors, $\alpha_\sigma, \alpha_\lambda, \alpha_\psi, \alpha_\mu = 0.001$ and $\beta_\sigma, \beta_\lambda, \beta_\psi, \beta_\mu = 0.001$.

We also enforce ordering of the eigenvalues, requiring the indicator $\mathbbm{1}(\lambda_1 \geq \ldots\geq \lambda_K)$.

\vspace{0.5em}
\underline{Prior on $w_\mu$:}

FAST implements a smoothing prior on $\mu(t)$ through a posterior penalty on spline coefficients $w_\mu$. For prior density $f(w_\mu)$, we have the following, where $\mathbf{R}(\cdot)$ indicates matrix rank.
$$f(w_\mu) \propto h_\mu^{\mathbf{R}(\mathbf{P}_\alpha)/2}\exp\left\{-\frac{h_{\mu}}{2} w^t_{\mu}\mathbf{P}_{\alpha} w_{\mu}\right\}$$

\vspace{0.5em}
\underline{Prior on FPCs:}

FAST implements a similar smoothing prior on the $\phi_k(t)$ through the spline coefficients $\psi_k$, For prior density $f(\psi_k)$, we have the following, where $\mathbf{R}(\cdot)$ indicates matrix rank.
$$f(\psi_k) \propto h_k^{\text{R}(\mathbf{P}_\alpha)/2}\exp\left\{-\frac{h_k}{2} \psi^t_k\mathbf{P}_{\alpha}\psi_k\right\}$$
For orthonormality of the FPCs, FAST requires the additional constraint that the matrix $\boldsymbol{\Psi} = [\psi_1| \ldots| \psi_K]$ is orthonormal. This requires introducing the indicator $\mathbbm{1}(\boldsymbol{\Psi} \in \mathcal{V}_{K,Q})$ to the prior distribution, where $\mathcal{V}_{K,Q}$ is the Stiefel manifold of dimension $Q \times K$.

\vspace{0.5em}
\underline{Prior on the Scores:}

Score priors are as defined in FPCA according to the Kosambi-Karhunen-Loève decomposition: $\xi_{ik} \sim N(0, \lambda_k)$ for $i \in \{1,2,\ldots, N\}$ and $k \in \{1,2,\ldots, K\}$.

\vspace{0.5em}
\underline{Model Likelihood:}

We begin with notation. Let $Y_i \in \mathbb{R}^M$ represent the vector of data for participant $i$ observed at the times $t \in \{t_1, \ldots, t_M\}$. Correspondingly, let $\mathbf{B} \in \mathbb{R}^{M \times Q}$ be the orthonormal basis matrix evaluated at the same set of observation points. Using these definitions, the FPCA model likelihood contribution is as follows.
$$Y_i \sim MVN\{\mathbf{B}(w_\mu + \sum_{k = 1}^K \xi_{ik} \psi_k), \sigma^2I_M\}$$

\vspace{0.5em}
\underline{Joint Posterior:}

Given the likelihood and priors previously described, the posterior density of all model parameters given the data is proportional to the following. Throughout, we use the notation $f(X|\Theta)$ to represent the evaluation of density $f(\cdot)$ at point $X$ with parameters $\Theta$.
\begin{equation*}
        \begin{split}
        &\prod_{i = 1}^N MVN\{Y_i|\mathbf{B}(w_{\mu} + \sum_{k = 1}^K \xi_{ik} \psi_{k}), \sigma^2 I_M\} \times \Gamma^{-1}(\sigma^2|\alpha_{\sigma}, \beta_{\sigma}) \\
         &\times \left\{\prod_{k = 1}^K N(\xi_{ik}|0, \lambda_k) \times \Gamma^{-1}(\lambda_k|\alpha_{\lambda}, \beta_{\lambda}) \times h_k^{\mathbf{R}(\mathbf{P}_\alpha)/2}\exp\left(-\frac{h_k}{2}\psi_k^t\mathbf{P}_{\alpha}\psi_k\right)\times \Gamma(h_k|\alpha_{\psi}, \beta_{\psi}) \right\}  \\
        &\times h_\mu^{\mathbf{R}(\mathbf{P}_\alpha)/2}\exp\left(-\frac{h_{\mu}}{2} w^t_{\mu}\mathbf{P}_{\alpha} w_{\mu}\right) \times \Gamma(h_{\mu}|\alpha_{\mu}, \beta_{\mu})\times \mathbbm{1}(\lambda_1 \geq \ldots \geq \lambda_K) \times \mathbbm{1}(\mathbf{\Psi} \in \mathcal{V}_{Q, K})
        \end{split}
        \label{EQ:Posterior}
\end{equation*}
Using this joint posterior form, we begin to derive the individual conditional posteriors for each component.

\vspace{0.5em}
\underline{Smoothing Parameters:}

First, we derive the conditional posterior for the smoothing parameter for the mean, $h_\mu$, which we denote $f(h_\mu|{\rm others})$.
\begin{align*}
    f(h_{\mu}|{\rm others}) & \propto h_\mu^{\mathbf{R}(\mathbf{P}_\alpha)/2}\exp\left(-\frac{h_{\mu}}{2} w^t_{\mu}\mathbf{P}_{\alpha} w_{\mu}\right) \times \Gamma(h_\mu|\alpha_\mu,\beta_\mu)\\
    & \propto h_\mu^{\mathbf{R}(\mathbf{P}_\alpha)/2}\exp\left(-\frac{h_{\mu}}{2} w^t_{\mu}\mathbf{P}_{\alpha} w_{\mu}\right) \times  h_{\mu}^{\alpha_\mu - 1} \exp(-\beta_\mu h_{\mu})\\
    & \propto h_{\mu}^{\{\mathbf{R}(\mathbf{P}_\alpha)/2 + \alpha_\mu\} - 1} \exp\{-\left(\frac{w_{\mu}^t\mathbf{P}_\alpha w_{\mu}}{2} + \beta_\mu\right) h_{\mu}\}
\end{align*}
One can recognize the form of the gamma distribution in the last line above.
$$[h_\mu|{\rm others}] \sim \Gamma(\mathbf{R}(\mathbf{P}_\alpha)/2 + \alpha_\mu, w_{\mu}^t\mathbf{P}_\alpha w_{\mu}/2 + \beta_\mu)$$

We can similarly derive the conditional posterior distribution for the general eigenfunction smoothing parameter $h_k$, which we denote $f(h_k|{\rm others})$.
\begin{align*}
    f(h_{k}|{\rm others}) & \propto h_k^{\mathbf{R}(\mathbf{P}_\alpha)/2}\exp\left(-\frac{h_k}{2}\psi_k^t\mathbf{P}_{\alpha}\psi_k\right) \times \Gamma(h_k|\alpha_\psi, \beta_\psi)\\
    & \propto h_k^{\mathbf{R}(\mathbf{P}_\alpha)/2}\exp\left(-\frac{h_k}{2}\psi_k^t\mathbf{P}_{\alpha}\psi_k\right) \times h_{k}^{\alpha_\psi - 1} \exp(-\beta_\psi h_{k})\\
    & \propto h_{k}^{\{\mathbf{R}(\mathbf{P}_\alpha)/2 + \alpha_\psi\} - 1} \exp\{-\left(\frac{\psi_k^t\mathbf{P}_\alpha \psi_k}{2} + 
    \beta_\psi\right)h_{k}\}
\end{align*}
Once more, the form of the gamma distribution is clear in the final line above.
$$[h_k|{\rm others}] \sim \Gamma(\mathbf{R}(\mathbf{P}_\alpha)/2 + \alpha_\psi, \psi_k^t \mathbf{P}_\alpha \psi_k/2 + \beta_\psi)$$

\vspace{0.5em}
\underline{Noise variance $\sigma^2$:}

We derive the conditional posterior of $\sigma^2$, denoted $f(\sigma^2|{\rm others})$, below.
\begin{align*}
    f(\sigma^2|{\rm others}) & \propto \prod_{i = 1}^N MVN\{Y_i|\mathbf{B}(w_{\mu} + \sum_{k = 1}^K \xi_{ik} \psi_{k}), \sigma^2 I_M\} \times \Gamma^{-1}(\sigma^2|\alpha_\sigma, \beta_\sigma)\\
    & \propto \prod_{i = 1}^N (\sigma^2)^{-M/2} \exp(-\frac{||Y_i - \mathbf{B}(w_{\mu} + \sum_{k = 1}^K \xi_{ik} \psi_{k})||^2}{2\sigma^2}) \times (\sigma^2)^{-\alpha_\sigma - 1}\exp(-\frac{\beta_\sigma}{\sigma^2})\\
    & \propto (\sigma^2)^{-(NM/2 + \alpha_\sigma) - 1} \exp\{-\frac{1}{\sigma^2}\left(\frac{1}{2}\sum_{i = 1}^n||Y_i - \mathbf{B}(w_\mu + \sum_{k = 1}^K \xi_{ik} \psi_k)||^2 + \beta_\sigma\right)\}
\end{align*}
The last line above is clearly the form of the inverse gamma distribution, with exact parameterization as follows:
$$[\sigma^2|{\rm others}] \sim \Gamma^{-1}(\frac{NM}{2} + \alpha_\sigma, \frac{1}{2}\sum_{i = 1}^n||Y_i - \mathbf{B}(w_\mu + \sum_{k = 1}^K \xi_{ik} \psi_k)||^2 + \beta_\sigma \bigr])$$

\vspace{0.5em}
\underline{Eigenvalues $\lambda_k$:}

The joint conditional posterior distribution of the eigenvalaues $\lambda_k$, denoted $f(\lambda_1,\ldots, \lambda_K|{\rm others})$, is derived below.
\begin{align*}
    f(\lambda_1,\ldots, \lambda_K|{\rm others}) & \propto \prod_{k = 1}^K \prod_{i = 1}^N N(\xi_{ik}|0, \lambda_k) \times \Gamma^{-1}(\lambda_k|\alpha_\lambda,\beta_\lambda) \times \mathbbm{1}(\lambda_1 \geq \ldots \geq \lambda_K) \\
    & \propto \prod_{k = 1}^K \lambda_k^{-N/2}\exp(-\frac{1}{2\lambda_k}\sum_{i = 1}^N \xi_{ik}^2)\times \lambda_k^{-\alpha_\lambda - 1} \exp(-\frac{\beta_\lambda}{\lambda_k}) \times \mathbbm{1}(\lambda_1 \geq \ldots \geq \lambda_K)\\
    & \propto \prod_{k = 1}^K \lambda_k^{-(N/2 + \alpha_\lambda) - 1} \exp\{-\frac{1}{\lambda_k}\left(\frac{1}{2}\sum_{i = 1}^N \xi_{ik}^2 + \beta_\lambda \right)\} \times \mathbbm{1}(\lambda_1 \geq \ldots \geq \lambda_K)
\end{align*}
The above joint distribution has the form of independent inverse gamma distributions ($[\lambda_k|{\rm others}] \sim \Gamma^{-1}(N/2 + \alpha_\lambda, \frac{1}{2}\sum_{i = 1}^N \xi_{ik}^2 + \beta_\lambda)$), with the additional constraint that the $\lambda_k$ be ordered. Sampling from this type of joint distribution is possible using ordered transforms and corresponding Jacobian transforms as described in the STAN documentation \citep{team_constraint}.

\vspace{0.5em}
\underline{Mean spline coefficients $w_\mu$:}

We derive the conditional posterior of $w_\mu$, denoted $f(w_\mu|{\rm others})$ as follows.
\begin{align*}
    f(w_\mu|{\rm others}) & \propto \prod_{i = 1}^N  MVN\{Y_i|\mathbf{B}(w_{\mu} + \sum_{k = 1}^K \xi_{ik} \psi_{k}), \sigma^2 I_M\} \times \exp\left(-\frac{h_{\mu}}{2} w^t_{\mu}\mathbf{P}_{\alpha} w_{\mu}\right) \\
    & \propto \prod_{i = 1}^N \exp\left\{-\frac{||\mathbf{B}w_\mu - (Y_i - \mathbf{B}\sum_{k = 1}^K \xi_{ik} \psi_k)||^2}{2\sigma^2} - \frac{h_{\mu}}{2} w^t_{\mu}\mathbf{P}_{\alpha} w_{\mu} \right\} \\
    & \propto \exp\left\{-\sum_{i = 1}^N\frac{||\mathbf{B}w_\mu||^2 - 2\langle Y_i - \mathbf{B}\sum_{k = 1}^K \xi_{ik} \psi_k, \mathbf{B}w_\mu \rangle}{2\sigma^2} - \frac{h_{\mu}}{2} w^t_{\mu}\mathbf{P}_{\alpha} w_{\mu} \right\}
\end{align*}
We now introduce the shorthand $D_i = Y_i - \mathbf{B}\sum_{k = 1}^K \xi_{ik} \psi_k$ to denote the residual between the data $Y_i$ and the cumulative random effect $\mathbf{B}\sum_{k = 1}^K \xi_{ik} \psi_k$. We continue using this notation.
\begin{align*}
    f(w_\mu|{\rm others}) & \propto \exp\left\{-\sum_{i = 1}^N\frac{||\mathbf{B}w_\mu||^2 - 2D_i^t \mathbf{B}w_\mu}{2\sigma^2} - w^t_{\mu}\frac{h_\mu \mathbf{P}_\alpha}{2} w_{\mu} \right\}\\
    & \propto \exp\left\{- w_\mu^t \left(\sum_{i = 1}^N \frac{ \mathbf{B}^t \mathbf{B}}{2\sigma^2}\right) w_\mu + \left(\sum_{i = 1}^N\frac{D_i^t \mathbf{B}}{\sigma^2} \right) w_\mu - w^t_{\mu}\frac{h_\mu \mathbf{P}_\alpha}{2} w_{\mu} \right\}\\
    & \propto \exp\left\{- w_\mu^t \left(\frac{N \mathbf{B}^t \mathbf{B}}{2\sigma^2}  + \frac{h_\mu \mathbf{P}_\alpha}{2}\right) w_\mu + \left(\sum_{i = 1}^N\frac{D_i^t \mathbf{B}}{\sigma^2} \right) w_\mu\right\}\\
    & \propto \exp\left[-\frac{1}{2}\{ w_\mu^t \left(\frac{N \mathbf{B}^t \mathbf{B}}{\sigma^2}  + h_\mu \mathbf{P}_\alpha\right) w_\mu - 2\left(\sum_{i = 1}^N\frac{D_i^t \mathbf{B}}{\sigma^2} \right) w_\mu\}\right]
\end{align*}
Completing the square, we find that $w_\mu$ has a multivariate normal distribution
$$[w_\mu|{\rm others}] \sim MVN\{\left(\frac{N\mathbf{B}^t\mathbf{B}}{\sigma^2} + h_\mu \mathbf{P}_\alpha \right)^{-1} \frac{1}{\sigma^2}\sum_{i =1}^N \mathbf{B}^tD_i,\left(\frac{N\mathbf{B}^t\mathbf{B}}{\sigma^2} + h_\mu \mathbf{P}_\alpha \right)^{-1}\}$$

\vspace{0.5em}
\underline{Scores $\xi_{ik}$:}

For arbitrary score $\xi_{ik}$ (study participant $i$ and FPC $k$), we derive the conditional posterior $f(\xi_{ik}|{\rm others})$ as follows.
\begin{align*}
    f(\xi_{ik}|{\rm others}) & \propto MVN\{Y_i|\mathbf{B}(w_{\mu} + \sum_{p = 1}^K \xi_{ip} \psi_{p}), \sigma^2 I_M\} \times N(\xi_{ik}|0, \lambda_k)\\
    & \propto \exp(-\frac{||Y_i - \mathbf{B}(w_\mu + \sum_{p  =1}^K \xi_{ip} \psi_p)||^2}{2\sigma^2} - \frac{\xi_{ik}^2}{2\lambda_k})\\
    & \propto \exp(-\frac{||Y_i - \mathbf{B}(w_\mu + \sum_{p  \neq k}^K \xi_{ip} \psi_p) - \mathbf{B}\psi_{k} \xi_{ik}||^2}{2\sigma^2} - \frac{\xi_{ik}^2}{2\lambda_k})
\end{align*}
For the sake of conciseness, we now define the residual quantity $P_{ik} = Y_i - \mathbf{B}(w_\mu + \sum_{p  \neq k}^K \xi_{ip} \psi_p)$, which can be thought of as the residual between $Y_i - \mathbf{B}w_\mu$ and the projection of $Y_i - \mathbf{B}w_\mu$ onto the FPCs other than $k$.
\begin{align*}
    f(\xi_{ik}|{\rm others}) & \propto \exp(-\frac{||P_{ik} - \mathbf{B}\psi_{k} \xi_{ik}||^2}{2\sigma^2} - \frac{\xi_{ik}^2}{2\lambda_k})\\
    & \propto \exp(-\frac{||\mathbf{B}\psi_{k}||^2 \xi_{ik}^2 - 2P_{ik}^t \mathbf{B}\psi_{k}\xi_{ik}}{2\sigma^2} - \frac{\xi_{ik}^2}{2\lambda_k})\\
    & \propto \exp[-\frac{1}{2}\left\{\xi_{ik}^2\left(\frac{||\mathbf{B}\psi_k||^2}{\sigma^2} + \frac{1}{\lambda_k}\right) - 2\frac{P_{ik}^t \mathbf{B}\psi_k}{\sigma^2}\xi_{ik}\right\}]
\end{align*}
Completing the square, we find that $\xi_{ik}$ has a normal distribution.
$$[\xi_{ik}|{\rm others}] \sim N\{\left(\frac{||\mathbf{B}\psi_k||^2}{\sigma^2} + \frac{1}{\lambda_k}\right)^{-1}\frac{P_{ik}^t \mathbf{B}\psi_k}{\sigma^2}, \left(\frac{||\mathbf{B}\psi_k||^2}{\sigma^2} + \frac{1}{\lambda_k}\right)^{-1}\}$$

\vspace{0.5em}
\underline{FPC Weights $\boldsymbol{\Psi}$:}

We derive the posterior distribution of the full matrix of FPC spline weights, denoted $f(\boldsymbol{\Psi}|{\rm others})$, below.
\begin{align*}
    f(\boldsymbol{\Psi}|{\rm others}) & \propto \prod_{i = 1}^N MVN\{Y_i|\mathbf{B}(w_{\mu} + \sum_{k = 1}^K \xi_{ik} \psi_{k}), \sigma^2 I_M\} \times \prod_{k = 1}^K h_k^{\mathbf{R}(\mathbf{P}_\alpha)/2}\exp\left(-\frac{h_k}{2}\psi_k^t\mathbf{P}_{\alpha}\psi_k\right) \times \mathbbm{1}(\boldsymbol{\Psi} \in \mathcal{V}_{K,Q})\\
    & \propto \exp(-\frac{1}{2\sigma^2}\sum_{i = 1}^N||Y_i - \mathbf{B}(w_\mu + \sum_{k  =1}^K \xi_{ik} \psi_k)||^2) \times \exp(-\frac{1}{2}\sum_{k = 1}^K h_k \psi_k^t \mathbf{P}_\alpha \psi_k) \times \mathbbm{1}(\boldsymbol{\Psi} \in \mathcal{V}_{K,Q})
\end{align*}
We now introduce the shorthand notation: score vector $\xi_i = \{\xi_{i1}, \ldots, \xi_{iK}\}^t \in \mathbb{R}^K$, diagonal smoothing parameter matrix $\mathbf{H} = \diag(h_1, \ldots, h_K)$, and residual vector $R_i = Y_i - \mathbf{B}w_\mu$. Let $\tr(\cdot)$ indicate the trace of a matrix.
\begin{align*}
    f(\boldsymbol{\Psi}|{\rm others}) & \propto \exp\left\{-\frac{1}{2\sigma^2}\sum_{i = 1}^N||R_i -  \mathbf{B}\boldsymbol{\Psi}\xi_i||^2 -\frac{1}{2}\tr(\mathbf{H} \boldsymbol{\Psi}^t \mathbf{P}_\alpha \boldsymbol{\Psi})\right\} \times \mathbbm{1}(\boldsymbol{\Psi} \in \mathcal{V}_{K,Q})\\
    & \propto \exp\left\{-\frac{1}{2\sigma^2}\sum_{i = 1}^N\left(||R_i||^2 - 2R_i^t \mathbf{B}\boldsymbol{\Psi}\xi_i + ||\mathbf{B}\boldsymbol{\Psi}\xi_i||^2\right) -\frac{1}{2}\tr(\mathbf{H} \boldsymbol{\Psi}^t \mathbf{P}_\alpha \boldsymbol{\Psi})\right\} \times \mathbbm{1}(\boldsymbol{\Psi} \in \mathcal{V}_{K,Q})\\
    & \propto \exp\left\{-\frac{1}{2\sigma^2}\sum_{i = 1}^N\left( - 2R_i^t \mathbf{B}\boldsymbol{\Psi}\xi_i + ||\mathbf{B}\boldsymbol{\Psi}\xi_i||^2\right) -\frac{1}{2}\tr(\mathbf{H} \boldsymbol{\Psi}^t \mathbf{P}_\alpha \boldsymbol{\Psi})\right\} \times \mathbbm{1}(\boldsymbol{\Psi} \in \mathcal{V}_{K,Q})\\
    & \propto \exp\left\{-\frac{1}{2\sigma^2}\left(- 2 \sum_{i = 1}^N R_i^t \mathbf{B}\boldsymbol{\Psi}\xi_i + \sum_{i = 1}^N||\mathbf{B}\boldsymbol{\Psi}\xi_i||^2\right) -\frac{1}{2}\tr(\mathbf{H} \boldsymbol{\Psi}^t \mathbf{P}_\alpha \boldsymbol{\Psi})\right\} \times \mathbbm{1}(\boldsymbol{\Psi} \in \mathcal{V}_{K,Q})
\end{align*}
We now introduce more notation: score matrix $\Xi \in \mathbb{R}^{K \times N}$, where each row is the score vector $\xi_i$, and residual matrix $\mathbf{R} \in \mathbb{R}^{N \times M}$, where each row is the residual vector $R_i^t$.
\begin{align*}
    f(\boldsymbol{\Psi}|{\rm others}) & \propto \exp\left[-\frac{1}{2\sigma^2}\left\{- 2 \tr(\mathbf{R} \mathbf{B}\boldsymbol{\Psi}\Xi) + \tr(\Xi^t \boldsymbol{\Psi}^t\mathbf{B}^t \mathbf{B} \boldsymbol{\Psi} \Xi)\right\} -\frac{1}{2}\tr\{\mathbf{H} \boldsymbol{\Psi}^t \mathbf{P}_\alpha \boldsymbol{\Psi}\}\right] \times \mathbbm{1}(\boldsymbol{\Psi} \in \mathcal{V}_{K,Q})\\
    & \propto \exp\left\{\frac{1}{\sigma^2}\tr(\mathbf{R} \mathbf{B} \boldsymbol{\Psi} \Xi) - \frac{1}{2\sigma^2}\tr(\Xi^t \boldsymbol{\Psi}^t \mathbf{B}^t \mathbf{B} \boldsymbol{\Psi} \Xi) - \frac{1}{2}\tr(\mathbf{H}\boldsymbol{\Psi}^t \mathbf{P}_\alpha \boldsymbol{\Psi}) \right\}\times \mathbbm{1}(\boldsymbol{\Psi} \in \mathcal{V}_{K,Q})\\
    & \propto \exp\left\{\frac{1}{\sigma^2}\tr(\Xi \mathbf{R} \mathbf{B} \boldsymbol{\Psi}) - \frac{1}{2\sigma^2}\tr(\Xi \Xi^t \boldsymbol{\Psi}^t \mathbf{B}^t \mathbf{B} \boldsymbol{\Psi}) - \frac{1}{2}\tr(\mathbf{H}\boldsymbol{\Psi}^t \mathbf{P}_\alpha \boldsymbol{\Psi}) \right\}\times \mathbbm{1}(\boldsymbol{\Psi} \in \mathcal{V}_{K,Q})\\
    & \propto \exp\left\{\tr \left( \frac{\Xi \mathbf{R} \mathbf{B} \boldsymbol{\Psi}}{\sigma^2} - \frac{\Xi \Xi^t \boldsymbol{\Psi}^t \mathbf{B}^t \mathbf{B} \boldsymbol{\Psi}}{2\sigma^2} - \frac{\mathbf{H}\boldsymbol{\Psi}^t \mathbf{P}_\alpha \boldsymbol{\Psi}}{2} \right)\right\}\times \mathbbm{1}(\boldsymbol{\Psi} \in \mathcal{V}_{K,Q})
\end{align*}
The final line above is the form found in Result~\ref{Thm:EF}, which is not of a known distributional family.

\section{Simulation results}\label{supp:Simulations}

\subsection{Background}\label{supp:simGen}
\begin{figure}[!ht]
\centering
\includegraphics[width=12cm]{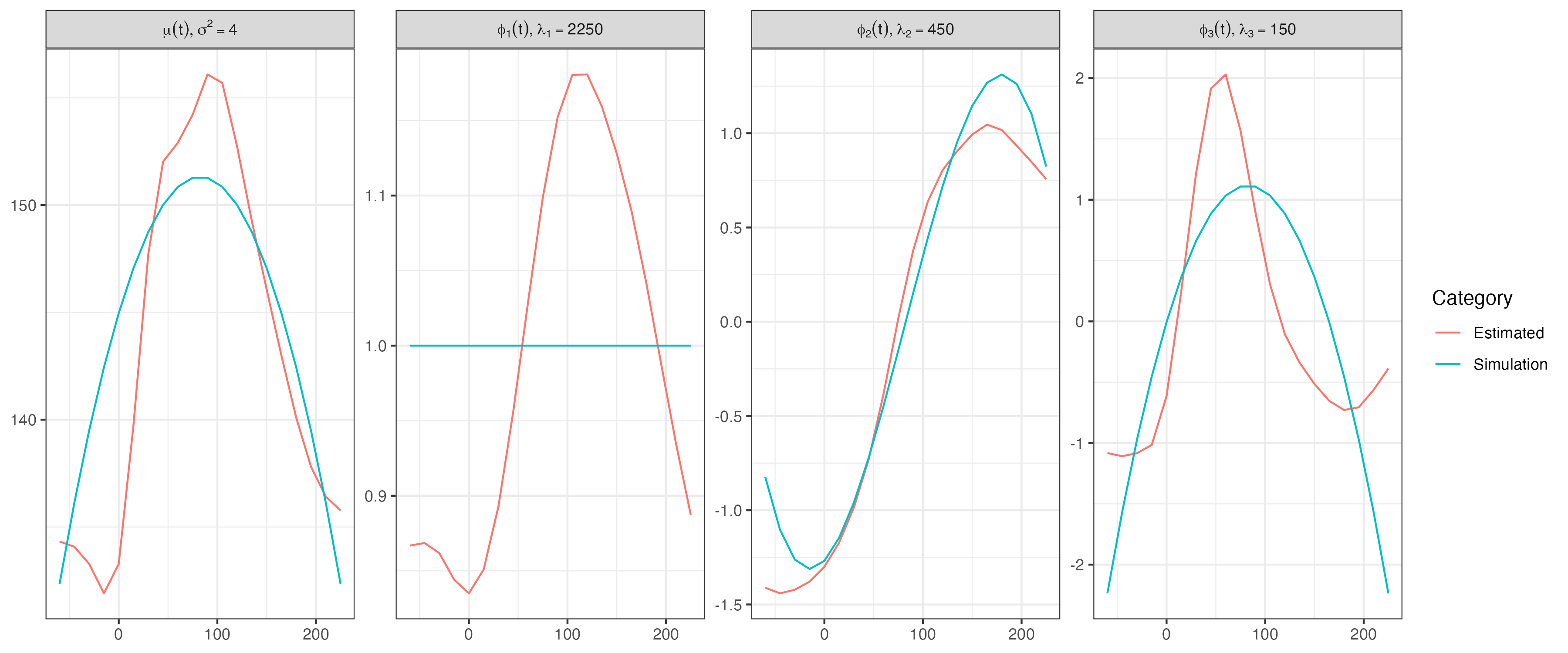}
\caption{Comparison between the generative FPCs used in simulations (``Simulation") and estimates from fitting FPCA to the CGM data (``Estimated"). The CGM data used was the by-participant mean postprandial response curves from the DASH4D study.}
\label{fig:FPCA_Sim_Source}
\end{figure}

All code used for data simulation and subsequent model fitting can be found at \href{https://github.com/JSartini/FAST-BFPCA}{this GitHub repository}. We compare the data generating model used in S1, meant to imitate the CGM data from DASH4D, to the actual FPCs estimated from the real CGM data in Figure~\ref{fig:FPCA_Sim_Source}.

While the FPCs used in our generative model for S1 (``Simulation") differ from the FPCs estimated from the mean postprandial CGM responses (``Estimated") in several non-trivial aspects, we choose them for their balance between adequate approximation and simplicity of generation. We built the ``Simulation" bases using linear combinations of the Legendre polynomials, a known orthonormal basis of $L^2([0,1])$, to enhance reproducibility.

\subsection{Fixed effects comparison}\label{supp:Sim_FE}

We compare the fixed effects estimates of $\mu(t)$ produced by FAST to those from GFSR and VMP. POLAR is not included, as it subtracts off the column means from teh data matrix in pre-processing rather than estimating the mean as part of the model. We compare ISE of the posterior estimate and aggregate coverage of equal-tailed $95\%$ credible intervals, each calculated using the same procedures outlined for the FPCs in Section~\ref{sec:simulation}.   

\begin{figure}[!ht]
\centering
\includegraphics[width=12cm]{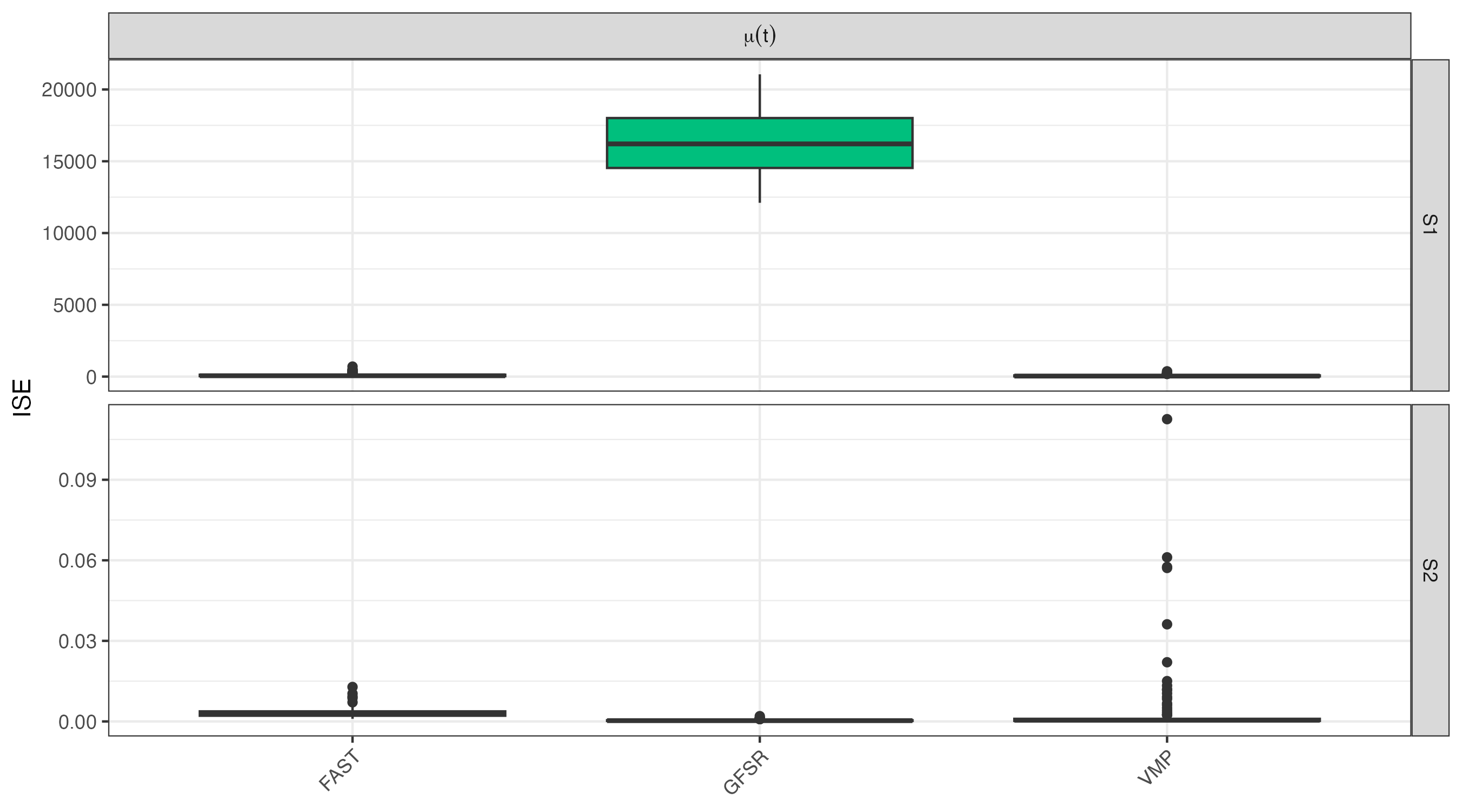}
\caption{Boxplots of ISE for the mean $\mu(t)$ by estimation method. Row indicates simulation scenario, S1 then S2.}
\label{fig:FPCA_Mu_ISE}
\end{figure}

Figure~\ref{fig:FPCA_Mu_ISE} demonstrates the ISE of $\mu(t)$ by method, where row indicates simulation scenario. FAST produces estimates of $\mu(t)$ with ISE similar or superior to those of the comparator methods. GFSR notably has substantial error for S1. Based upon further analysis of this phenomenon, we believe this is due to projection of the fixed effect onto the two FPCs $\phi_1(t), \phi_3(t)$, as described in Section~\ref{sec:simulation}.

\begin{figure}[!ht]
\centering
\includegraphics[width=12cm]{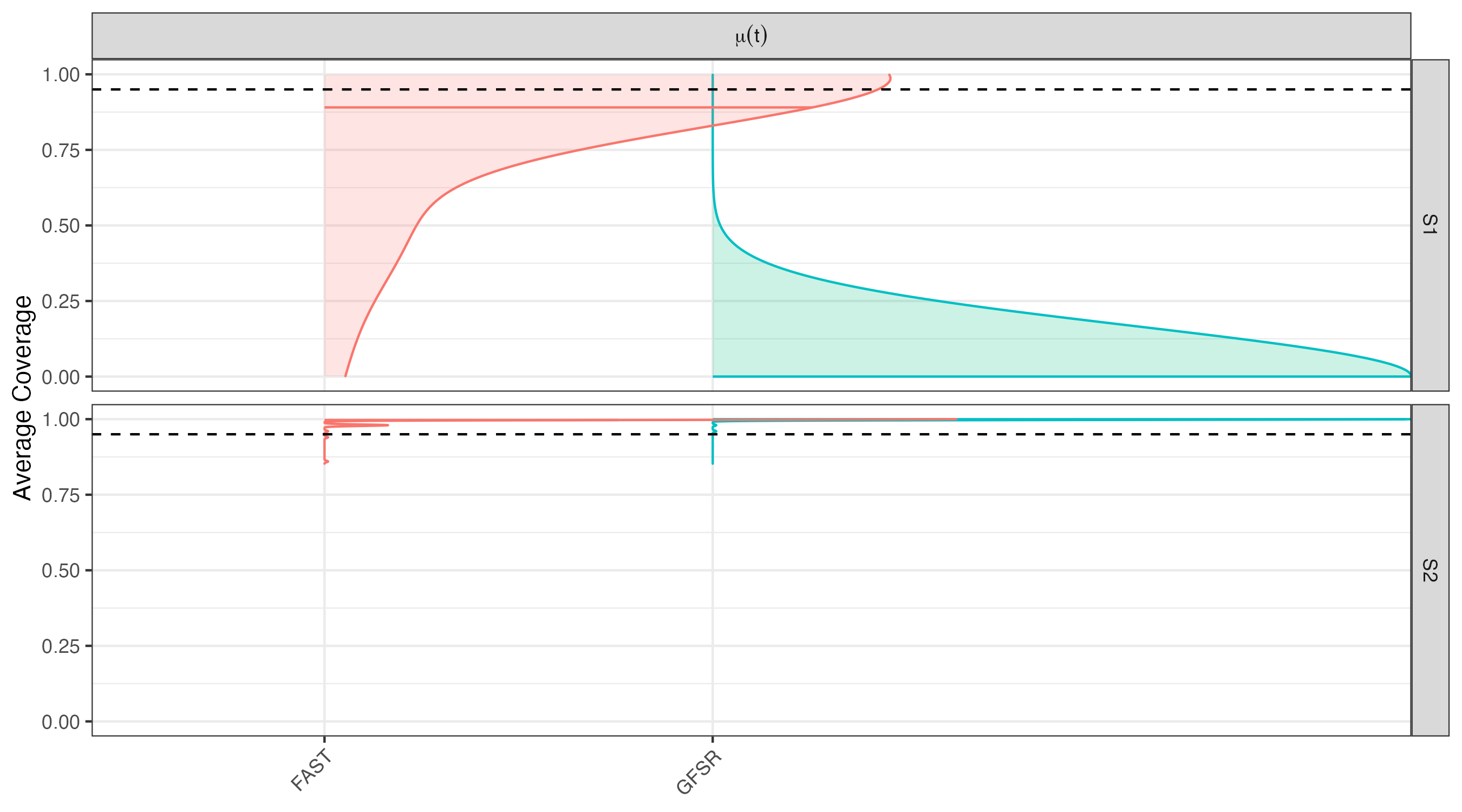}
\caption{Kernel smoother of coverage probabilities of $95$\% credible intervals of the true mean $\mu(t)$ for FAST and GFSR. Row indicates simulation scenario, S1 then S2. Distribution means: horizontal solid lines; nominal $95$\% level: horizontal dotted lines.}
\label{fig:FPCA_Mu_COV}
\end{figure}

Figure~\ref{fig:FPCA_Mu_COV} visualizes kernel smooths of equal-tailed $95\%$ credible interval coverage for FAST and GFSR, where row again indicates simulation scenario. We do not include POLAR and VMP in this comparison, as neither produces posterior inferences upon the mean function $\mu(t)$. Inference is similar between FAST and GFSR for the canonical scenario S2, but coverage is far closer to nominal for FAST in the CGM-based S1. 

\subsection{Score comparison}\label{supp:Sim_Scores}

We also compare FAST to existing Bayesian FPCA implementations in their estimation of the score $\xi_{ik}$. Metrics of interest include mean squared error (MSE) of posterior mean estimates and mean coverage of equal-tail $95\%$ credible intervals, each aggregated over curves $i$ for each combination of FPC $k$ and simulation.

\begin{figure}[!ht]
\centering
\includegraphics[width=12cm]{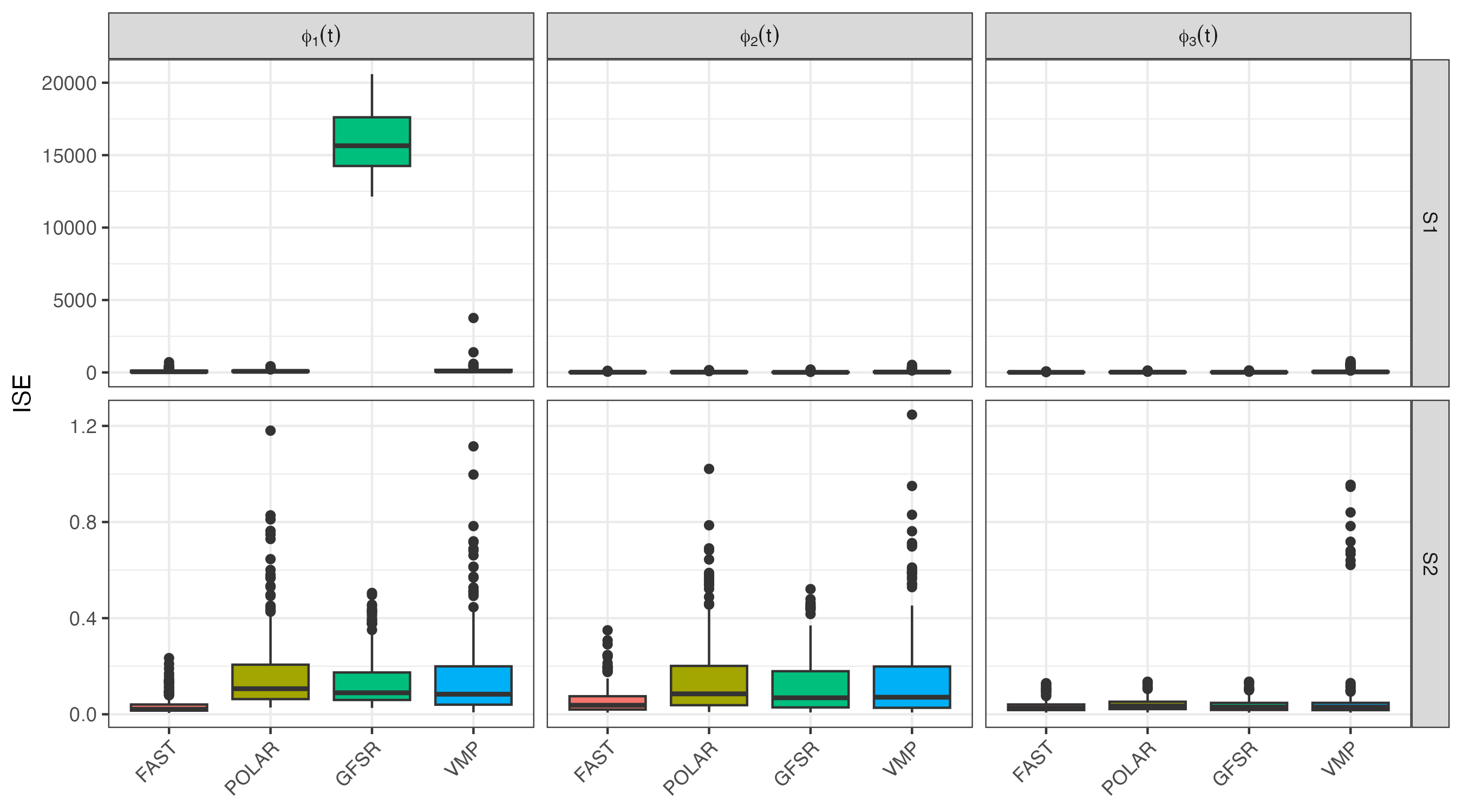}
\caption{Boxplots of MSE for the scores $\xi_{ik}$ by estimation method. Column indicates FPC, and row indicates simulation scenario, S1 then S2.}
\label{fig:FPCA_Score_ISE}
\end{figure}

Figure~\ref{fig:FPCA_Score_ISE} visualizes the distribution of score MSE over simulations. FAST is has consistently similar or lower MSE when compared to the existing methods. GFSR has much larger MSE for those scores associated with $\phi_1(t)$ in S1. The order of this MSE is close to $140^2 = 19600$, supporting our previous assertion that the mean $\mu(t)$ is being projected onto $\phi_1(t)$ and $\phi_3(t)$ by this implementation. 

\begin{figure}[!ht]
\centering
\includegraphics[width=12cm]{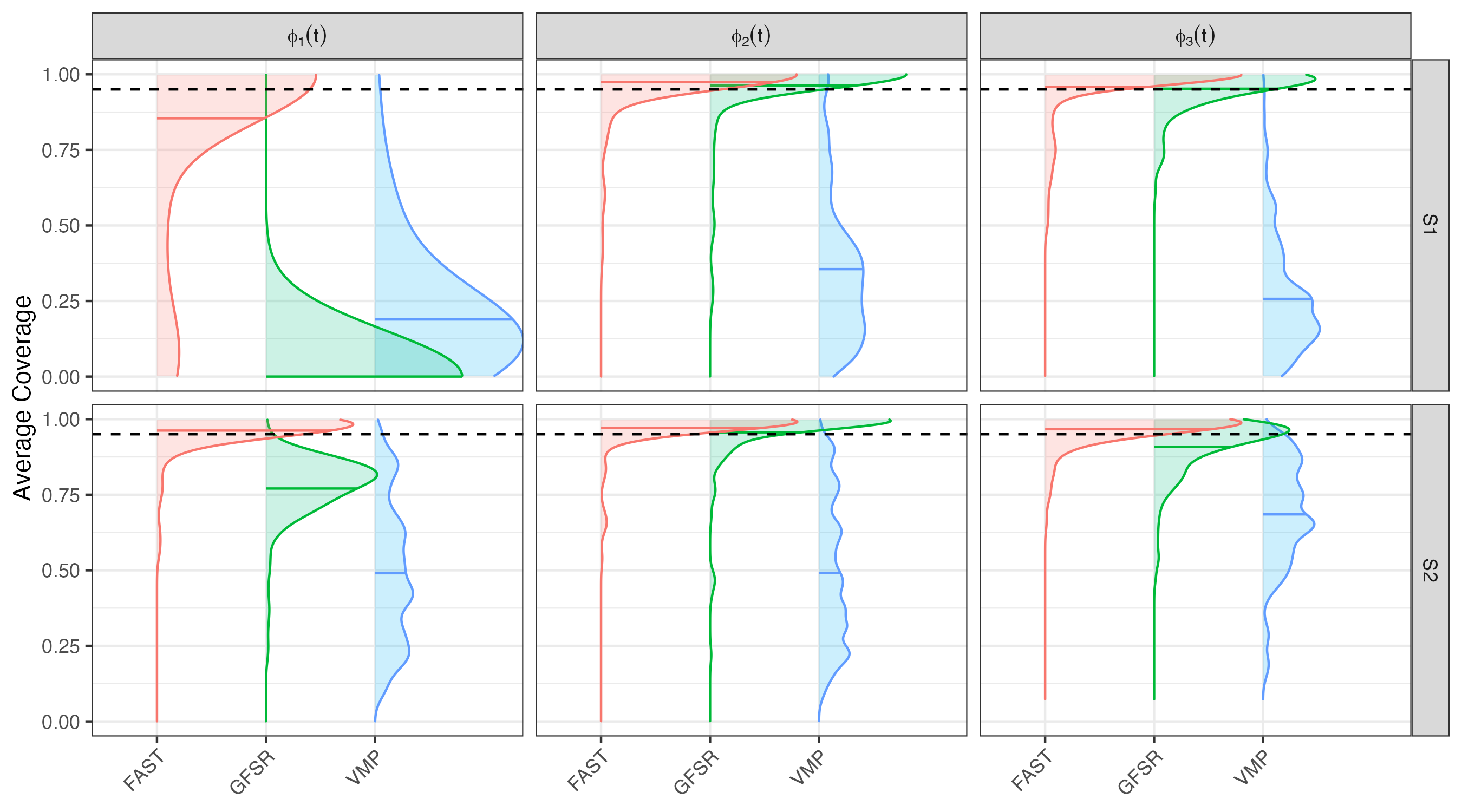}
\caption{Kernel smoother of coverage probabilities of $95$\% credible intervals of the true scores, $\xi_{ik}^b$, for FAST, GFSR, and VMP. First row: S1; second row: S2. Each column: corresponding eigenfunction. Distribution means: horizontal solid lines; nominal  $95$\% level: horizontal dotted lines.}
\label{fig:FPCA_Score_COV}
\end{figure}

Figure~\ref{fig:FPCA_Score_COV} displays the kernel smooth of the vectors of coverage proportions for FAST and relevant comparator methods by FPC (columns) and simulation scenario (rows). FAST is the only implementation which produces consistently close to nominal coverage for all FPCs under both scenarios.

\subsection{Multilevel extension}\label{supp:simMulti}

We perform a simple 2-level simulation study to assess performance of the multilevel extension of FAST in comparison to GFSR. This simulation includes 50 groups at the first level ($N = 50$), each with 5 functional observations ($\forall i$, $J_i = 5$) for a total of 250 functions. We are only able to compare against GFSR due to both POLAR and VMP having no available multilevel extension. We use a well-known and frequently used multilevel simulation scenario for our study, where the FPCs are orthogonal within level but not between. All explicit details are provided below \citep{cui_fast_2023}. 

\textit{Multilevel Scenario}
\begin{align*}
    \phi_k^{(1)}(t) & = \{\sqrt{2}\sin (2\pi t), \sqrt{2}\cos(2\pi t), \sqrt{2}\sin(4 \pi t), \sqrt{2} \cos(4 \pi t)\}; \quad \lambda^{(1)}_k = 0.5^{k - 1}\\
    \phi_l^{(2)}(t) & = \{1, \sqrt{3}(2t - 1), \sqrt{5}(6t^2 - 6t + 1), \sqrt{7}(20t^3 - 30t^2 + 12t - 1)\}; \quad \lambda^{(2)}_l = 0.5^{l - 1}\\
    \mu(t) & = 0; \quad \sigma^2 = 1
\end{align*}

We first present the integrated squared error (ISE) of the FPCs $\phi_k^{(1)}(t), \phi_l^{(2)}(t)$. We calculate this measure using the same procedure detailed in Section~\ref{sec:simulation}. Each panel corresponds to one of the eigenfunctions, with column indicating index and row indicating level.

\begin{figure}[!ht]
\centering
\includegraphics[width=12cm]{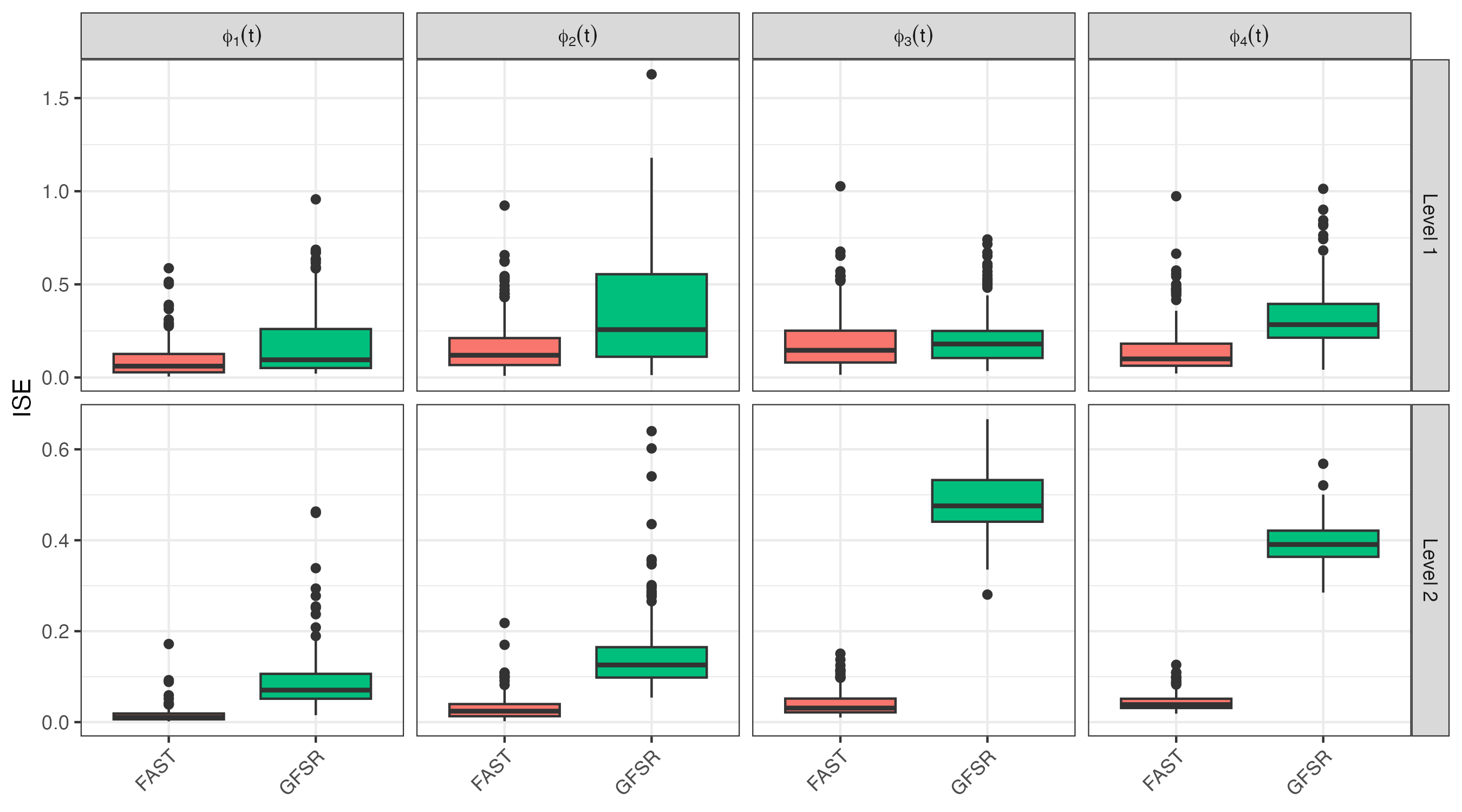}
\caption{Boxplots of FPC ISE for FAST and GFSR from the multilevel simulations. Row indicates hierarchical level, while columns correspond to FPCs.}
\label{fig:MFPCA_FPC_ISE}
\end{figure}

Figure~\ref{fig:MFPCA_FPC_ISE} indicates uniformly lower ISE for FAST as compared to GFSR. These differences can be extreme, see for example the third and fourth FPCs at the visit level, $\phi_3^{(2)}(t), \phi_4^{(2)}(t)$.

\begin{figure}[!ht]
\centering
\includegraphics[width=12cm]{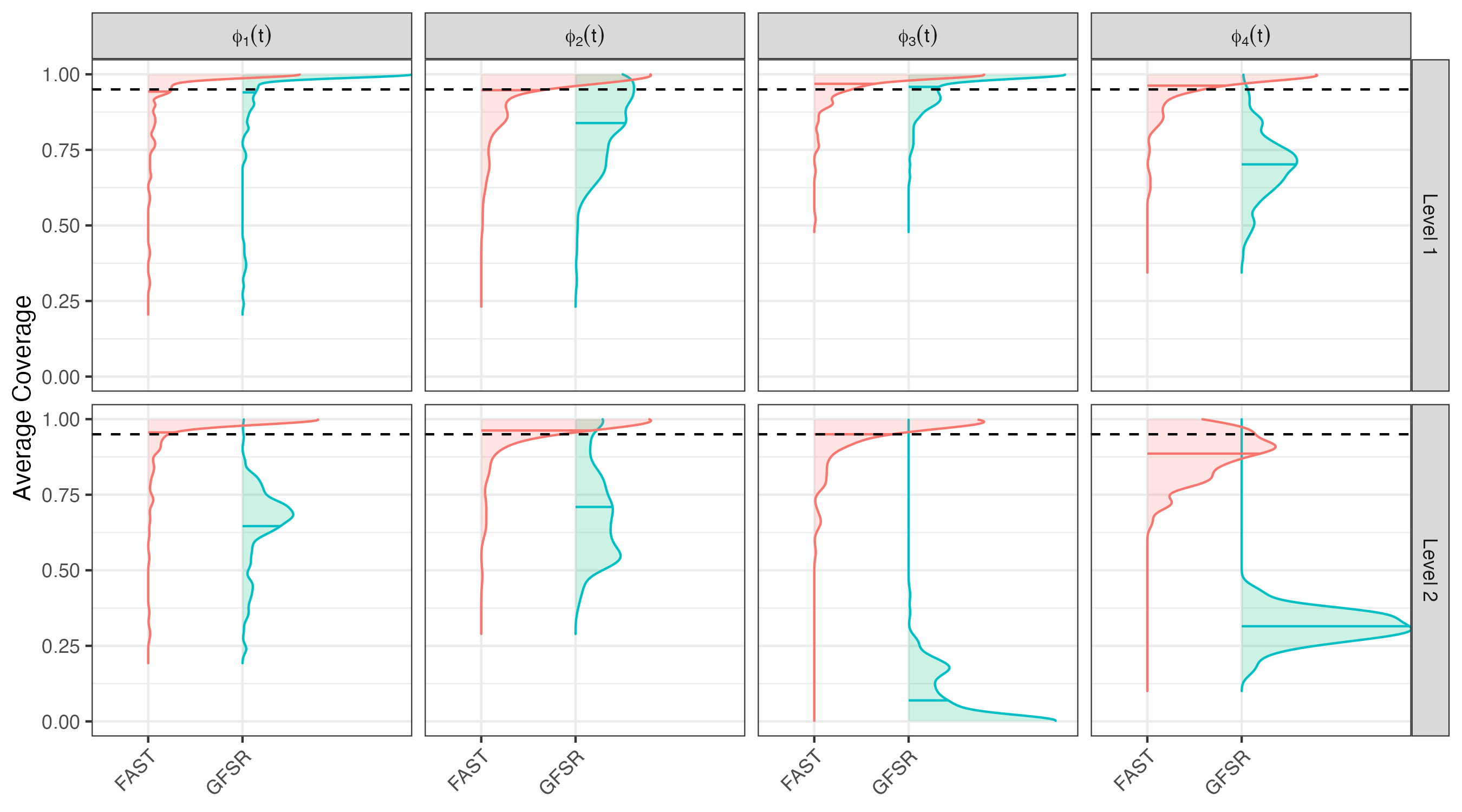}
\caption{Kernel smoother of coverage probabilities of $95$\% credible intervals of the true FPCs for FAST and GFSR from the multilevel simulations. Row indicates hierarchical level, while columns correspond to FPCs. The horizontal dotted lines indicate the nominal  $95$\% level.}
\label{fig:MFPCA_FPC_COV}
\end{figure}

We next consider FPC coverage according to the point-wise equal-tailed 95\% credible intervals produced by FAST and GFSR. As in Section~\ref{sec:simulation}, we estimate coverage probability using the proportion of time points which are covered for each simulated dataset, visualizing the result with a kernel smooth in Figure~\ref{fig:MFPCA_FPC_COV}. We include on this Figure the mean of each coverage distribution (horizontal solid lines) and the nominal coverage 95\% coverage level (horizontal dotted lines).

Similar to the single-level simulations, we find that FAST produces nearly nominal coverage for all FPCs. In comparison, GFSR does not cover the eigenfunctions at the visit level well. For these FPCs, mean coverage ranges from $\approx 0.15$ to $\approx 0.7$. 

\begin{figure}[!ht]
\centering
\includegraphics[width=12cm]{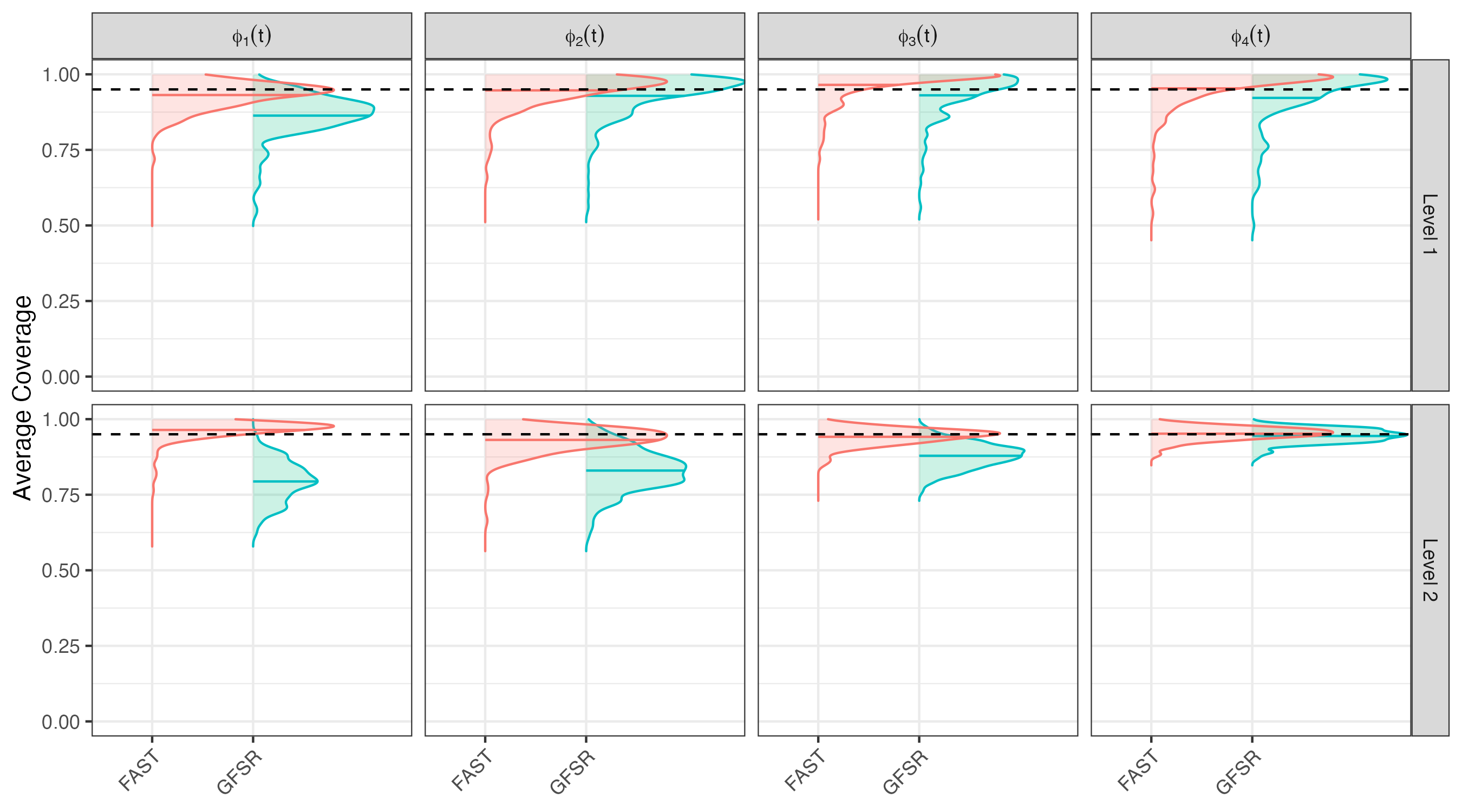}
\caption{Kernel smoother of coverage probabilities of $95$\% credible intervals of the true scores, $\xi_{ik}^b, \zeta_{ijk}^b$, for FAST and GFSR from the multilevel simulations. Row indicates hierarchical level, while columns correspond to FPCs. The horizontal dotted lines indicate the nominal  $95$\% level.}
\label{fig:MFPCA_Score_COV}
\end{figure}

This disparity in coverage extends to the scores at both levels, $\xi_{ik}$ and $\zeta_{ijk}$, as can be observed in Figure~\ref{fig:MFPCA_Score_COV}. We estimate score coverage in the same fashion illustrated in Section~\ref{sec:simulation}. For each simulated dataset $b \leq B$, we calculate the 95\% credible intervals for each individual $\xi_{ik}^b, \zeta_{ijk}^b$ from the posterior samples. We then estimate the coverage probability by aggregating the coverage indicators by the corresponding eigenfunction. Figure~\ref{fig:MFPCA_Score_COV} visualizes the kernel smooths of the corresponding vectors of coverage, complete with distribution mean (solid horizontal lines) and nominal level (dashed horizontal lines). As was the case for the functional components presented in Figure~\ref{fig:MFPCA_FPC_COV}, nominal coverage of the scores is uniquely achieved by FAST. GFSR achieves lower mean coverages, particularly for the FPCs $\phi_1(t)$ at each level.

\section{Simulation sensitivity analyses}

For each sensitivity analysis, we vary one of the core hyper-parameters from the set $K$ (number of FPCs), $Q$ (dimension of the spline basis), and $\alpha$ (proportion of the penalty which is absolute rather than second order). The default values are $K = 3$ (the true number of FPCs), $Q = 20$, and $\alpha = 0.1$. All sensitivity analyses are conducted within simulation scenarios S1 and S2 described in Section~\ref{sec:simulation}. We finally evaluate the effect of each of these sensitivity analyses on the computational efficiency of FAST in Section~\ref{supp:timing_sensitivity}. 

\subsection{Chosen number of FPCs $K$} \label{supp:Mis_K}

We first vary the fixed number of FPCs $K$ which FAST estimates. The true value is $K = 3$ for scenarios S1 and S2, so we vary between $K = 2$ and $K = 6$ to understand the effects of insufficient and excessive $K$. We evaluate FAST under each $K$ value using ISE and coverage of the FPCs as described in Section~\ref{sec:simulation}.

\begin{figure}[!ht]
\centering
\includegraphics[width=12cm]{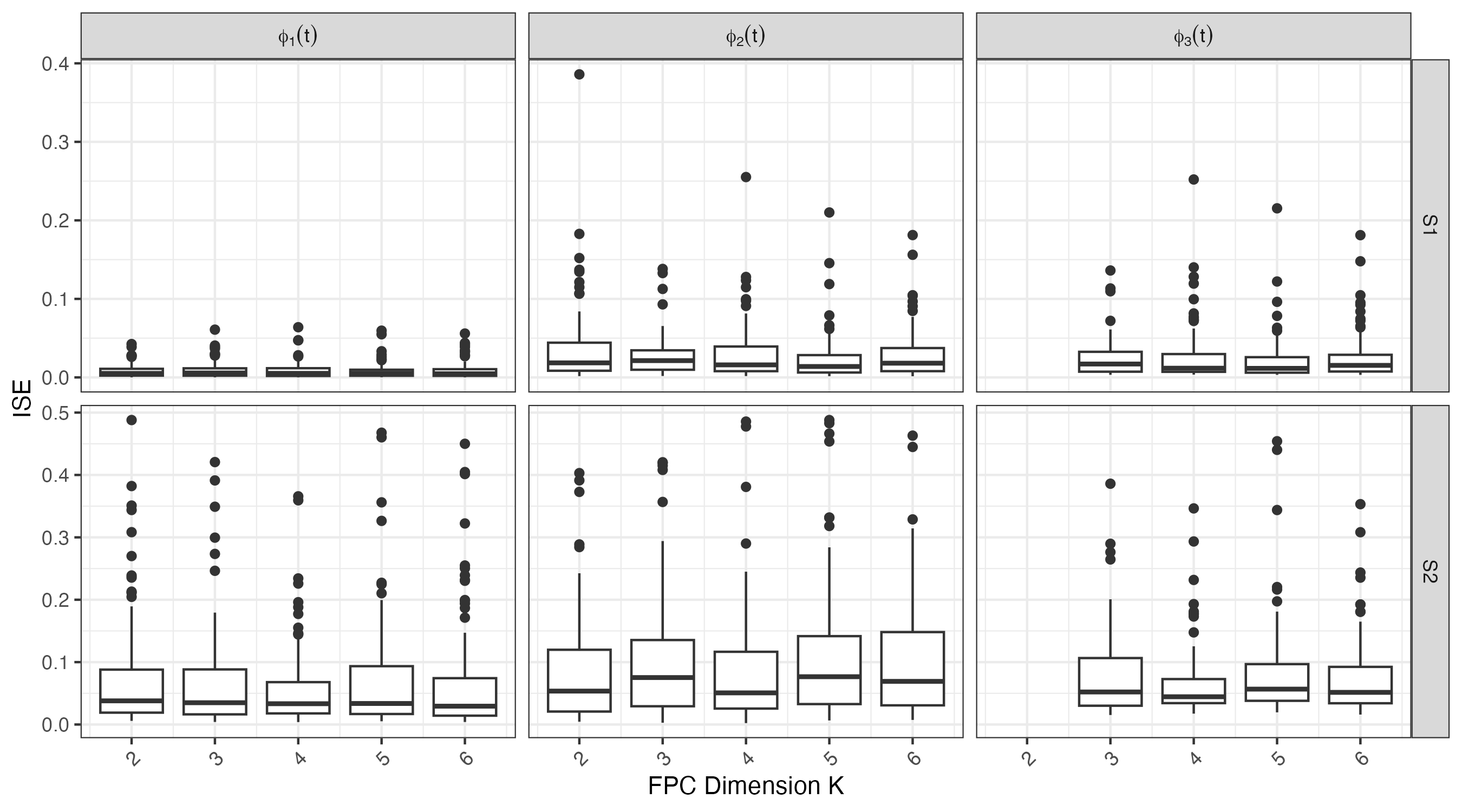}
\caption{Boxplots of FPC ISE from applying FAST by number of chosen FPCs $K$. Columns indicate FPC, and rows correspond to simulation scenario.}
\label{fig:K_ISE}
\end{figure}

Figure~\ref{fig:K_ISE} indicates that the value of $K$ does not appear to have a substantial impact on the accuracy of FPC estimates, outside of not estimating FPCs when they are present. When $K = 2$, the first two FPCs are still well estimated, and all 3 true FPCs are recovered with qualitatively similar accuracy whenever $K \geq 3$.

\begin{figure}[!ht]
\centering
\includegraphics[width=12cm]{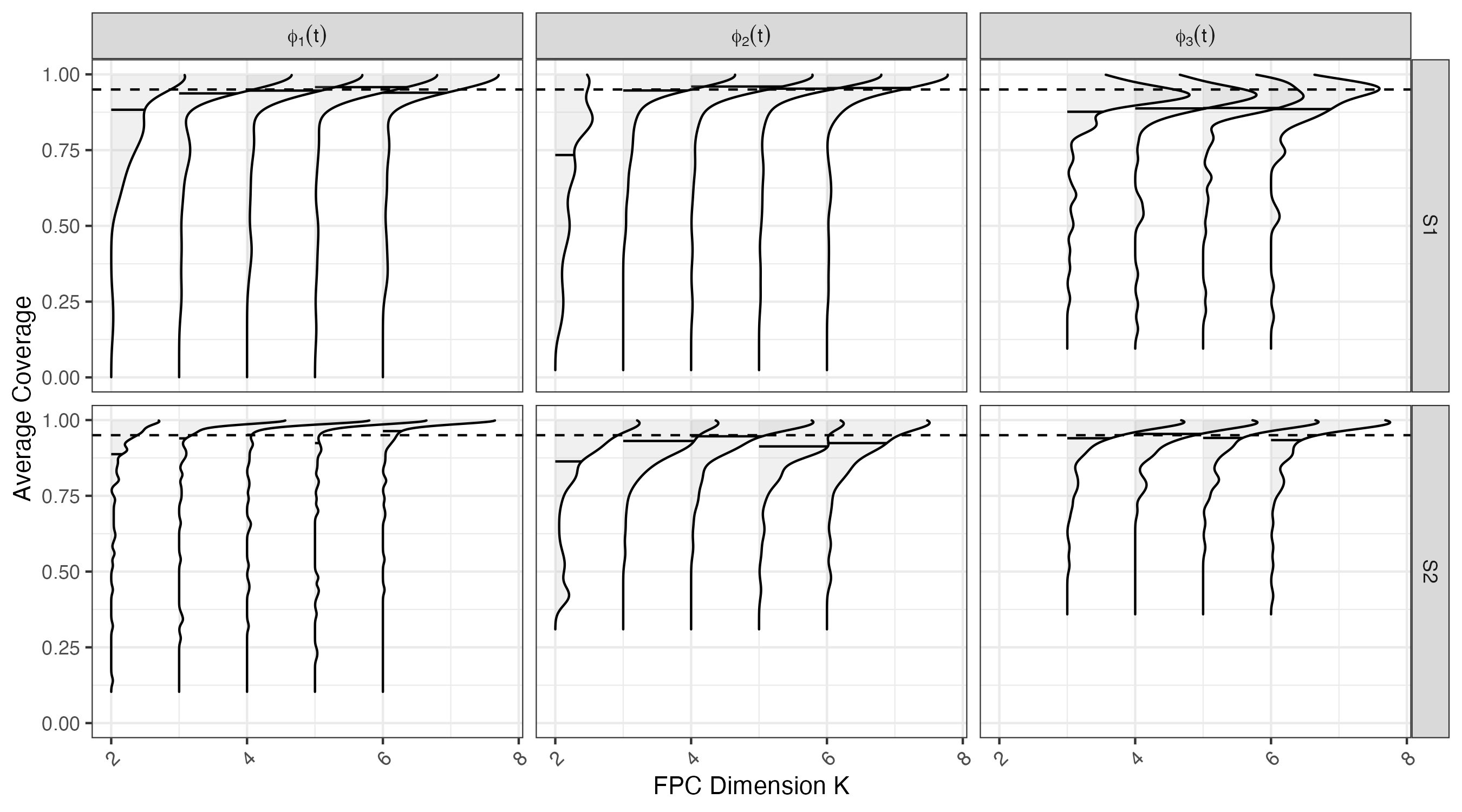}
\caption{Kernel smooths of true FPC $95\%$ credible interval coverage probabilities for FAST by number of chosen FPCs $K$. Columns indicate FPC, and rows correspond to simulation scenario.}
\label{fig:K_COV}
\end{figure}

Figure~\ref{fig:K_COV} shows that $K = 2$ produces non-optimal inference for $\phi_1(t), \phi_2(t)$, with inferences for the first 3 FPCs unaffected when $K \geq 3$.

\subsection{Chosen spline dimension $Q$}\label{supp:Q_Choice} 

We next vary the spline basis dimension $Q$ used by FAST to estimate the functional components of FPCA. We vary between $Q = 5$ and $Q = 40$ to understand the effects of both having a rather restrictive basis as well as having one which is very rich. We evaluate FAST under each $Q$ value using ISE and coverage of the FPCs as described in Section~\ref{sec:simulation}.

\begin{figure}[!ht]
\centering
\includegraphics[width=12cm]{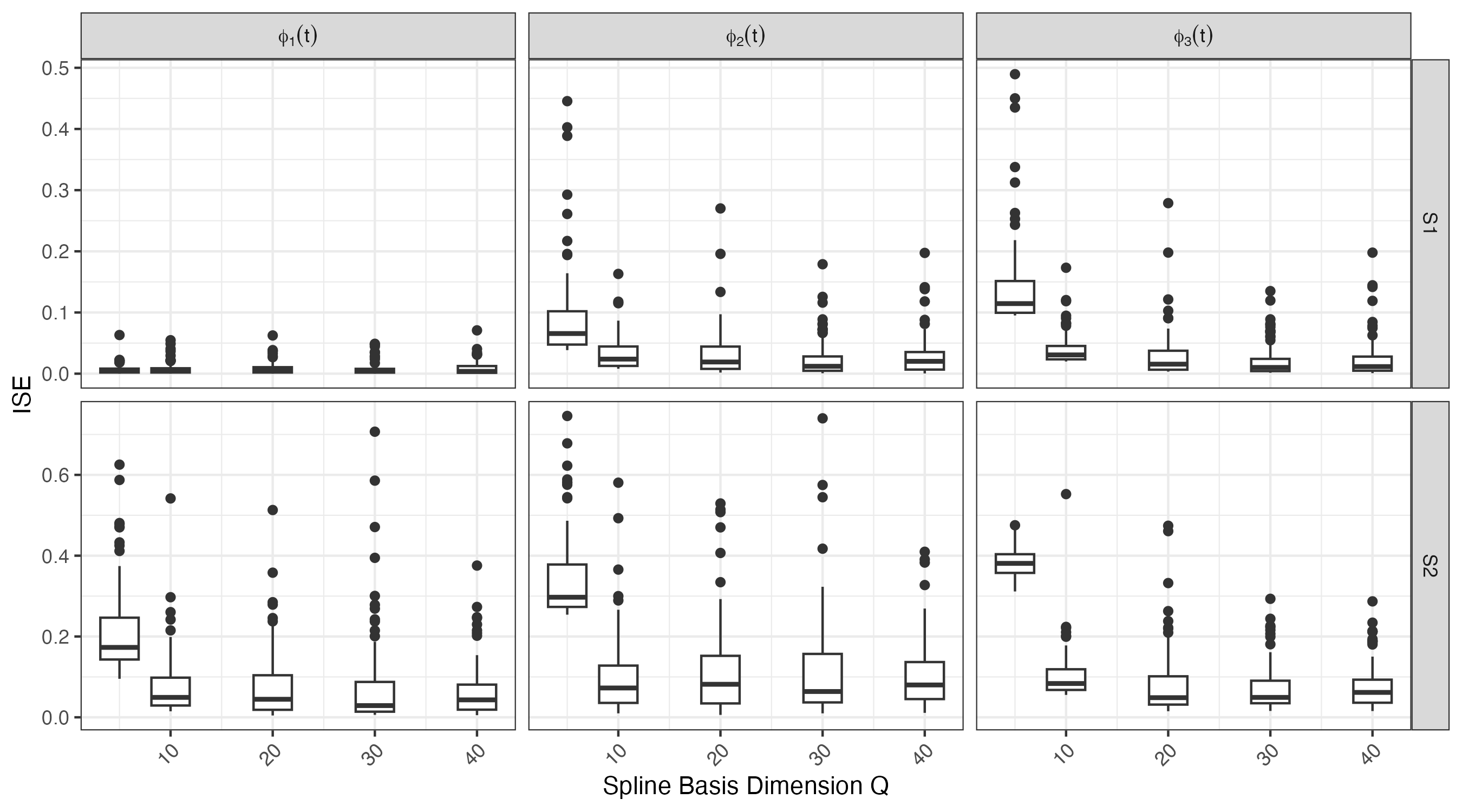}
\caption{Boxplots of FPC ISE from applying FAST by spline basis dimension $Q$. Columns indicate FPC, and rows correspond to simulation scenario.}
\label{fig:Q_ISE}
\end{figure}

Figure~\ref{fig:Q_ISE} indicates setting $Q = 5$ produces biased estimates of the FPCs, likely due to lack of flexibility in the basis. However, the FPC ISE uniformly reduces and stabilizes by $Q = 20$ across both simulation scenarios.

\begin{figure}[!ht]
\centering
\includegraphics[width=12cm]{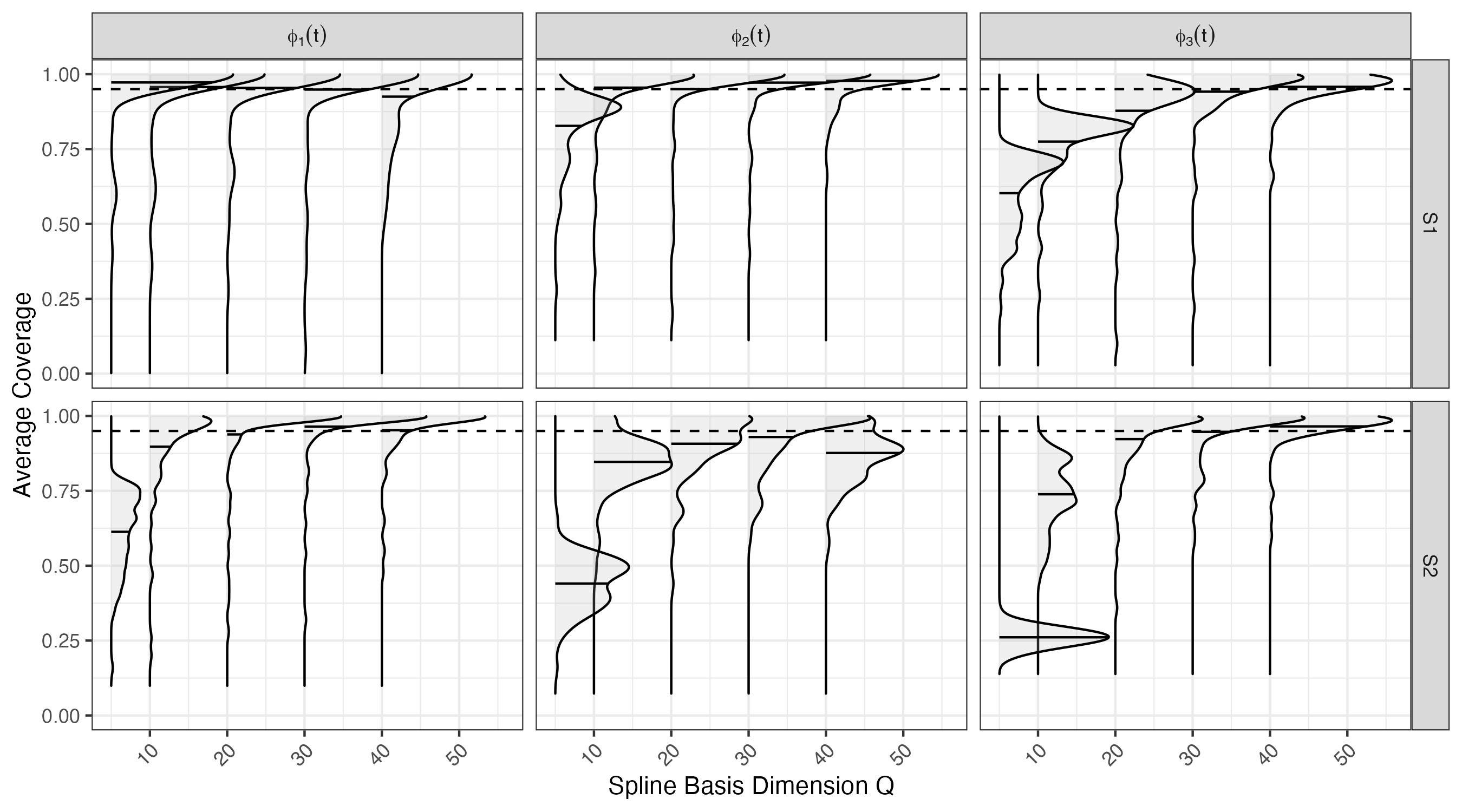}
\caption{Kernel smooths of true FPC $95\%$ credible interval coverage probabilities for FAST by spline basis dimension $Q$. Columns indicate FPC, and rows correspond to simulation scenario.}
\label{fig:Q_COV}
\end{figure}

Figure~\ref{fig:Q_COV} similarly shows that $Q < 20$ produces non-optimal inference, with coverage stabilizing at near nominal levels for $Q \geq 20$.

\subsection{Chosen penalty parameter $\alpha$}\label{supp:Alpha_Choice}

We finally vary penalty hyperparameter $\alpha$, which dictates what proportion of the smoothing penalty used by FAST will be allocated to absolute value rather than second-order variation. We vary between $\alpha = 0.01$ and $\alpha = 0.3$, keeping $\alpha< 0.5$ to ensure the majority of the penalty is focused on the "wiggliness" which should be the core of any smoothing penalty. We evaluate FAST under each $\alpha$ value using ISE and coverage of the FPCs as described in Section~\ref{sec:simulation}.

\begin{figure}[!ht]
\centering
\includegraphics[width=12cm]{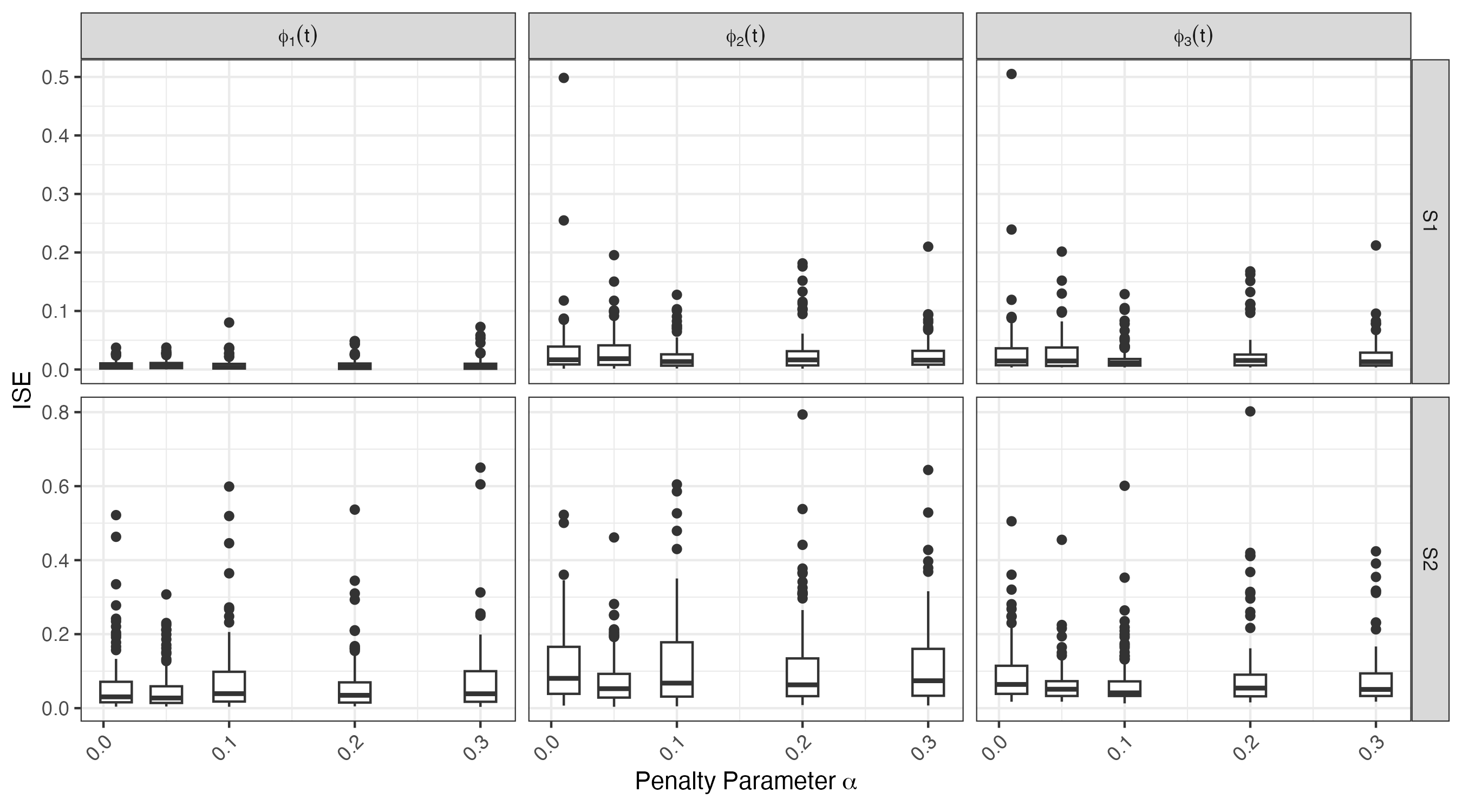}
\caption{Boxplots of FPC ISE from applying FAST by penalty parameter $\alpha$. Columns indicate FPC, and rows correspond to simulation scenario.}
\label{fig:Alpha_ISE}
\end{figure}

Figure~\ref{fig:Alpha_ISE} indicates that estimation accuracy, evaluated using ISE between the estimated and true FPCs, is consistent across the range of $\alpha$ values tested.

\begin{figure}[!ht]
\centering
\includegraphics[width=12cm]{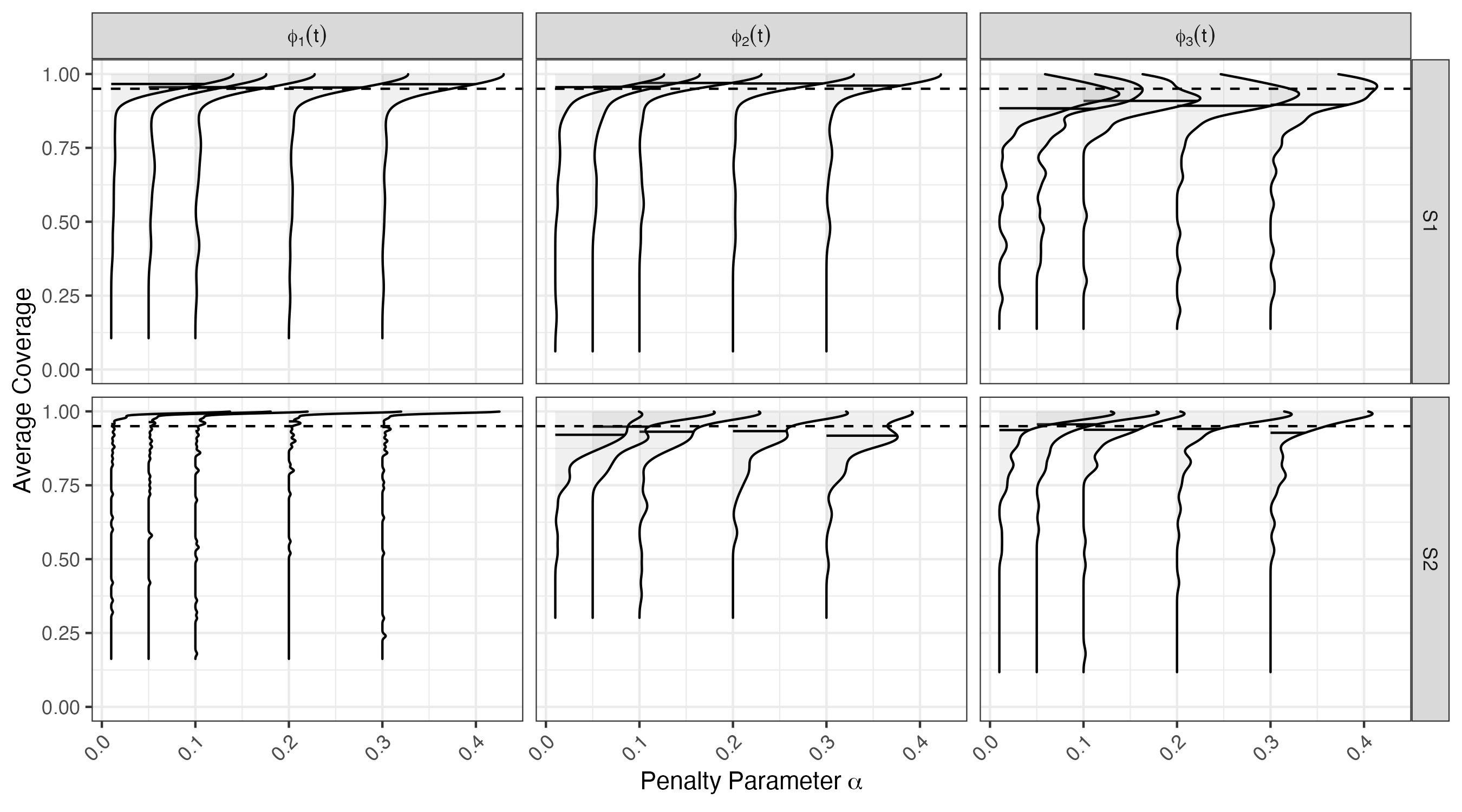}
\caption{Kernel smooths of true FPC $95\%$ credible interval coverage probabilities for FAST by penalty parameter $\alpha$. Columns indicate FPC, and rows correspond to simulation scenario.}
\label{fig:Alpha_COV}
\end{figure}

Figure~\ref{fig:Alpha_COV} indicates that inference,  evaluated using point-wise coverage of the true FPCs by equal-tailed 95\% credible intervals, is also consistent across the range of $\alpha$ values tested.

\subsection{Timing by $Q, K$}\label{supp:timing_sensitivity}

Given that it appears that estimation accuracy and inference are consistent when $K$ and $Q$ are sufficiently large, we evaluate the effect of increasing these quantities on the computation time of FAST. For this experiment, we focus on S1, the more computationally complex simulation scenario based upon CGM data. For each combination of $K \in \{3,4,5\}$ and $Q \in \{20, 30, 40\}$, we timed FAST on the same personal laptop described in Section~\ref{sec:simulation} (2023 MacBook Pro with Apple M2 Max$@$3.49 GHz and 32GB of memory).

\begin{table}[!ht]
\centering
\begin{tabular}{c|c|c|c}
 \multicolumn{1}{c|}{$K/Q$} & $Q = 20$ & $Q = 30$ & $Q = 40$ \\
 \hline
 $K = 3$ & 2.4 & 2.4 & 3.6 \\ 
 \hline
 $K = 4$ & 2.2 & 2.1 & 2.7 \\ 
 \hline
 $K = 5$ & 1.9 & 2.0 & 2.5
\end{tabular}
\caption{Table of FAST BayesFPCA computation times (in minutes) for each combination of spline dimension $Q$ and number of FPCs $K$. We fix the simulation to S1, number of time series to $N = 50$, and number of observations along the domain to $M = 50$.}
\label{tab:CompTime_Sense}
\end{table}

From Table~\ref{tab:CompTime_Sense}, we find that FAST does not require appreciable additional computation time when $Q,K$ are increased.

\section{Additional CGM Analyses/Details}

\subsection{Variability explained - Single Level}

\begin{figure}[!ht]
\centering
\includegraphics[width=12cm]{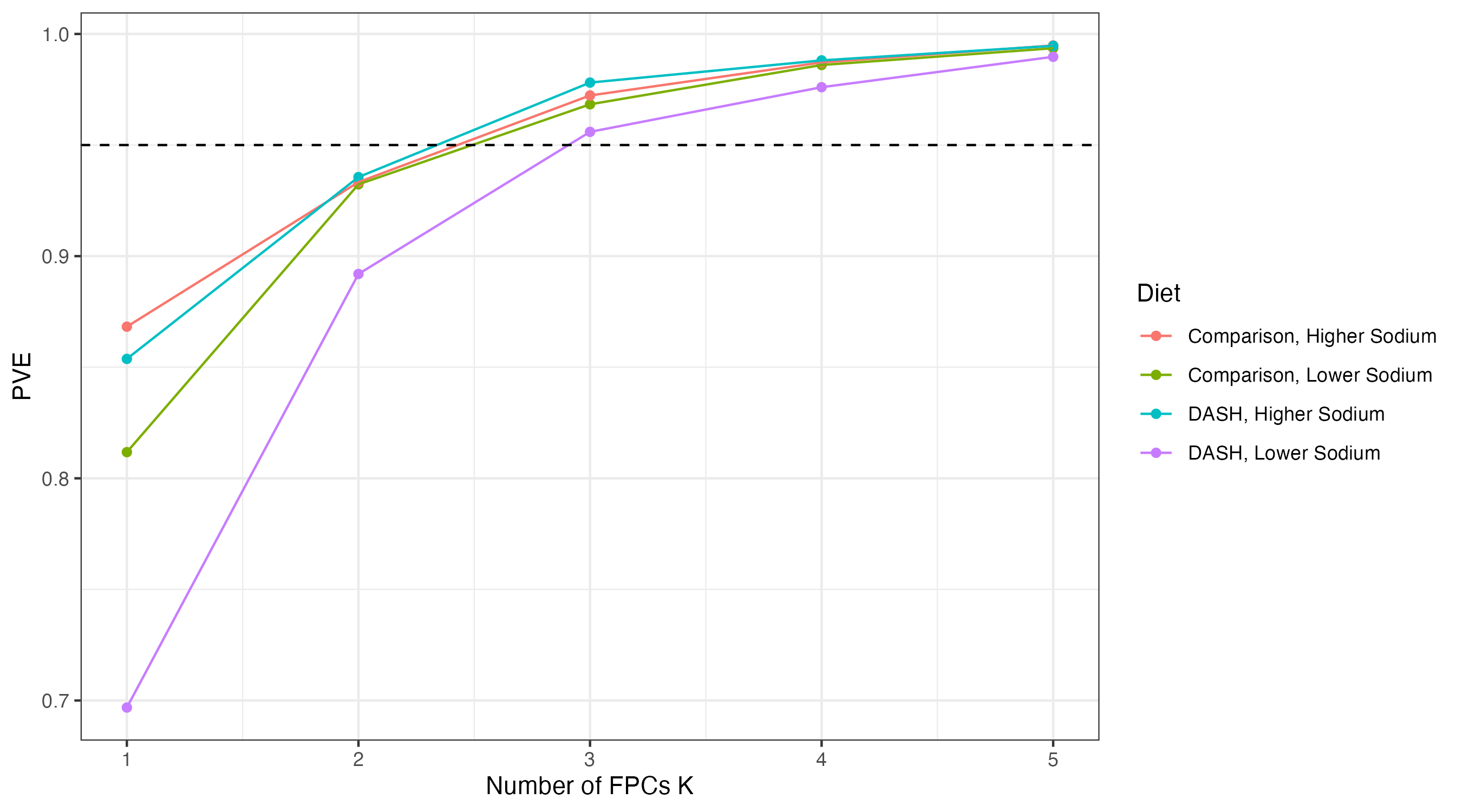}
\caption{Variability in the single-level mean CGM curve data explained by FPCA approximation over a range of $K$ values, stratified by diet.}
\label{supp:K_PVE}
\end{figure}

We plot the variability explained versus the number of FPCs used, $K$, for each of the 4 sub-diets in Figure~\ref{supp:K_PVE}. These values are calculated using frequentist FACE estimates of FPCA for each $K$ value.

Figure~\ref{supp:K_PVE} indicates that, for all diets, $K = 3$ principal components explains more than 95\% of the variability.

\subsection{Bayesian FPCA of randomly-chosen CGM within diets}\label{supp:Random_FPCA_Data}

\begin{figure}[!ht]
\centering
\includegraphics[width=12cm]{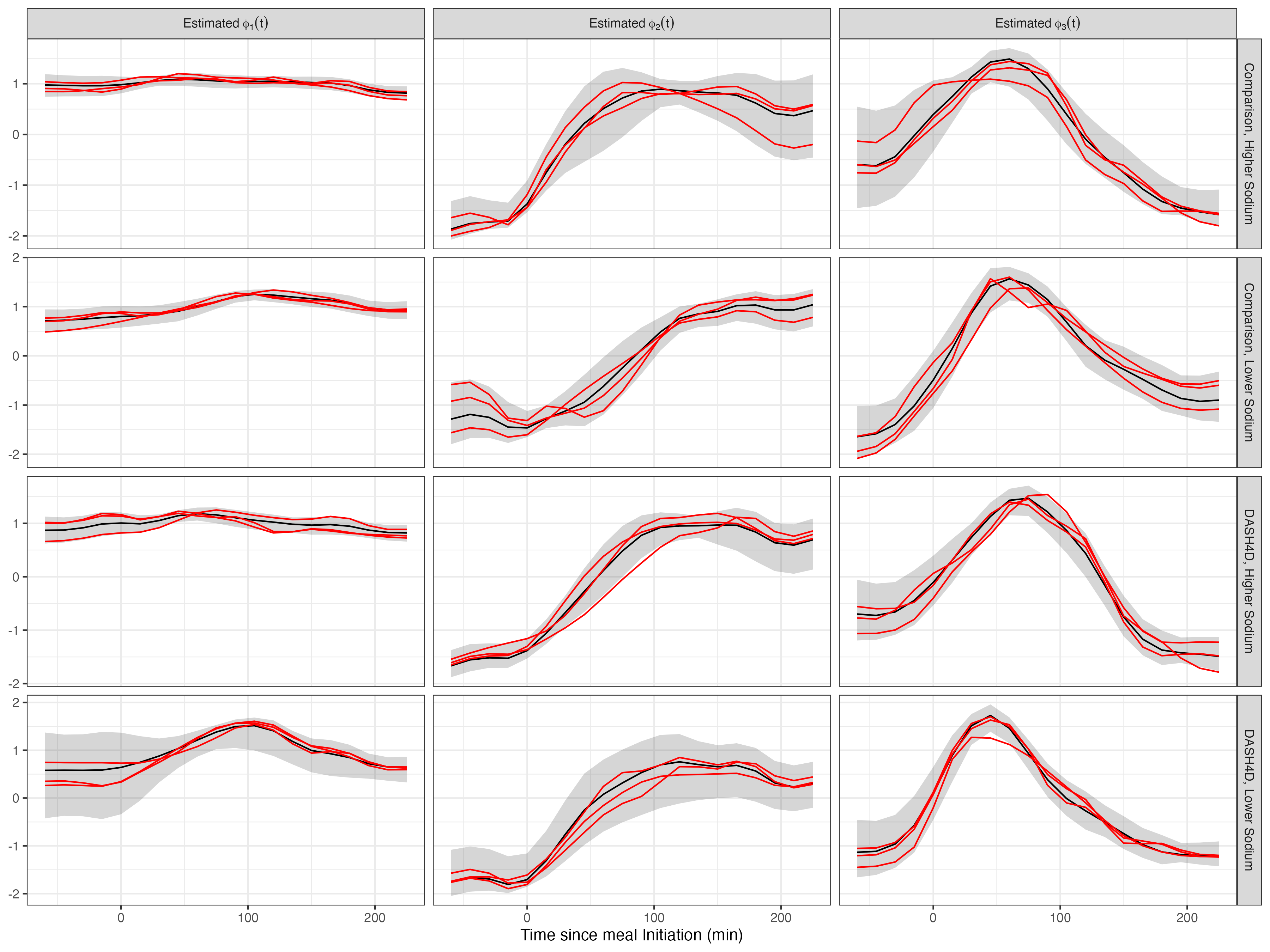}
\caption{Bayesian FPCA results for the first three PCs (each column corresponds to one FPC) for each of the four diets (each row corresponds to one diet). X-axis: time from the start of the meal. Black curves: posterior mean; red curves: three samples from the posterior of the PCs; shaded areas: pointwise $95$\% credible intervals.}
\label{fig:EF_SL_Sample}
\end{figure}

As an alternative single-level analysis, we randomly sample a function for each participant within each diet rather than aggregating. These functions each represent a single instantiation of the meal process for that participant within the particular diet, whereas the aggregate functions analyzed in Section~\ref{subsec:Bayes_FPCA_data} have varying levels of noise based upon the number of curves included in the average. However, when performing the by-diet FPCA analyses on these randomly chosen curves instead of the aggregated CGM, we observe qualitatively identical results. We first present the eigenfunctions with associated uncertainty in Figure~\ref{fig:EF_SL_Sample}. The primary difference we observe is in the additional sampling variability of $\widehat{\phi}_1(t)$ for the DASH, Lower Sodium diet, along with some of the posterior samples having greater curvature. This is indicative of differences in mean mealtime glucose being much lower for this population, reduced to be on similar scale to differences peakedness.

\begin{figure}[!ht]
\centering
\includegraphics[width=12cm]{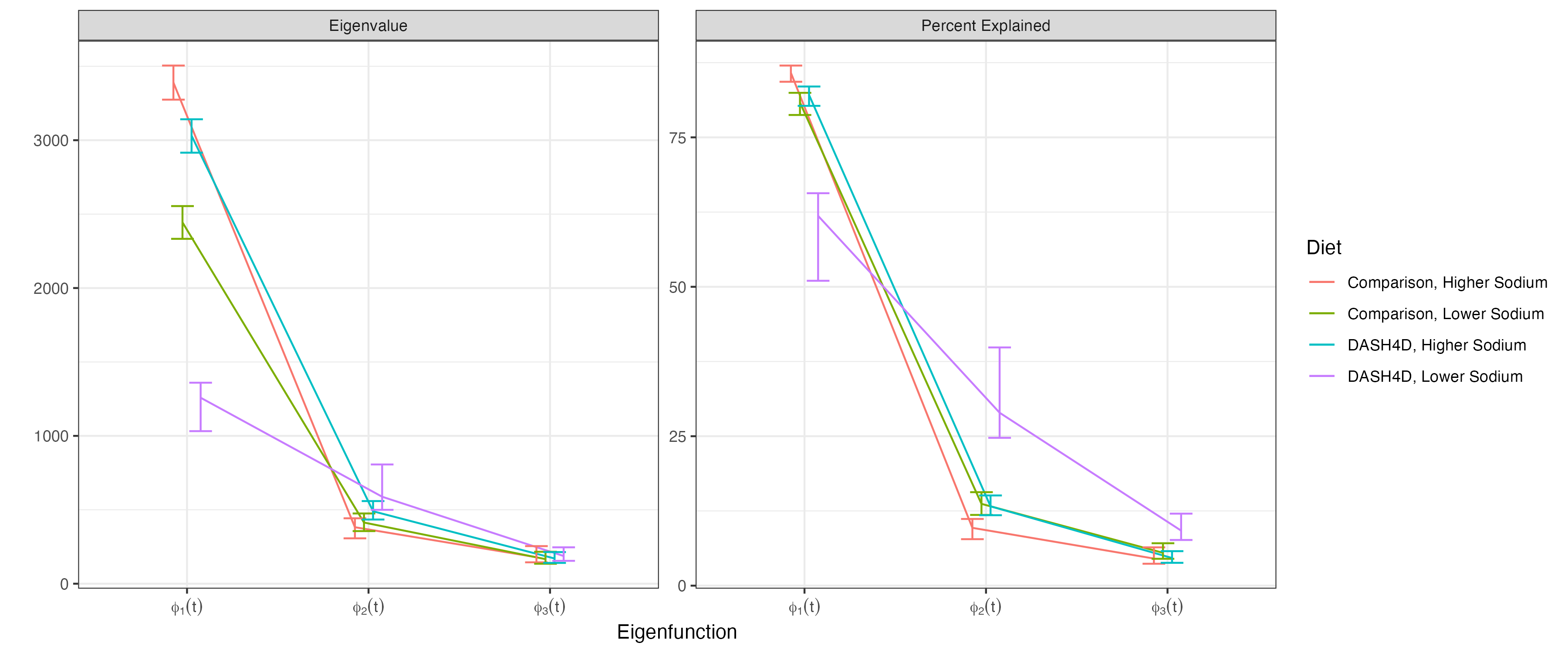}
\caption{Eigenvalue and percent-variability estimates from the Bayesian FPCA models fit to each of the four diets (each line and color corresponds to one diet). X-axis: eigenfunction corresponding to eigenvalue/percent variance explained. All estimates are presented with their corresponding $95$\% credible intervals.}
\label{fig:EV_SL_Sample}
\end{figure}

We present also the corresponding eigenvalues in Figure~\ref{fig:EV_SL_Sample}, where again one observes the unique properties of the DASH, Lower Sodium diet. In this analysis as well, there is lower overall variability for this diet (assessed by sum of eigenvalues). This difference is again driven by the first eigenvalue/eigenfunction pair in this analysis.

As for the aggregate data, we fit each model using $2000$ iterations with the first $1000$ discarded as burn-in. Employing the routine from Section~\ref{subsec:practical} to assess convergence of the scores and FPCs, we found final Gelman-Rubin statistics were $< 1.05$.

\end{document}